\documentclass{article}
\pdfoutput=1
\usepackage[affil-it]{authblk}
\usepackage{graphicx}
\usepackage{microtype}
\usepackage{amsmath,amssymb,amstext}
\usepackage{newtxtext,newtxmath}
\usepackage{enumitem}
\usepackage{subcaption}
\usepackage[square,comma,numbers,sort&compress]{natbib}
\usepackage[usenames,dvipsnames]{xcolor}
\graphicspath{{./}{./img/}}
\providecommand{\Vector}[1]{\boldsymbol{#1}}
\providecommand{\unitVector}[1]{\boldsymbol{\mathbf{#1}}}
\providecommand{\Tensor}[1]{\boldsymbol{\mathsf{#1}}}
\providecommand{\unitTensor}[1]{\boldsymbol{{\mathsf{#1}}}}
\providecommand{\Matrix}[1]{\boldsymbol{\mathbf{#1}}}
\providecommand{\Unit}[1]{\,\mathrm{#1}}
\providecommand{\Des}[1]{\mathrm{#1}}
\providecommand{\Loptr}{\operatorname{\mathcal{L}}}
\providecommand{\Toptr}{\operatorname{\mathcal{LL}}}
\providecommand{\Soptr}{\operatorname{\mathcal{S}}}
\begin{document}

\title{A simple field function for solving complex and dynamic fluid-solid system on Cartesian grid}

\author[1]{Huangrui Mo%
\thanks{Email: \texttt{huangrui.mo@uwaterloo.ca}}}

\author[1]{Fue-Sang Lien}

\author[2]{Fan Zhang}

\author[1]{Duane S. Cronin}

\affil[1]{Department of Mechanical Engineering, University of Waterloo, 200 University Avenue West, Waterloo, ON N2L 3G1, Canada}
\affil[2]{Defence Research and Development Canada, P.O. Box 4000, Station Main, Medicine Hat, AB T1A 8K6, Canada}

\maketitle

\begin{abstract}
    \normalsize
    In this paper, a simple field function is presented for facilitating the solution of complex and dynamic fluid-solid systems on Cartesian grids with interface-resolved fluid-fluid, fluid-solid, and solid-solid interactions. For a Cartesian-grid-discretized computational domain segmented by a set of solid bodies, this field function explicitly tracks each subdomain with multiple resolved interfacial node layers. As a result, the presented field function enables low-memory-cost multidomain node mapping, efficient node remapping, fast collision detection, and expedient surface force integration. Implementation algorithms for the field function and its described functionalities are also presented. Equipped with a deterministic multibody collision model, numerical experiments involving fluid-solid systems with flow conditions ranging from subsonic to supersonic states are conducted to validate and illustrate the applicability of the proposed field function.
\end{abstract}

\section{Introduction}\label{sec:intro}

Modeling complex and dynamic fluid-solid systems such as fluidized beds \citep{glowinski2001fictitious, van2008numerical}, blood flow \citep{peskin1972flow}, and particle-added explosives \citep{rodriguez2013solid, mo2017numerical} has been receiving increasing attention in recent years. These systems usually involve multiscale interactions that comprise the coupled motions of solid bodies and fluid flow. When a predictive modeling approach that resolves the fluid-solid interfaces is employed, in addition to addressing the coupled fluid-fluid, fluid-solid, and solid-solid interactions, one additional challenge can be the numerical discretization related to a set of irregular and moving geometries. Advances in numerical methods, for instance, the development of Cartesian-grid-based boundary treatment methods \citep{peskin1972flow, fedkiw1999non, fedkiw2002coupling, uhlmann2005immersed, kempe2012improved}, have provided a feasible way to simulate these systems using Cartesian grids that do not conform to solid boundaries, greatly simplifying the grid generation for irregular geometries and grid regeneration for moving geometries \citep{mittal2005immersed}.

In solving a fluid-solid system involving multiple irregular and moving solid bodies on a Cartesian grid, the solids immersed in the computational grid can be described by the STereoLithography (STL) representation, which approximates an object as a closed triangulated surface and is a standard format for rapid prototyping and computer-aided design (CAD) systems. When equipped with a suitable Cartesian-grid-based numerical framework, the STL represented solids can be directly inputted into the numerical solver without the need of CAD/CFD geometric translations \citep{iaccarino2003immersed}. Nonetheless, as each immersed solid occupies a corresponding spatial region and segments the computational domain, classifying the ownership of computational nodes and identifying numerical boundaries are then prerequisites for numerical discretization implementation and interface condition enforcement. This node classification and boundary identification procedure regarding multiple solids is referred to as a multidomain node mapping problem herein.

A binary node map that distinguishes fluid and solid domains through flagging the nodes inside any solid as $0$ and nodes outside the solids as $1$, or vice versa, is popularly utilized in the literature \citep{iaccarino2003immersed, sambasivan2009ghostb, kapahi2013three}, and works properly for applications in which there is little need of differentiating a solid domain from the other solid domains. However, when there are interactions such as collisions among solids or different material properties/boundary conditions for some solid domains, uniquely tracking and identifying each solid domain are then useful and even necessary. In addition, when the solids are movable, the requirement for node remapping also arises during the solution process.

In a dynamic fluid-solid system with moving solids, the motions of solids usually depend on the local flow conditions. Therefore, the integration of surface forces exerted on solids via fluids is an essential part of the solution process. However, to the authors' knowledge, surface force integration for immersed solids on non-body-conformal Cartesian grids has not been specifically addressed in literature.

Meanwhile, when multiple solids are presented in the system, the interactions among solids (collision forces) can exert a strong influence on the stresses in the fluid-solid mixture \citep{walton1993numerical}. Therefore, collision modeling has an important role in computing a fluid-solid system with dense solids. Modeling multibody collisions is undoubtedly complex and challenging. Compared with models based on experimental correlations \citep{walton1993numerical, glowinski1999distributed, moreno20123d} or short-range repulsive-force collisions \citep{hu2001direct, glowinski2001fictitious}, an interface-resolved collision model provides a more accurate representation of the practical problems and enhances physical reality. However, additional challenges from collision detection and response are introduced and need to be addressed.

In collision detection, when the surfaces of solids are explicitly represented by triangulated meshes, checking every solid against every other solid is very inefficient if the number of solids is large and the geometry is complex. Considerable research has been devoted to optimizing the problem with strategies focusing on hierarchical object representation, orientation-based pruning criteria, spatial partitioning schemes, and distance computation algorithms \citep{lin1998collision, jimenez20013d, guendelman2003nonconvex, wald2007ray}. One efficient approach for convex rigid-body collision detection is a multilevel algorithm that integrates temporal coherence exploitation, pairwise pruning, and exact pairwise collision-detection techniques to minimize the collision-detection operations for a dense solid system \citep{jimenez20013d, ericson2004real}.

Employing implicit surfaces defined by field functions has shown success in collision-related modeling \citep{jones20063d, guendelman2003nonconvex, kapahi2013three}. In the simulation of non-convex rigid-body interactions, \citet{guendelman2003nonconvex} developed a dual geometry representation in which a solid is described by both a Lagrangian triangulated surface and a signed distance function defined on an Eulerian grid. While triangulated surface representation allows accurate normal calculations and maintains sharp interfaces, a signed distance function \citep{osher2001level} for each object permits convenient point-inclusion tests. For instance, using the layer of grid points that is nearest to the zero isocontour of a signed distance function as sample points, one can determine the collision status by testing the values of the sample points with regard to other signed distance functions. Since the surface resolution of implicitly defined surfaces is proportional to the grid resolution for field functions, high accuracy can be achieved when a well-resolved grid is employed \citep{guendelman2003nonconvex}.

However, using signed distances as field functions for defining implicit surfaces for solid objects generally requires one signed-distance function per object. As each signed distance function needs to be defined on an individual grid, this requirement consumes memory that is proportional to the number of represented solids. In addition, classifying the ownership of computational nodes and identifying numerical boundaries involve active fetching and comparing of signed-distance data scattered in the memory storage, which may result in an undesirable computational load.

In the motions of a solid system, simultaneous multibody collisions are much less common than single collisions. Nonetheless, when the system contains dense solids with special geometries or with sustained contacts among solids to transmit impulses, the presence of multibody collisions can increase greatly. Due to the ill-posedness of simultaneous multibody collisions, unless additional assumptions are imposed, multibody collision in general is an unsolvable problem \citep{ivanov1995multiple}. Under weak perturbations in the pre-collision states, a multibody collision can be decomposed into a sequence of single collisions. In reference \citep{guendelman2003nonconvex}, a sequential pairwise collision approach is proposed to resolve multibody collision among solids with arbitrary shapes, in which pairwise collision sequentially repeats among interfering solids until all solids are separating at least once. Due to the application of sequential collision, a temporal priority of the pairwise collisions is introduced. Although a random sampling of the collision queue can alleviate the temporal priority issue, the deterministic modeling of the multibody collision process is considered difficult to achieve in this approach.

To facilitate the solution of complex and dynamic fluid-solid systems on Cartesian grids, an integer-type field function that solves multidomain node mapping is proposed. For a Cartesian-grid-discretized computational domain segmented by a set of solid bodies, this field function explicitly tracks each subdomain with multiple resolved interfacial node layers. As a result, the presented field function enables low-memory-cost multidomain node mapping, efficient node remapping, fast collision detection, and expedient surface force integration. Implementation algorithms for the field function and its described functionalities are also presented. Equipped with a deterministic multibody collision model, numerical experiments involving fluid-solid systems with flow conditions ranging from subsonic to supersonic states are conducted to validate and illustrate the applicability of the proposed field function.

\section{Method development}

\subsection{Field function description}

\begin{figure}[!htbp]
    \centering
    \includegraphics[width=0.8\textwidth]{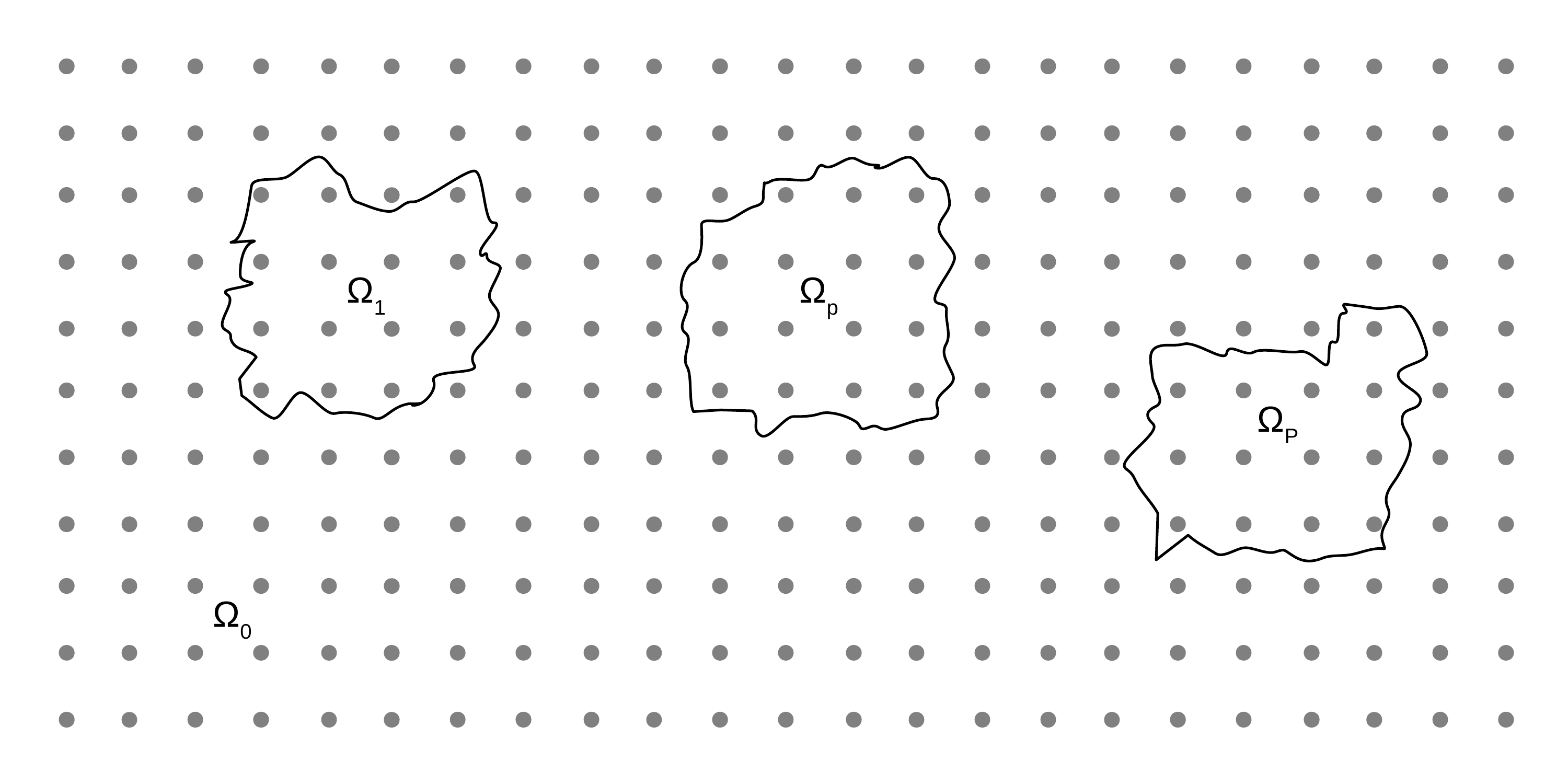}
    \caption{A 2D schematic diagram illustrating a set of polyhedron domains distributed in a spatial domain discretized by a Cartesian grid.}
    \label{fig:n_domain_basis_demo}
\end{figure}

As illustrated in Fig.~\ref{fig:n_domain_basis_demo}, for a set of solids represented by triangulated polyhedrons $\{\Omega_p: \ p = 1, \dotsc, P\}$ and distributed in a spatial domain $\Omega$, an additional subdomain $\Omega_0$ can be introduced as
\begin{equation}
    \Omega_0 = \{\Vector{x} \in \Omega: \Vector{x} \notin \cup_{p=1}^P\Omega_p\}
\end{equation}

When the spatial domain $\Omega$ is discretized by a Cartesian grid $I \times J \times K$, to classify computational nodes inside $\Omega_m$, $m = 0, \dotsc, P$, while identifying $R$ layers of interfacial nodes, a two-component integer-type field function is introduced as
\begin{equation}
    \Vector{\Phi} = \{(\phi, \varphi): \phi \in \{0, \dotsc, P\}, \varphi \in \{0, \dotsc, R\}\}
\end{equation}
in which $\phi$ is the domain identifier determined by
\begin{equation} \label{eq:phicondition}
    \phi_{i,j,k} = m, \ \text{if}\ \Vector{x}_{i,j,k} \in \Omega_m
\end{equation}
and $\varphi$ is the interfacial node layer identifier determined by
\begin{equation} \label{eq:varphicondition}
    \varphi_{i,j,k} = 
    \begin{cases}
        r, &\ \text{if}\ \exists \, \phi_{i',j',k'} \neq \phi_{i,j,k} \ \text{for condition}\ $Q$\\
        0, &\ \text{if}\ r > R 
    \end{cases}
\end{equation}
where $\Vector{x}_{i,j,k}$ is the position vector of the node $(i, j, k)$ in the Cartesian grid $I \times J \times K$, $(i', j', k')$ denotes a neighboring node of the processing node $(i, j, k)$, $R$ is the maximum number of identified interfacial node layers, and the condition $Q$ is described as
\begin{equation}
    \begin{cases}
        |i'-i| = r, \, |j'- j| = 0, \, |k'- k| = 0 &\ \text{or}\\
        |i'-i| = 0, \, |j'- j| = r, \, |k'- k| = 0 &\ \text{or}\\
        |i'-i| = 0, \, |j'- j| = 0, \, |k'- k| = r &\ \text{or}\\
        |i'-i| = r-1, \, |j'- j| = r-1, \, |k'- k| = 0 &\ \text{or}\\
        |i'-i| = r-1, \, |j'- j| = 0, \, |k'- k| = r-1 &\ \text{or}\\
        |i'-i| = 0, \, |j'- j| = r-1, \, |k'- k| = r-1
    \end{cases}
\end{equation}

The criterion for the domain identifier $\phi$ (Eq.~\eqref{eq:phicondition}) is established on the point inclusion results. The criterion for the interfacial node layer identifier $\varphi$ (Eq.~\eqref{eq:varphicondition}) is based on the existence of a heterogeneous node $(i', j', k')$ (here heterogeneous nodes refer to nodes with different $\phi$ values) on the discretization stencils of the processing node $(i, j, k)$. Therefore, the value of $\varphi$ depends on the type of differential operators involved in the governing equations as well as the type and order of the employed spatial discretization schemes. 

In order to avoid either excessive or insufficient classification of interfacial nodes, the condition $Q$ in Eq.~\eqref{eq:varphicondition} requires adapting to the specific numerical discretization scenarios. For instance, if no mixed derivatives are discretized, then only line-type stencils will be involved in spatial discretization. As a result, the condition $Q$ in Eq.~\eqref{eq:varphicondition} can be reduced to
\begin{equation}
    \begin{cases}
        |i'-i| = r, \, |j'- j| = 0, \, |k'- k| = 0 &\ \text{or}\\
        |i'-i| = 0, \, |j'- j| = r, \, |k'- k| = 0 &\ \text{or}\\
        |i'-i| = 0, \, |j'- j| = 0, \, |k'- k| = r
    \end{cases}
\end{equation}

\subsection{Multidomain node mapping}

\begin{figure}[!htbp]
    \centering
    \includegraphics[width=0.8\textwidth]{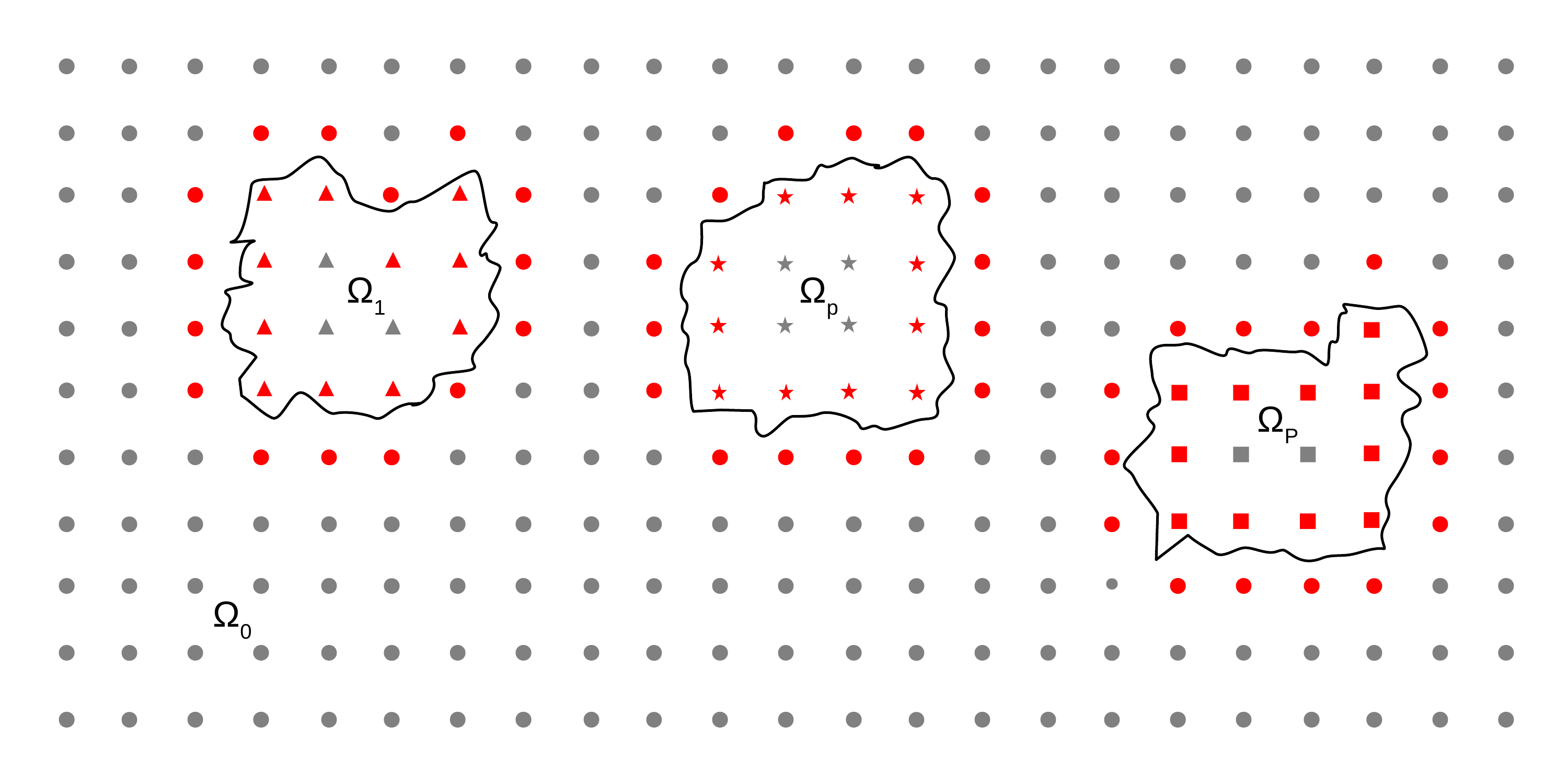}
    \caption{A 2D schematic diagram of applying the field function $\Vector{\Phi}(\phi,\varphi)$ for multidomain node mapping. \emph{Geometric shapes} represent the value of $\phi$: \emph{circle}, $0$; \emph{triangle}, $1$; \emph{star}, $p$; \emph{square}, $P$. \emph{Colors} represent the value of $\varphi$: \emph{grey}, $0$; \emph{red}, $1$; $R=1$ is assumed here for the purpose of clarification.}
    \label{fig:n_domain_demo}
\end{figure}

The field function $\Vector{\Phi}(\phi,\varphi)$ is able to generate a node map for computing complex fluid-solid systems. As illustrated in Fig.~\ref{fig:n_domain_demo}, for a computational node $(i,j,k)$, $\phi_{i,j,k}$ provides the domain inclusion state of the node, and $\varphi_{i,j,k}$ gives the interfacial state of the node.

In general, when $\Omega_m$ is a solution domain, two approaches are available to compute the solutions in $\Omega_m$. 1) a non-ghost-cell approach, in which $\Vector{\Phi}_{i,j,k}(\phi=m, \varphi=0)$ describes a normal computational node, while $\Vector{\Phi}_{i,j,k}(\phi=m, \varphi>0)$ describes that the node $(i,j,k)$ locates on the numerical boundaries of the solution domain $\Omega_m$. 2) a ghost-cell approach, in which $\Vector{\Phi}_{i,j,k}(\phi=m, \varphi \ge 0)$ describes a normal computational node, while $\Vector{\Phi}_{i,j,k}(\phi \neq m, \varphi>0)$ and a neighboring node $\Vector{\Phi}_{i',j',k'}(\phi = m, \varphi \ge 0)$ existing on the discretization stencils of node $(i,j,k)$ describe that the node $(i,j,k)$ locates on the numerical boundaries of the solution domain $\Omega_m$.

For a computational domain segmented by a set of solid bodies, the proposed field function $\Vector{\Phi}(\phi,\varphi)$ explicitly tracks each subdomain with multiple resolved interfacial node layers. Therefore, it is straightforward to enforce designated governing equations, constitutive models, numerical schemes, and interface conditions for each subdomain. In the practical implementation, the two-component $\Vector{\Phi}(\phi,\varphi)$ can be mapped onto a single scalar $\Phi = \phi + \varphi * (P+1)$, from which individual components can be extracted through $\varphi = \Phi \Des{\,mod\,} (P+1)$ and $\phi = \Phi - \varphi * (P+1)$, respectively. As an integer-type field function that can be defined on a single grid, this field function enables low-memory-cost multidomain node mapping and consumes memory that is independent from the number of represented objects. For instance, assume that the number of nodes for defining a field function is $M$, the number of represented objects is $P$, and the memory size ratio of "float" to "int" is $H$, the estimated memory consumption for using the proposed field function is then about $1/(HP)$ of that for using signed distance functions.

\subsection{Node remapping}

During the solution process, when the positions of the polyhedrons are changeable, the requirement for node remapping arises. As captured in Fig.~\ref{fig:node_remapping_demo}, the field function $\Vector{\Phi}(\phi,\varphi)$ enables efficient node remapping: From time $t^n$ to $t^{n+1}$, suppose the domain occupied by $\Omega_p$ changing from $\Omega_p^n$ to $\Omega_p^{n+1}$. When the computational time step size is restricted by a stability condition such as the Courant--Friedrichs--Lewy (CFL) condition \citep{courant1927partial}, the boundary of $\Omega_p$ will correspondingly have restricted travelling distances. If the stability condition restricts the value of travelling distance to no more than one grid size, and the maximum number of identified interfacial node layers has $R \ge 2$, one can safely assume that a node $(i,j,k)$ with $\Vector{\Phi}_{i,j,k}(\phi=p,\varphi=0)$ in $\Omega_p^n$ will remain in $\Omega_p^{n+1}$ and then only reset the interfacial nodes. As a result, nodes with $\Vector{\Phi}_{i,j,k}(\phi=p,\varphi=0)$ can be exempted from future point-inclusion tests. As these non-interfacial nodes constitute the major fraction of the computational nodes in a practical grid, this exemption from the expensive point-inclusion test can significantly reduce the costs of node remapping. An efficient node remapping algorithm established on the described observation is proposed as the following:
\begin{figure}[!htbp]
    \centering
    \includegraphics[width=0.8\textwidth]{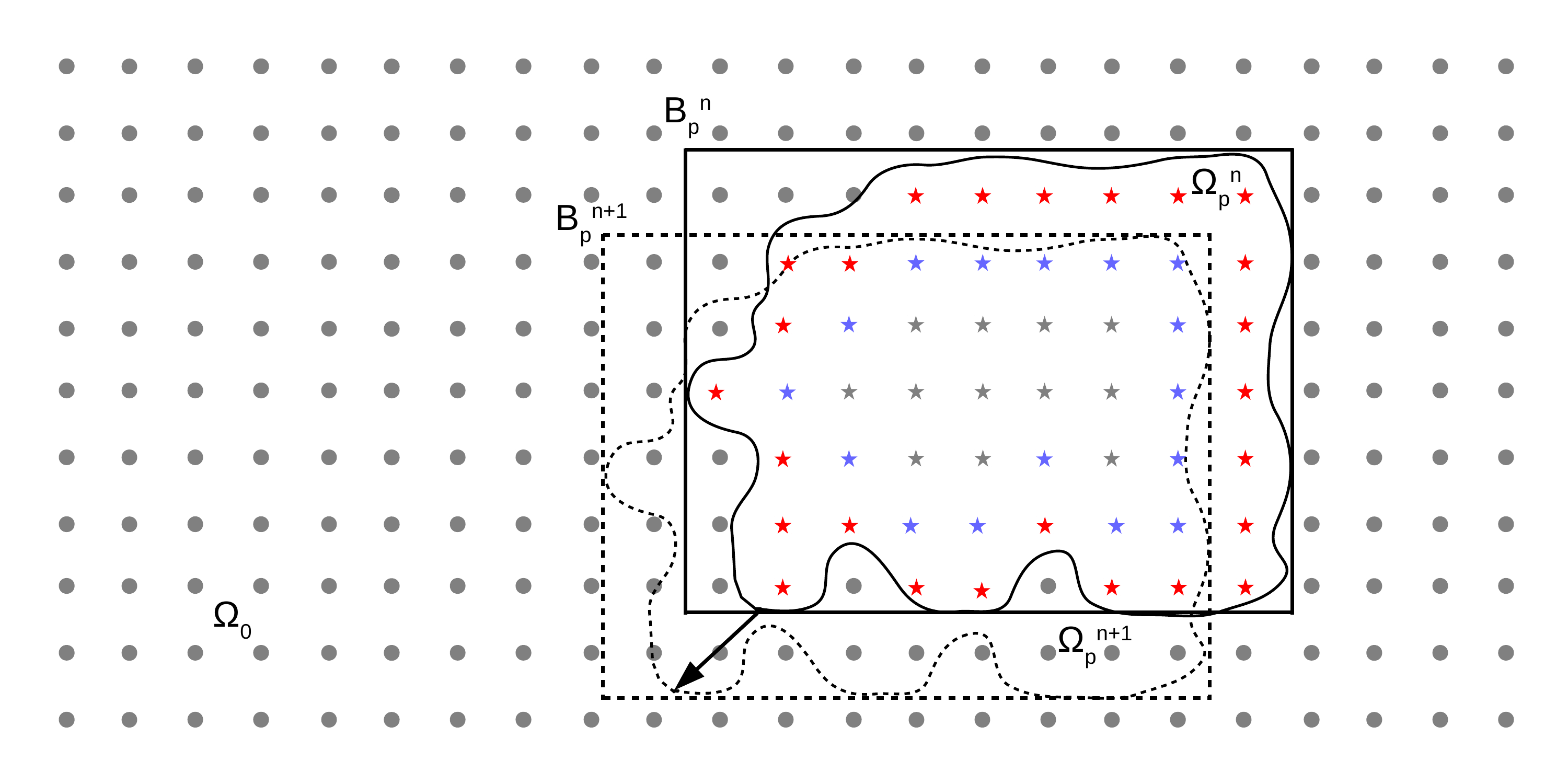}
    \caption{A 2D schematic diagram of applying the field function $\Vector{\Phi}(\phi,\varphi)$ for efficient node remapping. \emph{Geometric shapes} represent the value of $\phi$: \emph{circle}, $0$; \emph{star}, $p$. \emph{Colors} represent the value of $\varphi$: \emph{grey}, $0$; \emph{red}, $1$; \emph{blue}, $2$; $R=2$ is assumed, and the interfacial nodes of $\Omega_0$ are deactivated for the purpose of clarification.}
    \label{fig:node_remapping_demo}
\end{figure}
\begin{enumerate}
    \item \emph{Initialization}. Sweep each node $(i,j,k)$ in $I \times J \times K$: if $\varphi_{i,j,k} > 0$, set $\Vector{\Phi}_{i,j,k}(\phi,\varphi) = (0,0)$.
    \item \emph{Compute the domain identifier $\phi$}. Sweep each polyhedron $\Omega_p$ in $\{\Omega_p: \ p = 1, \dotsc, P\}$:
        \begin{enumerate}
            \item Find the bounding box $B_p=[I_{\Des{min}}, I_{\Des{max}}] \times [J_{\Des{min}}, J_{\Des{max}}] \times [K_{\Des{min}}, K_{\Des{max}}]$. 
            \item Sweep each node $(i,j,k)$ in $B_p$: if $\phi_{i,j,k} = 0$, do point-in-polyhedron test for the node $(i,j,k)$ over $\Omega_p$ to determine the value of $\phi_{i,j,k}$ using Eq.~\eqref{eq:phicondition}.
        \end{enumerate}
    \item \emph{Compute the interfacial node layer identifier $\varphi$}. Sweep each node $(i,j,k)$ in $I \times J \times K$: determine the value of $\varphi_{i,j,k}$ using Eq.~\eqref{eq:varphicondition}.
\end{enumerate}

In the described algorithm, the point-in-polyhedron test for the node $(i,j,k)$ over $\Omega_p$ is a point-inclusion test with regard to a single polyhedron. A variety of established methods, such as the ray-crossing methods \citep{o1998computational}, angular methods \citep{carvalho1995point}, winding number methods \citep{haines1994point}, and signed distance methods \citep{jones20063d}, are available. The angle weighted pseudonormal signed distance computation method \citep{baerentzen2005signed} is employed herein, as it provides good balance of efficiency and robustness. Meanwhile, it finds the closest point and the corresponding normal for a computational node, which is essential for implementing a Cartesian-grid-based boundary treatment method.

The proposed algorithm herein effectively solves the node classification and boundary identification of a Cartesian grid segmented by a set of polyhedrons in space, which involves a set of points together with a set of polyhedrons and represents a generalized point-in-polyhedron problem. In addition, since $\Vector{\Phi}(\phi,\varphi) = (0,0)$ is true initially, the presented algorithm successfully unifies the procedures of the initial multidomain node mapping and the subsequent node remapping. This unification can simplify the code structure and reduce the complexity of implementation.

\subsection{Collision detection}

\begin{figure}[!htbp]
    \centering
    \includegraphics[width=0.8\textwidth]{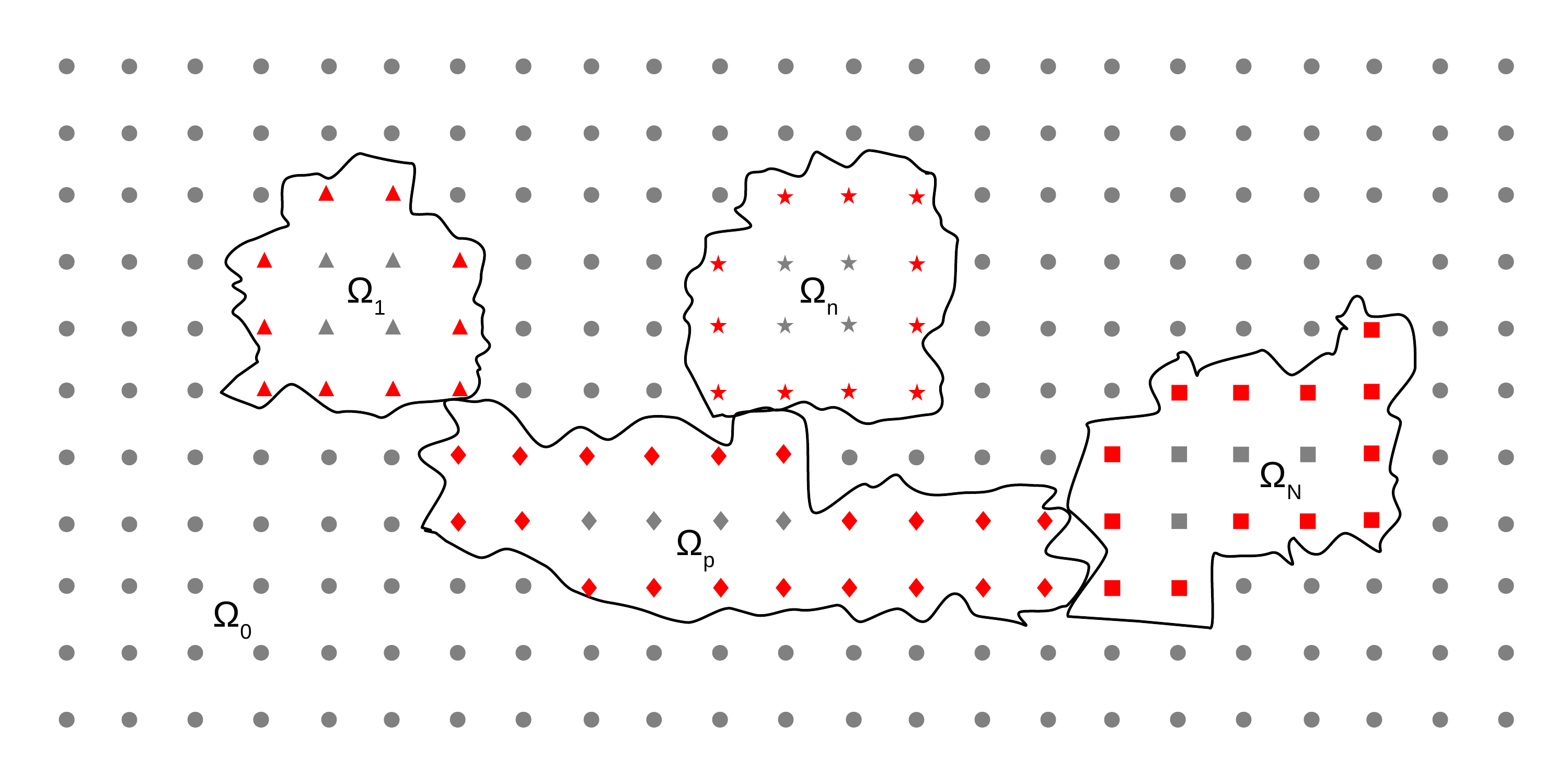}
    \caption{A 2D schematic diagram of applying the field function $\Vector{\Phi}(\phi,\varphi)$ for collision detection. \emph{Geometric shapes} represent the value of $\phi$: \emph{circle}, $0$; \emph{triangle}, $1$; \emph{star}, $n$; \emph{square}, $N$; \emph{lozenge}, $p$; \emph{Colors} represent the value of $\varphi$: \emph{grey}, $0$; \emph{red}, $1$; $R=1$ is assumed, and the interfacial nodes of $\Omega_0$ are deactivated for the purpose of clarification.}
    \label{fig:n_domain_collide_demo}
\end{figure}

In addition to multidomain node mapping, the field function $\Vector{\Phi}$ also enables fast collision detection. As illustrated in Fig.~\ref{fig:n_domain_collide_demo}, in the bounding box of $\Omega_p$, by sweeping through the nodes $(i,j,k)$ with $\Vector{\Phi}_{i,j,k}(\phi=p,\varphi=1)$ to find neighboring nodes $(i',j',k')$ with $\Vector{\Phi}_{i',j',k'}(\phi \neq p, \varphi=1)$, where $|i-i'| \leq 1$, $|j-j'| \leq 1$, and $|k-k'| \leq 1$, all the polyhedrons $\Omega_{n}$ colliding with the polyhedron $\Omega_p$ can be detected in an easy and efficient way. 

When the polyhedron $\Omega_p$ collides with a polyhedron $\Omega_n$, it is possible that multiple elements such as vertices, edges, and faces on $\Omega_p$ will contact with several elements on $\Omega_n$, in which an element on $\Omega_p$ may contact with either multiple elements or a portion of an element on $\Omega_n$. This multicontact issue imposes great difficulties in the determination of the line of impact. 

Instead of finding the common normal of the contacting geometric elements, an alternative approach suggested herein is to approximate the line of impact via the Cartesian grid. Suppose a number of $C$ computational nodes $(i_c,j_c,k_c)$, $c = 1, \dotsc, C$, in $\Omega_p$ satisfying $\Vector{\Phi}_{i_c,j_c,k_c}(\phi = p, \varphi=1)$, and each $(i_c,j_c,k_c)$ has a number of $D$ neighboring nodes $(i'_d,j'_d,k'_d)$, $d = 1, \dotsc, D$, with $\Vector{\Phi}_{i'_d,j'_d,k'_d}(\phi = n, \varphi=1)$, the suggested approximation of the line of impact between $\Omega_p$ and $\Omega_n$ is defined as
\begin{equation}
    \unitVector{e}_{pn} = \frac{\sum^{C}_{c=1}\sum^{D}_{d=1} [(i'_d - i_c)\unitVector{e}_1 + (j'_d - j_c)\unitVector{e}_2 + (k'_d - k_c)\unitVector{e}_3]}{|\sum^{C}_{c=1}\sum^{D}_{d=1} [(i'_d - i_c)\unitVector{e}_1 + (j'_d - j_c)\unitVector{e}_2 + (k'_d - k_c)\unitVector{e}_3]|}
\end{equation}
where $\unitVector{e}_1$, $\unitVector{e}_2$, $\unitVector{e}_3$ are the unit direction vectors of the coordinate $x$, $y$, and $z$, respectively. This approximation of the line of impact via the Cartesian grid treats the multicontact issue with great simplicity while providing adequate accuracy, as demonstrated in the numerical experiments.

\subsection{Collision response}

Suppose $\Omega_p$ collides with $N$ polyhedrons $\Omega_n$, $n=1,\dotsc,N$. Denote the pre- and post-collision velocity of $\Omega_p$ as $\Vector{V}_p$ and $\Vector{V}_p'$, respectively, and the pre-collision velocity of $\Omega_n$ as $\Vector{V}_n$. To approximate multibody collision while avoiding introducing temporal priority, the following collision model is employed herein:
\begin{enumerate}
    \item Conduct pairwise collision $(\Omega_p, \Omega_n)$ with the pre-collision velocity state $(\Vector{V}_p,\Vector{V}_n)$ to predict the post-collision velocity $\Vector{V}_{p,n}'$ and the velocity change $\Delta \Vector{V}_{p,n}'$ of $\Omega_p$ for the $n$-th pair collision:
        \begin{equation}
            \Delta \Vector{V}_{p,n}' = \Vector{V}_{p,n}' - \Vector{V}_p = - \frac{m_n}{m_p+m_n} (1+C_{\Des{R}}) (\Vector{V}_{pn} \cdot \unitVector{e}_{pn}) \unitVector{e}_{pn} - C_{\Des{f}} [\Vector{V}_{pn} - (\Vector{V}_{pn} \cdot \unitVector{e}_{pn}) \unitVector{e}_{pn}]
        \end{equation}
        where $\Vector{V}_{pn}= (\Vector{V}_p - \Vector{V}_n)$ is the relative velocity before collision, $C_{\Des{R}}$ is the coefficient of restitution in the normal direction ($C_{\Des{R}} = 0$, $0 < C_{\Des{R}} < 1$, and $C_{\Des{R}} = 1$ are corresponding to perfectly inelastic collision, partially inelastic collision, and elastic collision, respectively), $C_{\Des{f}}$ is a coefficient used to mimic the effect of sliding friction, $m_p$ and $m_n$ are the mass of $\Omega_p$ and $\Omega_n$, respectively.

    \item Then, approximate the post-collision velocity of $\Omega_p$ under the multibody collision via a vector summation of the pre-collision velocity and velocity changes:
        \begin{equation}
            \Vector{V}_p' = \Vector{V}_p + \sum_{n=1}^{N} \Delta \Vector{V}_{p,n}'
        \end{equation}

    \item Apply the above procedures to each $\Omega_p$, $p=1,\dotsc,P$, in the solid system to obtain a corresponding post-multibody-collision velocity $\Vector{V}_p'$.
    \item Update the velocity state of the solid system simultaneously. 
\end{enumerate}

\subsection{Surface force integration}

In the interface-resolved modeling of fluid-solid interactions, surface force integration is an essential part. The proposed field function $\Vector{\Phi}(\phi,\varphi)$ can expediently facilitates the surface force integration for irregular solids immersed in a Cartesian grid.

\begin{figure}[!htbp]
    \centering
    \begin{subfigure}[b]{0.48\textwidth}
        \includegraphics[trim = 10mm 0mm 10mm 0mm, clip, width=\textwidth]{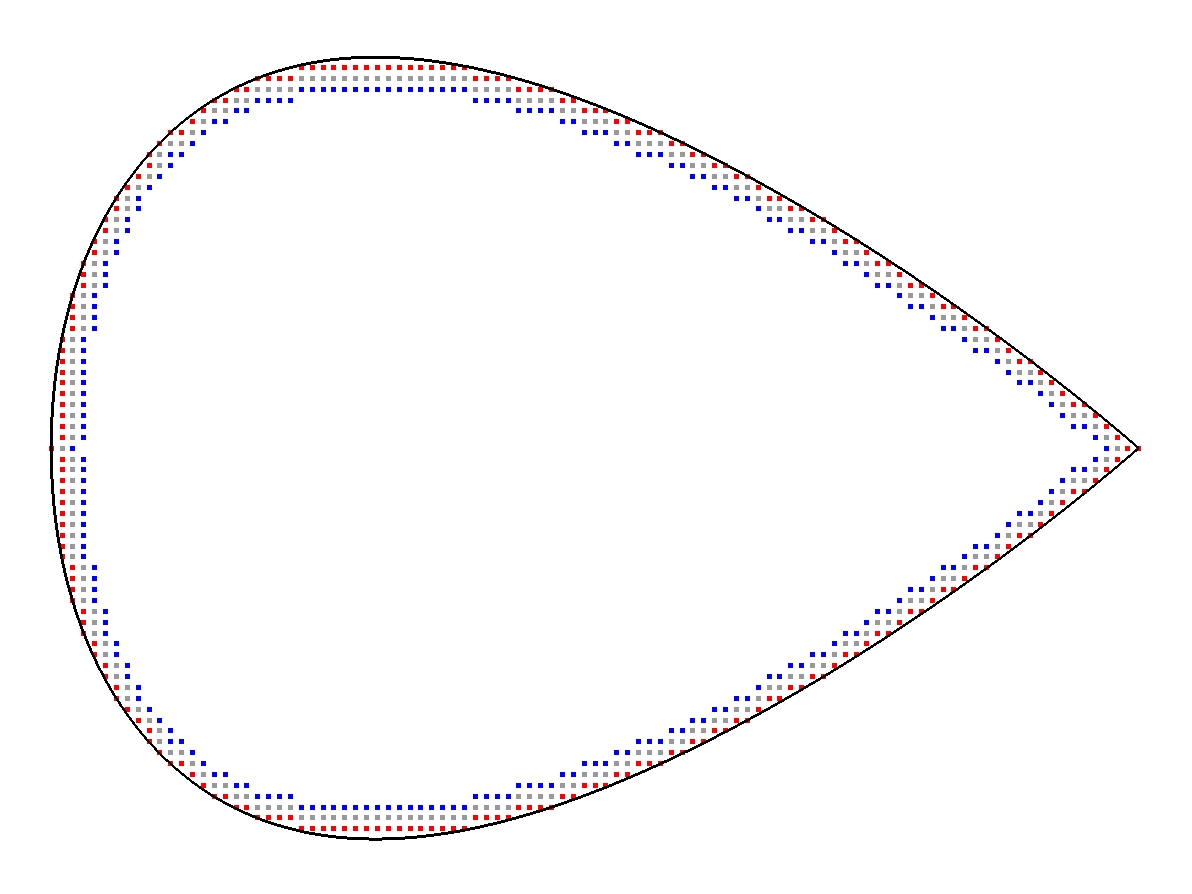}
        \caption{}
        \label{fig:wall_shear_stress_layer}
    \end{subfigure}%
    ~
    \begin{subfigure}[b]{0.48\textwidth}
        \includegraphics[width=\textwidth]{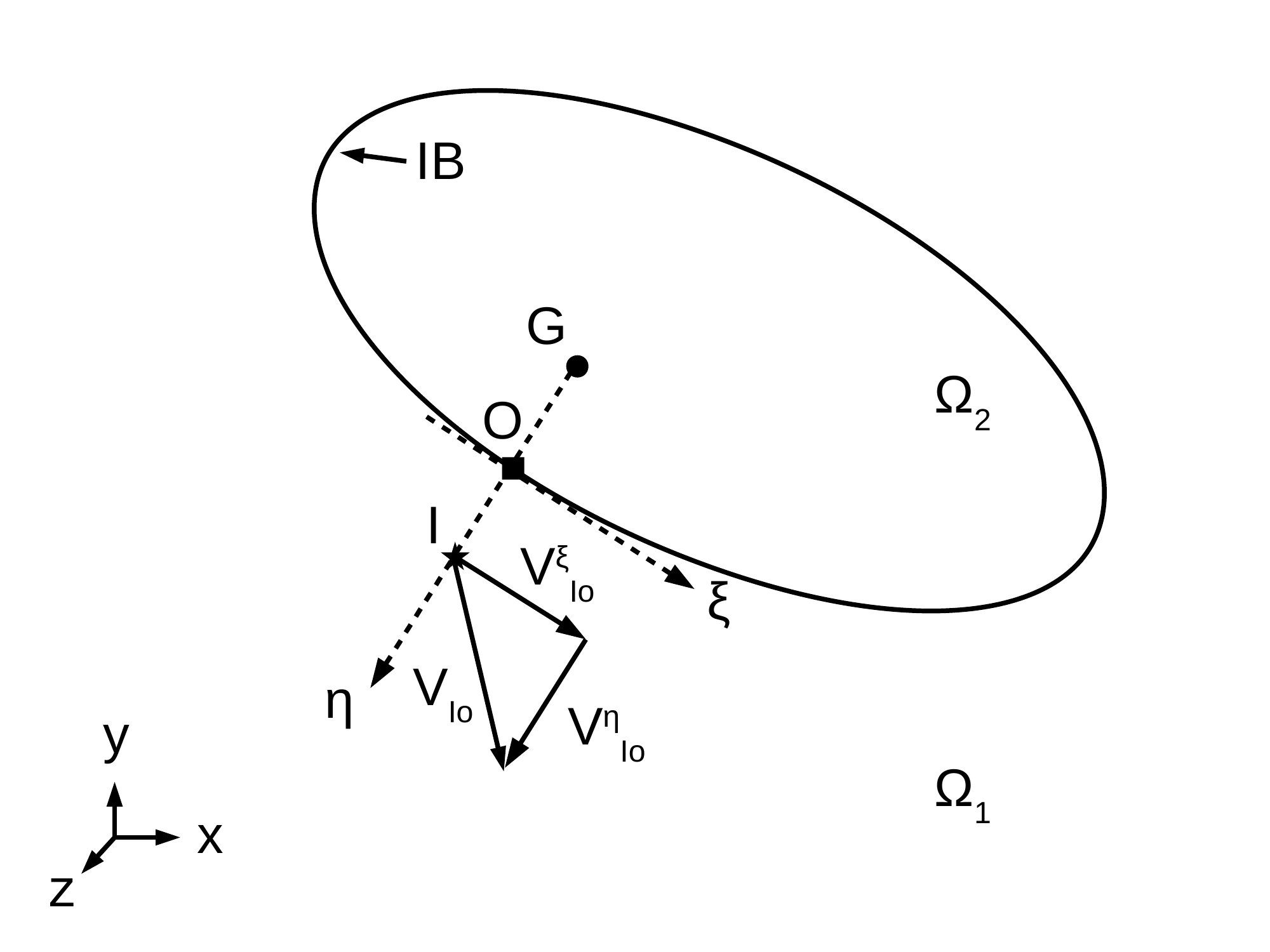}
        \caption{}
        \label{fig:wall_shear_stress_demo}
    \end{subfigure}%
    \caption{Diagrams of applying the field function $\Vector{\Phi}(\phi,\varphi)$ for surface force integration over immersed boundaries. (a) Computational results showing the interfacial node layer identifier $\varphi$ for an irregular solid immersed in a Cartesian grid. \emph{Colors} represent the value of $\varphi$: \emph{red}, $1$; gray, $2$; \emph{blue}, $3$. Nodes with $\varphi=0$ are deactivated for the purpose of clarification. (b) A schematic diagram illustrating wall shear stress calculation. [Nomenclature: $G$, ghost node; $O$, boundary point; $I$, image point; $\Omega_1$, fluid domain; $\Omega_2$, solid domain; $IB$, immersed boundary.]}
    \label{fig:wall_shear_stress}
\end{figure}

Employing the presented field function and its implementation algorithm, the computed three inner interfacial layers of an irregular solid immersed in a Cartesian grid are shown in Fig.~\ref{fig:wall_shear_stress_layer}. One can observe that the $\varphi=1$ layer conforms with the solid boundary very closely, in which the distance discrepancy reduces with mesh refinement and is in the interval $[0, \Delta s)$, where $\Delta s=\max(\Delta_x, \Delta_y, \Delta_z)$. Therefore, for a non-ghost-cell approach, the integration of surface forces can be conducted on the $\varphi=2$ node layer of the fluid domain surrounding the solid, of which the flow values are known and the distance discrepancy with solid boundary is in the interval $[\Delta s, 2\Delta s)$. Meanwhile, for a ghost-cell approach, the integration of surface forces can be properly conducted on the corresponding point layer formed by the image points of the $\varphi=2$ nodes of the solid.

Here, only the surface force integration in the ghost-cell approach is discussed, as the method can be easily transformed for a non-ghost-cell approach, whose surface force integration is more straightforward. By exploring the relation between the ghost node $G$, the boundary point $O$, and the image point $I$, the surface force integration in a ghost-cell approach can be simplified greatly as discussed below.

\paragraph{Wall pressure}

Pressure at the wall can generally be approximated via the zero normal gradient assumption
\begin{equation}
    \left. \frac{\partial p}{\partial n} \right|_O = 0
\end{equation}
Therefore,
\begin{equation}
    p_O = p_I = p_G
\end{equation}
Thus, the pressure component in the surface stress vector is
\begin{equation}
    -p_G \unitVector{n}
\end{equation}

\paragraph{Wall shear stress}

As illustrated in Fig.~\ref{fig:wall_shear_stress_demo}, suppose a natural coordinate system, $\eta-\xi$, is locally established at the boundary point $O$ and is located in the plane defined by the normal vector $(\unitVector{n} \times \Vector{V}_{IO})$, where $\Vector{V}_{IO} = \Vector{V}_I - \Vector{V}_O$ is the relative velocity of the image point $I$ to the boundary point $O$. Then, the wall shear stress is defined as
\begin{equation}
    \Vector{\tau}_{\Des{w}} \equiv \Vector{\tau}(\eta =0) = \mu \left. \frac{\partial \Vector{V}^{\xi}_{IO}}{\partial \eta}\right|_{\eta=0}
\end{equation}
A second-order central differencing approximation gives
\begin{equation}
    \Vector{\tau}_{\Des{w}} = \mu \frac{\Vector{V}^{\xi}_{IO} - \Vector{V}^{\xi}_{GO}}{2||\Vector{x}_I - \Vector{x}_O||}
\end{equation}
By using $\Vector{V}_{IO} = - \Vector{V}_{GO}$, $||\Vector{x}_I - \Vector{x}_O|| = ||\Vector{x}_G - \Vector{x}_O||$, and $\Vector{V}^{\xi}_{GO} = \Vector{V}_{GO} - (\Vector{V}_{GO} \cdot \unitVector{n}) \unitVector{n}$, it gives
\begin{equation}
    \Vector{\tau}_{\Des{w}} = - \mu \frac{\Vector{V}_{GO} - (\Vector{V}_{GO} \cdot \unitVector{n}) \unitVector{n}}{||\Vector{x}_G - \Vector{x}_O||}
\end{equation}
Hence, the surface stress vector at the boundary point $O$ is obtained as
\begin{equation}
    \Vector{T}_{O}^{(\unitVector{n})} = -p_G \unitVector{n} + \Vector{\tau}_{\Des{w}} = -p_G \unitVector{n} - \mu \frac{\Vector{V}_{GO} - (\Vector{V}_{GO} \cdot \unitVector{n}) \unitVector{n}}{||\Vector{x}_G - \Vector{x}_O||}
\end{equation}

By employing the derived relations, the surface force integration for irregular solids can be directly computed and expressed in the global Eulerian coordinate system without involving coordinate transformation. In addition, both the wall pressure and the wall shear stress are explicitly computed on the ghost node $G$, which is a computational node with known flow values. As a result, the proposed surface force integration method is straightforward to implement.

\section{Governing equations and discretization}

\subsection{Fluid-solid coupling}

The coupling between fluid and solid motions is modeled by a partitioned fluid-solid interaction algorithm, which is derived through applying Strang splitting \citep{strang1968construction}:
\begin{equation}
    \Vector{U}^{n+1} = \Soptr_{\Des{s}}(\frac{\Delta t}{2})\Soptr_{\Des{f}}(\frac{\Delta t}{2})\Soptr_{\Des{f}}(\frac{\Delta t}{2})\Soptr_{\Des{s}}(\frac{\Delta t}{2}) \Vector{U}^n
\end{equation}
where $\Vector{U}^{n}$ and $\Vector{U}^{n+1}$ are the solution vectors at time $t^n$ and $t^{n+1}$, respectively; $\Soptr_{\Des{s}}$ and $\Soptr_{\Des{f}}$ are the solution operators of solid dynamics and fluid dynamics, respectively.

\subsection{Fluid dynamics}

The motion of fluids is governed by the Navier--Stokes equations in Cartesian coordinates: 
\begin{equation}
    \frac{\partial \Vector{U}}{\partial t}+\frac{\partial \Vector{F}_i}{\partial x_i} = \frac{\partial \Vector{F}^{\Des{v}}_i}{\partial x_i}+\Vector{\Phi}
\end{equation}
where the vectors of conservative variables $\Vector{U}$, convective fluxes $\Vector{F}_i$, diffusive fluxes $\Vector{F}^{\Des{v}}_i$, and source terms $\Vector{\Phi}$ are as follows:
\begin{equation}
    \Vector{U} =
    \begin{pmatrix}
        \rho\\
        \rho V_j\\
        \rho e_{\Des{T}}
    \end{pmatrix}
    ,\,\,
    \Vector{F}_i =
    \begin{pmatrix}
        \rho V_i\\
        \rho V_i V_j + p \delta_{ij}\\
        (\rho e_{\Des{T}}+p) V_i
    \end{pmatrix}
    ,\,\,
    \Vector{F}^{\Des{v}}_i =
    \begin{pmatrix}
        0\\
        \tau_{ij}\\
        k \frac{\partial T}{\partial x_i} + \tau_{il} V_l
    \end{pmatrix}
    ,\,\,
    \Vector{\Phi} =
    \begin{pmatrix}
        0\\
        f^{\Des{b}}_j\\
        f^{\Des{b}}_l V_l
    \end{pmatrix}
\end{equation}
in which $\rho$ is the density, $\Vector{V}$ is the velocity, $e_{\Des{T}} = e +\Vector{V}\cdot\Vector{V}/2$ is the specific total energy, $e$ is the specific internal energy, $p$ is the thermodynamic pressure, $\Tensor{\tau}$ is the viscous stress tensor, $T$ is the temperature, $k$ is the thermal conductivity, $\Vector{f}^{\Des{b}}$ represents external body forces such as gravity, $i$ is a free index, $j$ is an enumerator, $l$ is a dummy index. The closure of the system is through supplying the Newtonian fluid relation with the Stokes hypothesis
\begin{equation}
    \tau_{ij} = \mu\left(\frac{\partial V_i}{\partial x_j} + \frac{\partial V_j}{\partial x_i} - \frac{2}{3}(\nabla\cdot\Vector{V}) \delta_{ij}\right)
\end{equation}
and the perfect gas law
\begin{equation}
    \begin{gathered}
        p = \rho R T \\
        e = C_v T
    \end{gathered}
\end{equation}
where $R$ is the specific gas constant, $C_v$ is the specific heat capacity at constant volume, and $\mu$ is the dynamic viscosity and is determined by the Sutherland viscosity law:
\begin{equation}
    \mu = \frac{C_1T^{\frac{3}{2}}}{T+C_2}
\end{equation}
with $C_1=1.458\times10^{-6} \Unit{kg \cdot m^{-1} \cdot s^{-1} \cdot K^{-1/2}}$, and $C_2=110.4 \Unit{K}$.

The temporal integration of the Navier--Stokes equations is achieved via the third-order SSP Runge--Kutta method \citep{shu1988efficient, gottlieb2001strong}:
\begin{equation}
    \begin{aligned}
        &\Vector{U}^{(1)} = \Toptr\Vector{U}^n\\
        &\Vector{U}^{(2)} = 3/4\Vector{U}^n+1/4\Toptr\Vector{U}^{(1)}\\
        &\Vector{U}^{n+1} = 1/3\Vector{U}^n+2/3\Toptr\Vector{U}^{(2)}\\
        &\Toptr = (\Matrix{I} + \Delta t \Loptr)
    \end{aligned}
\end{equation}
where $\Matrix{I}$ is the identity matrix, operator $\Loptr = \Loptr_x + \Loptr_y + \Loptr_z$, $\Loptr_x$, $\Loptr_y$, and $\Loptr_z$ represent the spatial operators of $x$, $y$, and $z$ dimension, respectively.

Dimensional-splitting \citep{strang1968construction} approximation is adopted to treat the system of conservation laws in multidimensional space. To ensure discrete mass conservation, conservative discretization is applied for both the convective fluxes and diffusive fluxes. Using the $x$ dimension as an example, the flux derivative at a node $i$ is approximated as
\begin{equation}
    \left. \frac{\partial \Vector{F}}{\partial x} \right|_i = \frac{1}{\Delta x}\left[\hat{\Vector{F}}_{i+\frac{1}{2}}-\hat{\Vector{F}}_{i-\frac{1}{2}}\right]
\end{equation}
where $\Vector{F}$ represents either the convective flux vector or the diffusive flux vector, $\hat{\Vector{F}}_{i+{1}/{2}}$ is a numerical flux at the interface between the discretization interval $\Omega_i=[x_{i-{1}/{2}},x_{i+{1}/{2}}]$ and $\Omega_{i+1}=[x_{i+{1}/{2}},x_{i+{3}/{2}}]$.

In the discretization of convective fluxes, local characteristic splitting is used to transform the vector system into a set of scalar conservation laws. In addition, scalar flux splitting is then applied to ensure upwinding property
\begin{equation}
    f(u) = f^+(u) + f^-(u), \,\, \frac{\mathrm{d} f^+(u)}{\mathrm{d} u} \ge 0, \,\, \frac{\mathrm{d} f^-(u)}{\mathrm{d} u} \le 0
\end{equation}
where $f$ is a scalar characteristic flux.

When both the forward and backward fluxes are discretized in conservative form, the discretization of a scalar flux derivative has the form
\begin{equation}
    \left. \frac{\partial f}{\partial x} \right|_i = \frac{1}{\Delta x} \left[ \hat{f}_{i+\frac{1}{2}} - \hat{f}_{i-\frac{1}{2}} \right], \,\, \hat{f}_{i+\frac{1}{2}} = \hat{f}_{i+\frac{1}{2}}^+ + \hat{f}_{i+\frac{1}{2}}^-, \,\, \hat{f}_{i-\frac{1}{2}} = \hat{f}_{i-\frac{1}{2}}^+ + \hat{f}_{i-\frac{1}{2}}^-
\end{equation}

The fifth-order WENO scheme \citep{jiang1996efficient} is then used for the reconstruction of the scalar numerical fluxes:
\begin{equation}
    \hat{f}_{i+\frac{1}{2}}^{+} = \sum_{n=0}^{N} \omega_{n}^+ q_{n}^+(f_{i+n-N}^+, \dotsc, f_{i+n}^+), \,\, N=(r-1)=2
\end{equation}
where
\begin{equation}
    \begin{gathered}
        q_0^+(f_{i-2}^+, \dotsc, f_{i}^+) = (2f_{i-2}^+ - 7f_{i-1}^+ + 11f_{i}^+) / 6\\
        q_1^+(f_{i-1}^+, \dotsc, f_{i+1}^+) = (-f_{i-1}^+ + 5f_{i}^+ + 2f_{i+1}^+) / 6\\
        q_2^+(f_{i}^+, \dotsc, f_{i+2}^+) = (2f_{i}^+ + 5f_{i+1}^+ - f_{i+2}^+) / 6\\
        \omega_{n}^+ = \frac{\alpha_n^+}{\alpha_0^+ + \dotsb + \alpha_N^+}, \,\, \alpha_n^+ = \frac{C_n}{(\varepsilon + IS_n^+)^2}, \,\, \varepsilon = 10^{-6}\\
        C_0 = \frac{1}{10}, \,\, C_1 = \frac{6}{10}, \,\, C_2 = \frac{3}{10}\\
        IS_{0}^+ = \frac{13}{12}(f_{i-2}^+ - 2f_{i-1}^+ + f_{i}^+)^2 + \frac{1}{4}(f_{i-2}^+ - 4f_{i-1}^+ + 3f_{i}^+)^2\\
        IS_{1}^+ = \frac{13}{12}(f_{i-1}^+ - 2f_{i}^+ + f_{i+1}^+)^2 + \frac{1}{4}(f_{i-1}^+ - f_{i+1}^+)^2\\
        IS_{2}^+ = \frac{13}{12}(f_{i}^+ - 2f_{i+1}^+ + f_{i+2}^+)^2 + \frac{1}{4}(3f_{i}^+ - 4f_{i+1}^+ + f_{i+2}^+)^2
    \end{gathered}
\end{equation}
in which $r$ is the number of candidate stencils, $q_n$ are the $r$-th order approximations of $\hat{f}_{i+1/2}$ on the candidate stencils $S_n=(x_{i+n-N},\dotsc,x_{i+n})$, $\omega_{n}$ are the actual weights of $q_n$, which are determined by the smoothness of solution in the candidate stencils $S_n$, as measured by $IS_n$, and $C_n$ are optimal weights to ensure that the convex combination of $q_n$ converges to a $(2r-1)$-th order approximation of $\hat{f}_{i+1/2}$ on the undivided stencil $S=(x_{i-N},\dotsc,x_{i+N})$ in smooth regions.

A conservative second-order central difference scheme is employed for diffusive flux discretization, in which the interfacial flux $\hat{\Vector{F}}^{\Des{v}}_{i+{1}/{2}}$ is reconstructed on the discretized space $[i, i+1] \times [j-1, j+1] \times [k-1, k+1]$. Suppose $\phi$ representing a physical quantity in $\Vector{F}^{\Des{v}}$, the reconstruction of the interfacial values of $\phi$ and its derivatives has the form: 
\begin{equation}
    \begin{aligned}
        \phi_{i+\frac{1}{2}, j, k} &= \frac{\phi_{i, j, k} + \phi_{i+1, j, k}}{2} \\
        \left.\frac{\partial \phi}{\partial x}\right|_{i+\frac{1}{2}, j, k} &= \frac{\phi_{i+1, j, k} - \phi_{i, j, k}}{\Delta x} \\
        \left.\frac{\partial \phi}{\partial y}\right|_{i+\frac{1}{2}, j, k} &= \frac{\phi_{i, j+1, k} + \phi_{i+1, j+1, k} - \phi_{i, j-1, k} - \phi_{i+1, j-1, k}}{4\Delta y} \\
        \left.\frac{\partial \phi}{\partial z}\right|_{i+\frac{1}{2}, j, k} &= \frac{\phi_{i, j, k+1} + \phi_{i+1, j, k+1} - \phi_{i, j, k-1} - \phi_{i+1, j, k-1}}{4\Delta z}
    \end{aligned}
\end{equation}

The interface boundary treatment is conducted via an immersed boundary method \citep{mo2016immersed}, which employs a second-order three-step flow reconstruction scheme to enforce the Dirichlet, Neumann, Robin, and Cauchy boundary conditions in a straightforward and consistent manner. A description of the integration of the employed numerical techniques and its application to the direct simulation of explosively dispersed granular flows are available in the reference \citep{mo2017numerical}.

\subsection{Solid dynamics}

The motion of solids is governed by the equation system comprising the Newton's second law of translational motion and the Euler equations of rotational motion:
\begin{equation}
    \frac{\mathrm{d} \Vector{U}}{\mathrm{d} t} = \Vector{\Phi}
    ,\,\,
    \Vector{U} =
    \begin{pmatrix}
        \Vector{V}\\
        \Vector{x}_{\Des{c}}\\
        \Matrix{I}_{\Des{c}} \Vector{\omega}\\
        \Vector{\theta}
    \end{pmatrix}
    ,\,\,
    \Vector{\Phi} =
    \begin{pmatrix}
        \frac{1}{m}\int\limits_{\partial\Omega} \unitVector{n} \cdot (-p \unitTensor{I} + \Tensor{\tau}) \, \mathrm{d}S + \Vector{g} \\
        \Vector{V}\\
        \int\limits_{\partial\Omega} (\Vector{x} - \Vector{x}_{\Des{c}}) \times [\unitVector{n} \cdot (-p \unitTensor{I} + \Tensor{\tau})] \, \mathrm{d}S\\
        \Vector{\omega}
    \end{pmatrix}
\end{equation}
where $\Vector{x}$ is the position vector of spatial points, $\Omega$ is the spatial domain occupied by a solid, $\Vector{x}_{\Des{c}}$ is the position vector of the solid centroid, $\Vector{\theta}$ is the orientation (vector of Euler angles) of the solid, $\Vector{V}$ and $\Vector{\omega}$ are the translational and angular velocities of the solid, respectively, $m$ is the mass of the solid, $\Matrix{I}_{\Des{c}}$ is the moment of inertia matrix, $\unitVector{n}$ is the unit outward surface normal vector, $p$ and $\Tensor{\tau}$ are the pressure and viscous stress tensor field exerted on the solid surface via fluid, respectively, and $\Vector{g}$ is the body force per unit mass, such as gravitational acceleration, exerted by external fields. The time integration of this ordinary differential equation system is via a second-order Runge--Kutta scheme:
\begin{equation}
    \begin{aligned}
        &\Vector{k}_1 = \Vector{\Phi}(t^n, \Vector{U}^n)\\
        &\Vector{k}_2 = \Vector{\Phi}(t^n + \Delta t, \Vector{U}^n + \Delta t \Vector{k}_1)\\
        &\Vector{U}^{n+1} = \Vector{U}^n + \Delta t(\Vector{k}_1 + \Vector{k}_2) / 2
    \end{aligned}
\end{equation}

\section{Numerical experiments}\label{sec:result}

\subsection{Subsonic flow around a cylinder}

The subsonic viscous flow around a cylinder resulting from the passing of a uniform free-stream flow $(\rho_{\infty}, u_{\infty}, v_{\infty}, p_{\infty})=(1.176 \Unit{kg/m^3}, 34.7 \Unit{m/s}, 0, 101327 \Unit{Pa})$ is computed to validate the proposed field function applied to subsonic fluid-solid interactions. For flow around a cylinder, when the Reynolds number of the free-stream flow, $Re$, is less than about $47$, the flow is steady, and two symmetric recirculating vortices form behind the cylinder. As $Re$ increases to a higher value, the flow becomes unsteady with vortex shedding \citep{gao2007improved}. In the present calculation, flow conditions with $Re=20$, $40$, $100$, and $200$ are considered, which cover both the steady and unsteady flow regimes and have a considerable amount of published data for comparison.

\begin{figure}[!htbp]
    \centering
    \begin{subfigure}[b]{0.40\textwidth}
        \includegraphics[width=\textwidth]{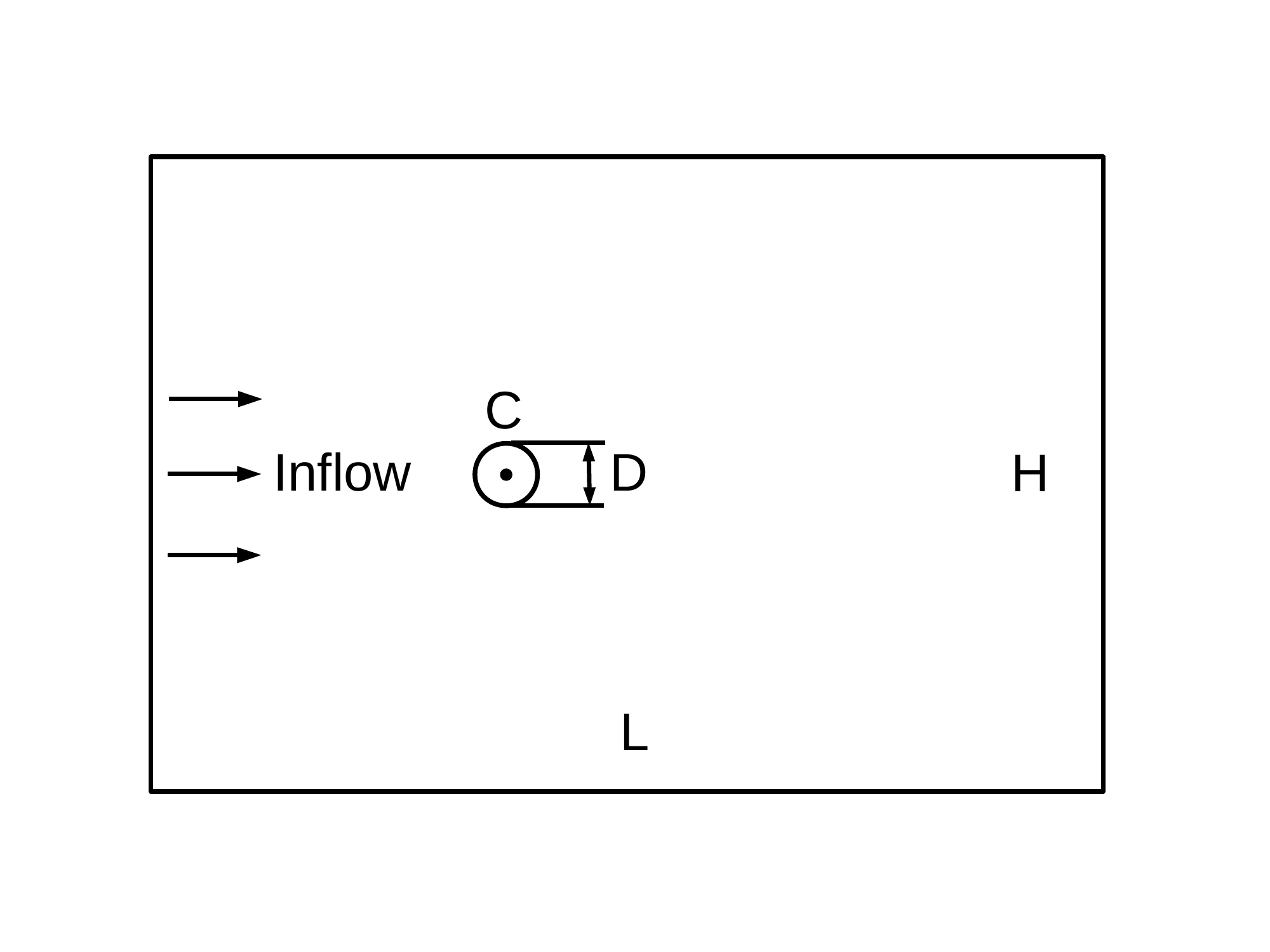}
        \caption{}
        \label{fig:flow_cylinder_demo_a}
    \end{subfigure}%
    ~
    \begin{subfigure}[b]{0.40\textwidth}
        \includegraphics[width=\textwidth]{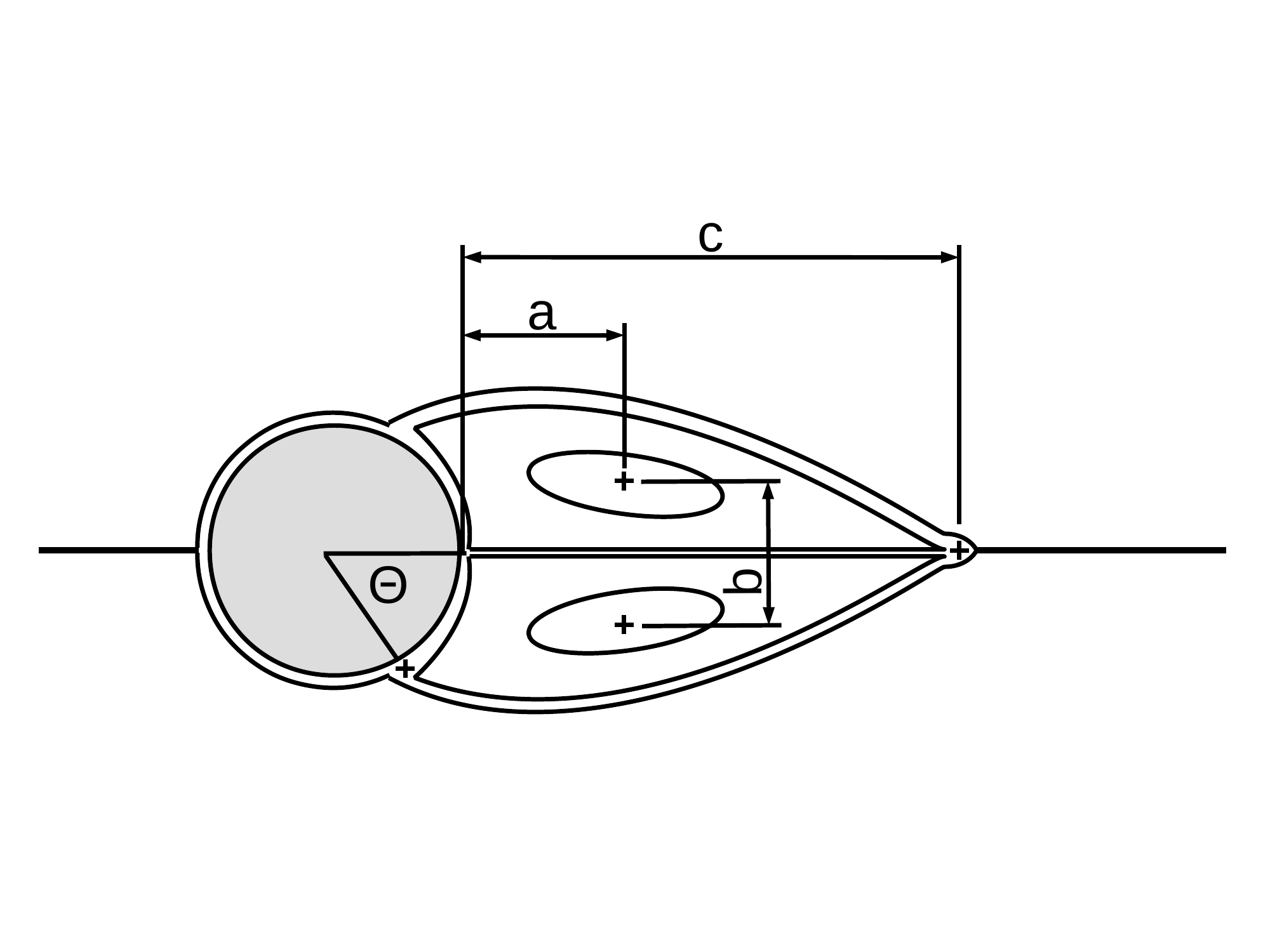}
        \caption{}
        \label{fig:flow_cylinder_demo_b}
    \end{subfigure}%
    \caption{Schematic diagrams illustrating the flow around a cylinder problem. (a) Computational setup. (b) Vortex parameters for steady flows. [Nomenclature: $L$, domain length; $H$, domain height; $C$, cylinder center; $D$, cylinder diamter; $a$, horizontal distance from vortex center to cylinder rear; $b$, vertical distance between vortex centers; $c$, length of separation; $\theta$, angle of separation.]}
    \label{fig:flow_cylinder_demo}
\end{figure}

As illustrated in Fig.~\ref{fig:flow_cylinder_demo_a}, a cylinder with diameter $D=1 \Unit{m}$ is positioned in a domain of size $L \times H = [0, 45D] \times [0, 30D]$, and the center of the cylinder is located at $C(10D, 15D)$. A subsonic inflow condition is imposed at the left domain boundary, in which the flow states have the same values as the free-stream ones except that the density is extrapolated, and the outflow condition is used for the right domain boundary. In addition, the slip-wall boundary condition is applied for the lower and upper domain boundaries, while the no-slip wall boundary condition is enforced on the cylinder. The dynamic viscosity of the flow is set to a constant value determined from $\mu = \rho_{\infty} u_{\infty} D / Re$. After a grid convergence test, a $1350\times900$ grid was found to be sufficient for producing reliable results, which has a grid resolution of about $0.03D$ ($30$ nodes per diameter) and is comparable to the grid resolutions used in references \citep{gao2007improved, brehm2015locally}.

As illustrated in Fig.~\ref{fig:flow_cylinder_demo_b}, the steady flows can be quantified through a variety of vortex parameters such as the vortex location $a$, vortex distance $b$, length of separation $c$, and angle of separation $\theta$. In addition, the drag coefficient $C_{\Des{D}} = F_x / (0.5\rho_{\infty} u_{\infty}^2 D)$ can be used to validate the force behaviour, where $F_x$ is the $x$-component of the total force acting on the cylinder.

\begin{figure}[!htbp]
    \centering
    \begin{subfigure}[b]{0.45\textwidth}
        \includegraphics[width=\textwidth]{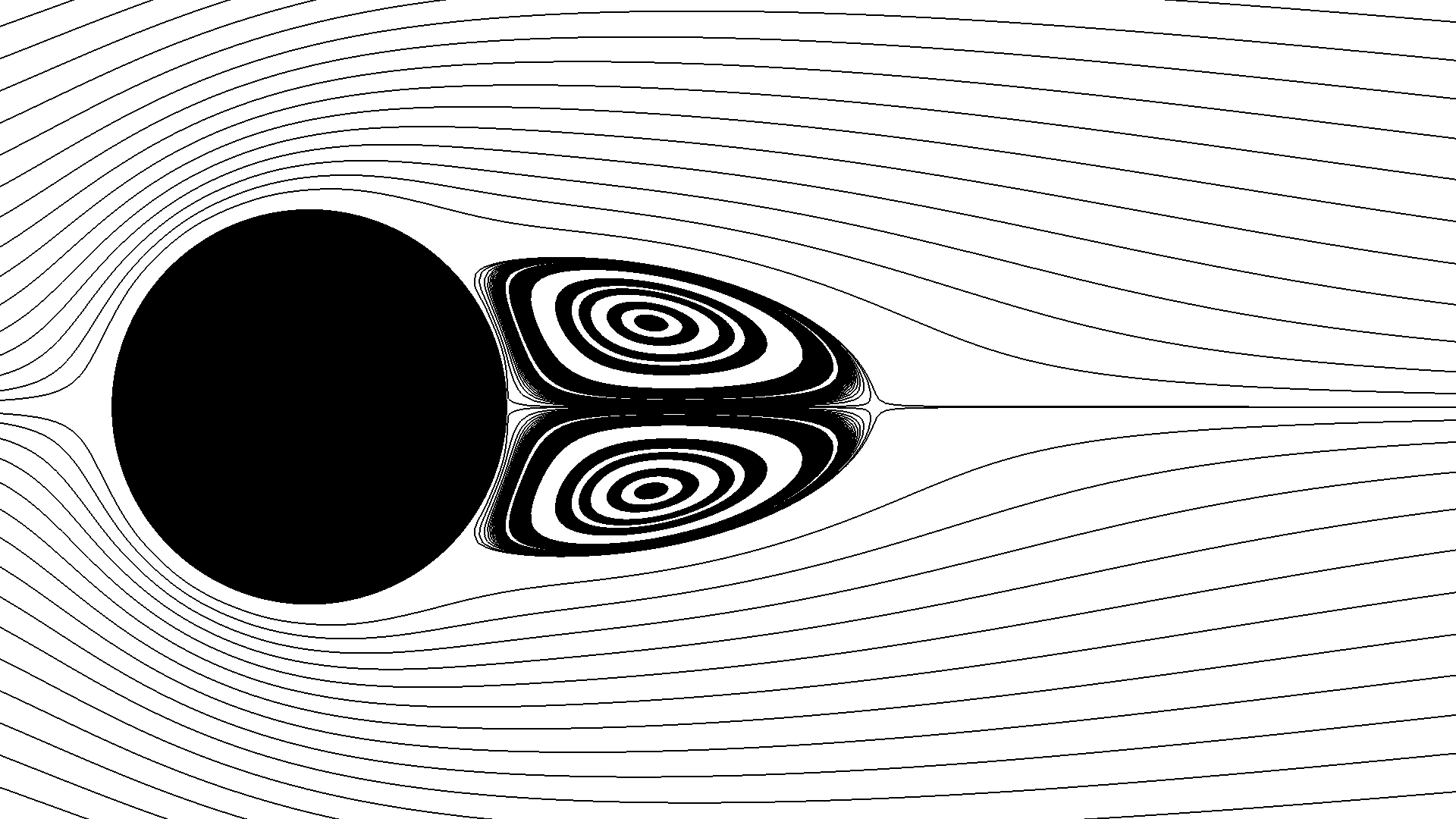}
        \caption{}
        \label{fig:1_cyn_vortex_stream_re020}
    \end{subfigure}%
    ~
    \begin{subfigure}[b]{0.45\textwidth}
        \includegraphics[width=\textwidth]{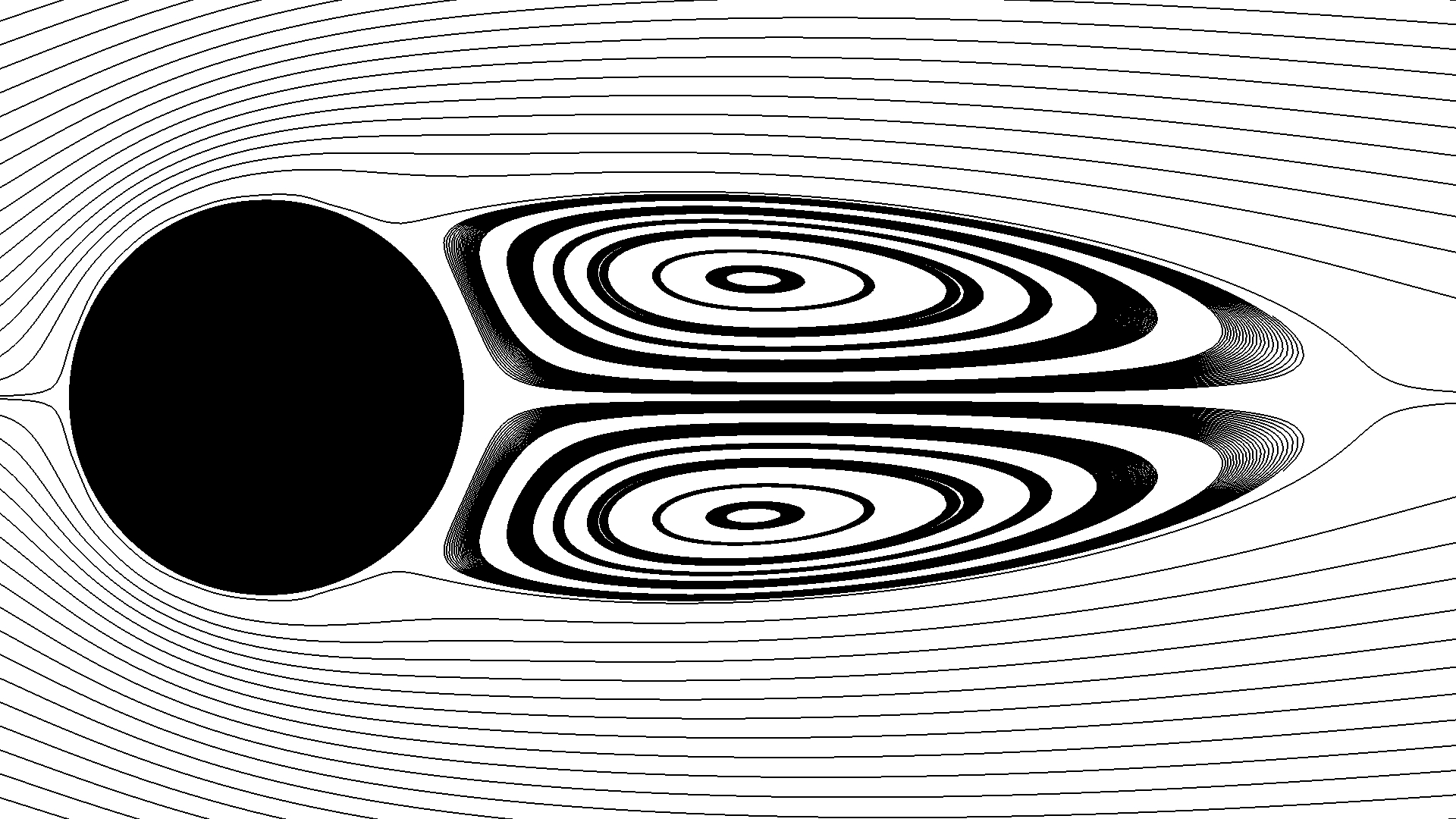}
        \caption{}
        \label{fig:1_cyn_vortex_stream_re040}
    \end{subfigure}%
    \caption{Local streamlines of the steady flows around a cylinder at different Reynolds numbers. (a) $Re=20$. (b) $Re=40$.}
    \label{fig:1_cyn_vortex_stream_steady}
\end{figure}

The predicted local streamlines of the steady flows at Reynolds numbers $Re=20$ and $Re=40$ are shown in Fig.~\ref{fig:1_cyn_vortex_stream_steady}. As the Reynolds number increases, the attached, steady, and symmetric vortex pair grow in length. A comparison of the predicted flow properties with the published ones in the literature is presented in Table~\ref{tab:1_cyn_vortex_para_steady}, in which excellent agreement is presented among all the quantities of interest.
\begin{table}[!htbp]
    \centering
    \begin{tabular}{lccccc}
        \hline\hline
         & $a$ & $b$ & $c$ & $\theta$ & $C_{\Des{D}}$ \\
        \hline
        $Re=20$ & & & & & \\
        \citet{coutanceau1977experimental}$^*$ & $0.33$ & $0.46$ & $0.93$ & $45.0^{\circ}$ & $--$\\
        \citet{linnick2005high} & $0.36$ & $0.43$ & $0.93$ & $43.5^{\circ}$ & $2.06$\\
        \citet{brehm2015locally} & $0.36$ & $0.42$ & $0.96$ & $44.0^{\circ}$ & $2.02$\\
        \emph{Present study} & $0.36$ & $0.42$ & $0.93$ & $43.8^{\circ}$ & $2.01$\\
        $Re=40$ & & & & & \\
        \citet{coutanceau1977experimental}$^*$ & $0.76$ & $0.59$ & $2.13$ & $53.8^{\circ}$ & $--$\\
        \citet{linnick2005high} & $0.72$ & $0.60$ & $2.28$ & $53.6^{\circ}$ & $1.54$\\
        \citet{brehm2015locally} & $0.72$ & $0.58$ & $2.26$ & $52.9^{\circ}$ & $1.51$\\
        \emph{Present study} & $0.73$ & $0.60$ & $2.27$ & $52.4^{\circ}$ & $1.50$\\
        \hline\hline
    \end{tabular}
    \caption{Comparison of results for steady flows around a cylinder. [Nomenclature: $a$, horizontal distance from vortex center to cylinder rear; $b$, vertical distance between vortex centers; $c$, length of separation; $\theta$, angle of separation; $C_{\Des{D}}$, drag coefficient; $*$, experimental results.]}
    \label{tab:1_cyn_vortex_para_steady}
\end{table}

\begin{figure}[!htbp]
    \centering
    \begin{subfigure}[b]{0.48\textwidth}
        \includegraphics[trim = 0mm 80mm 0mm 80mm, clip, width=\textwidth]{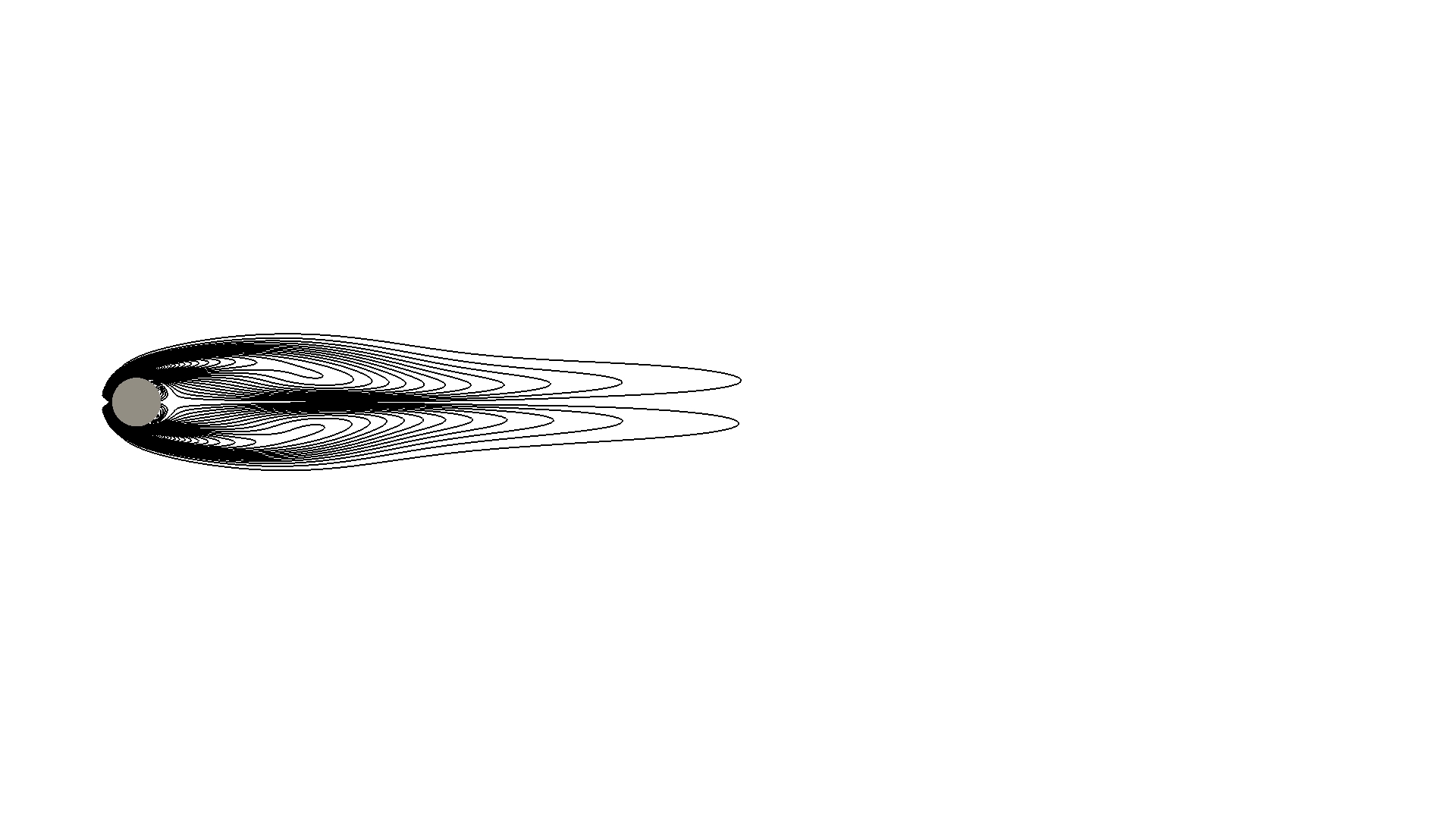}
        \caption{$0.5 \Unit{s}$}
        \label{fig:1_cyn_vortex_re100_t0500ms}
    \end{subfigure}%
    ~
    \begin{subfigure}[b]{0.48\textwidth}
        \includegraphics[trim = 0mm 80mm 0mm 80mm, clip, width=\textwidth]{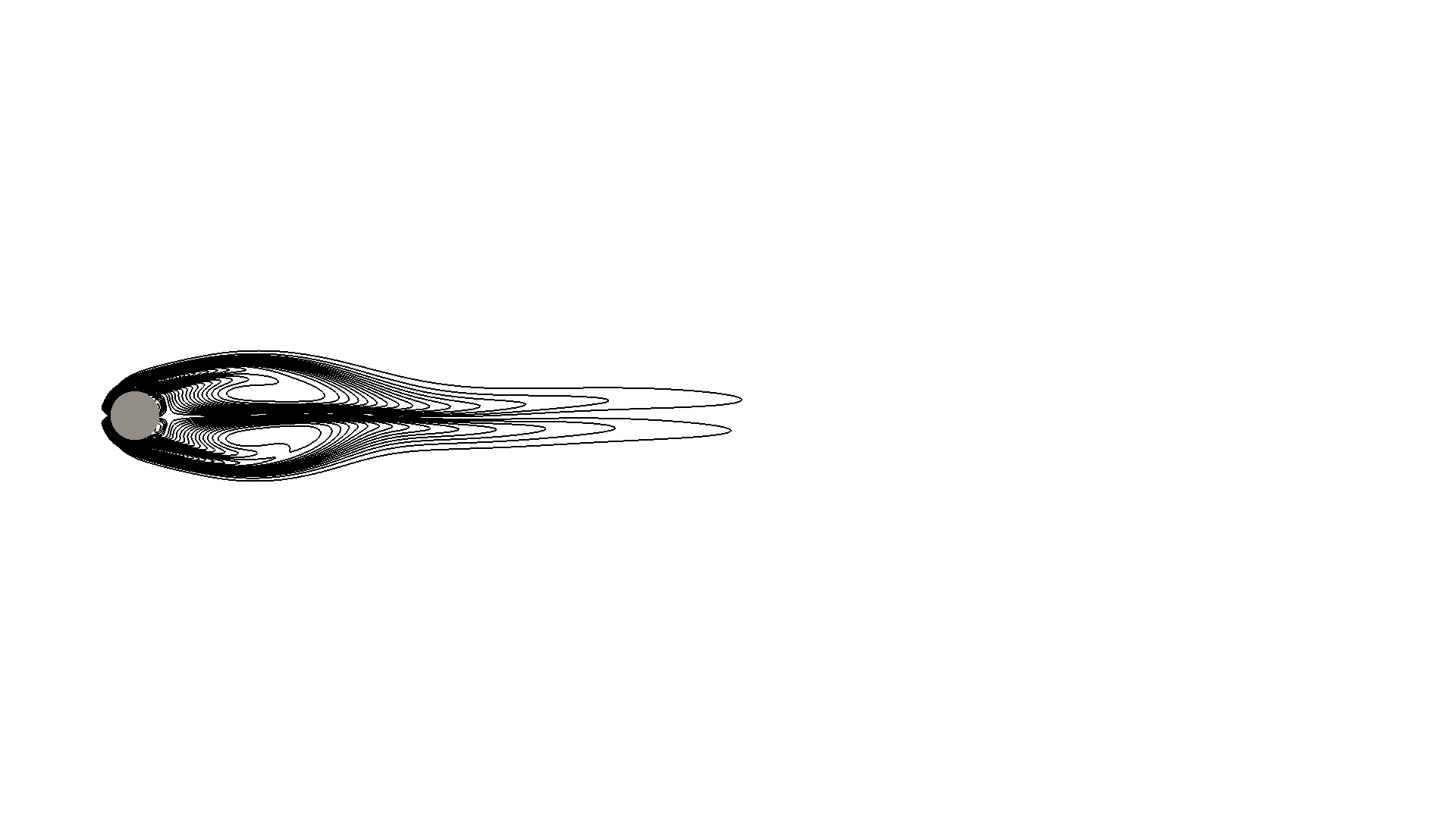}
        \caption{$0.5 \Unit{s}$}
        \label{fig:1_cyn_vortex_re200_t0500ms}
    \end{subfigure}%
    \\
    \begin{subfigure}[b]{0.48\textwidth}
        \includegraphics[trim = 0mm 80mm 0mm 80mm, clip, width=\textwidth]{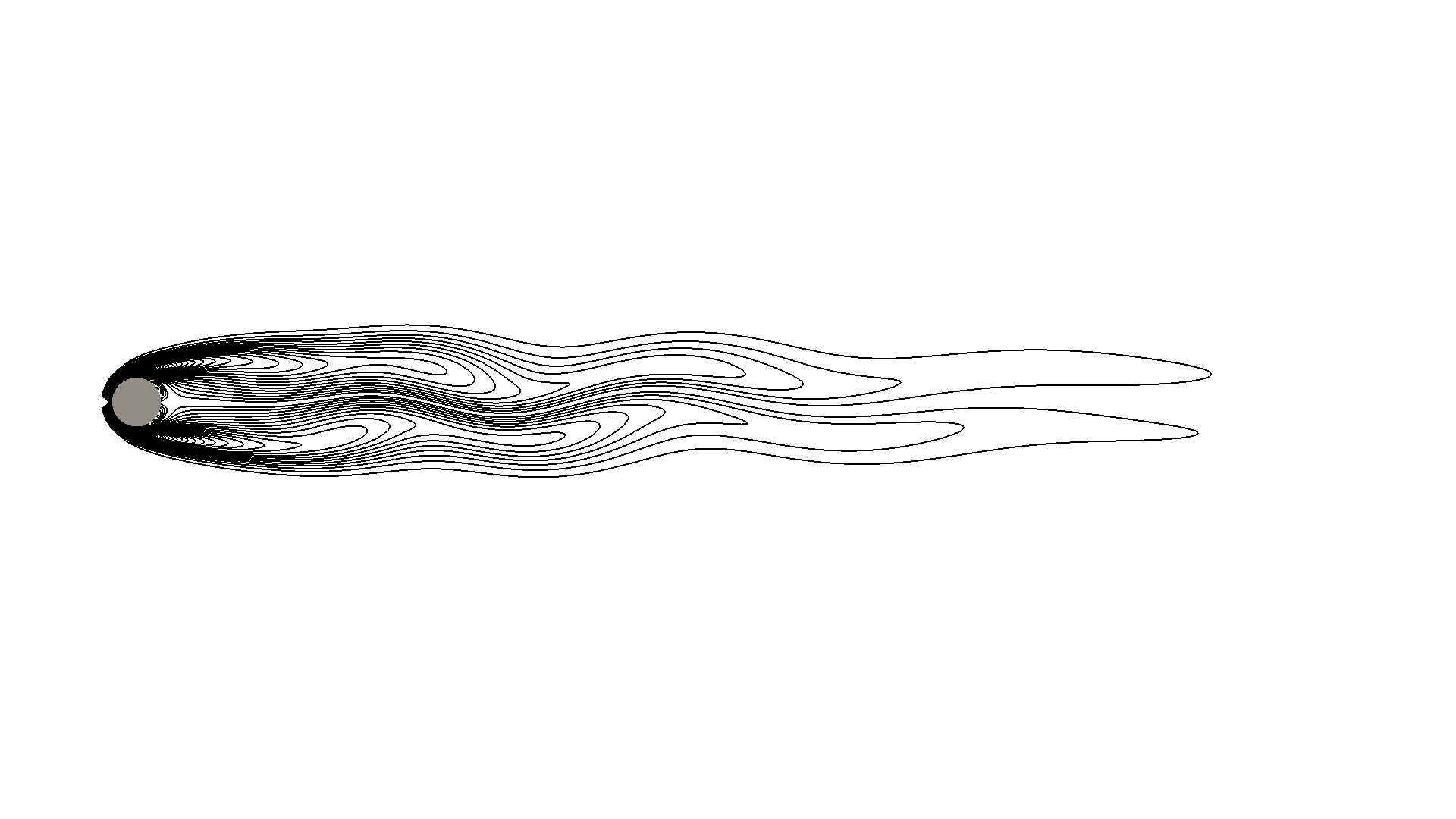}
        \caption{$1.0 \Unit{s}$}
        \label{fig:1_cyn_vortex_re100_t1000ms}
    \end{subfigure}%
    ~
    \begin{subfigure}[b]{0.48\textwidth}
        \includegraphics[trim = 0mm 80mm 0mm 80mm, clip, width=\textwidth]{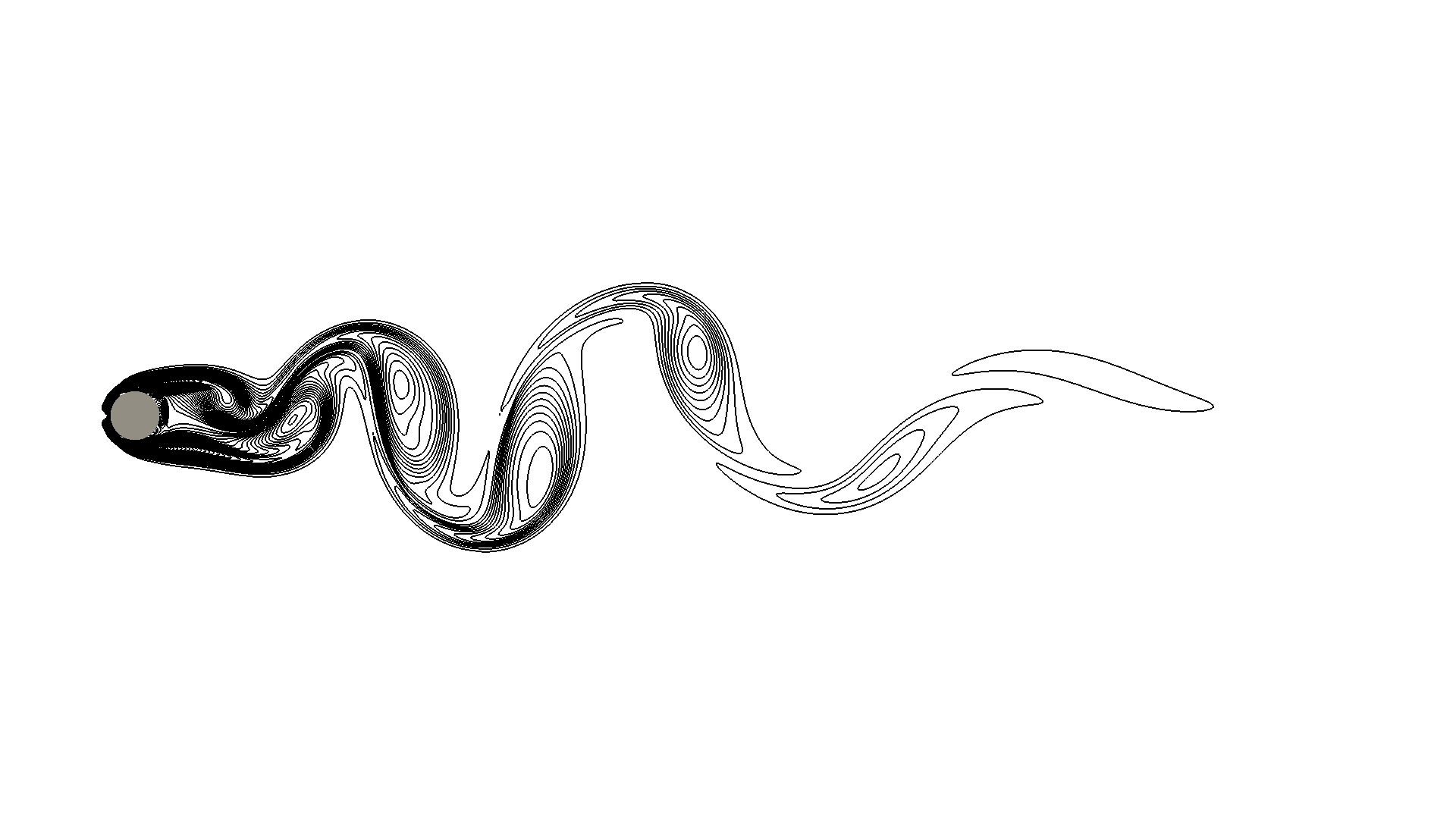}
        \caption{$1.0 \Unit{s}$}
        \label{fig:1_cyn_vortex_re200_t1000ms}
    \end{subfigure}%
    \\
    \begin{subfigure}[b]{0.48\textwidth}
        \includegraphics[trim = 0mm 80mm 0mm 80mm, clip, width=\textwidth]{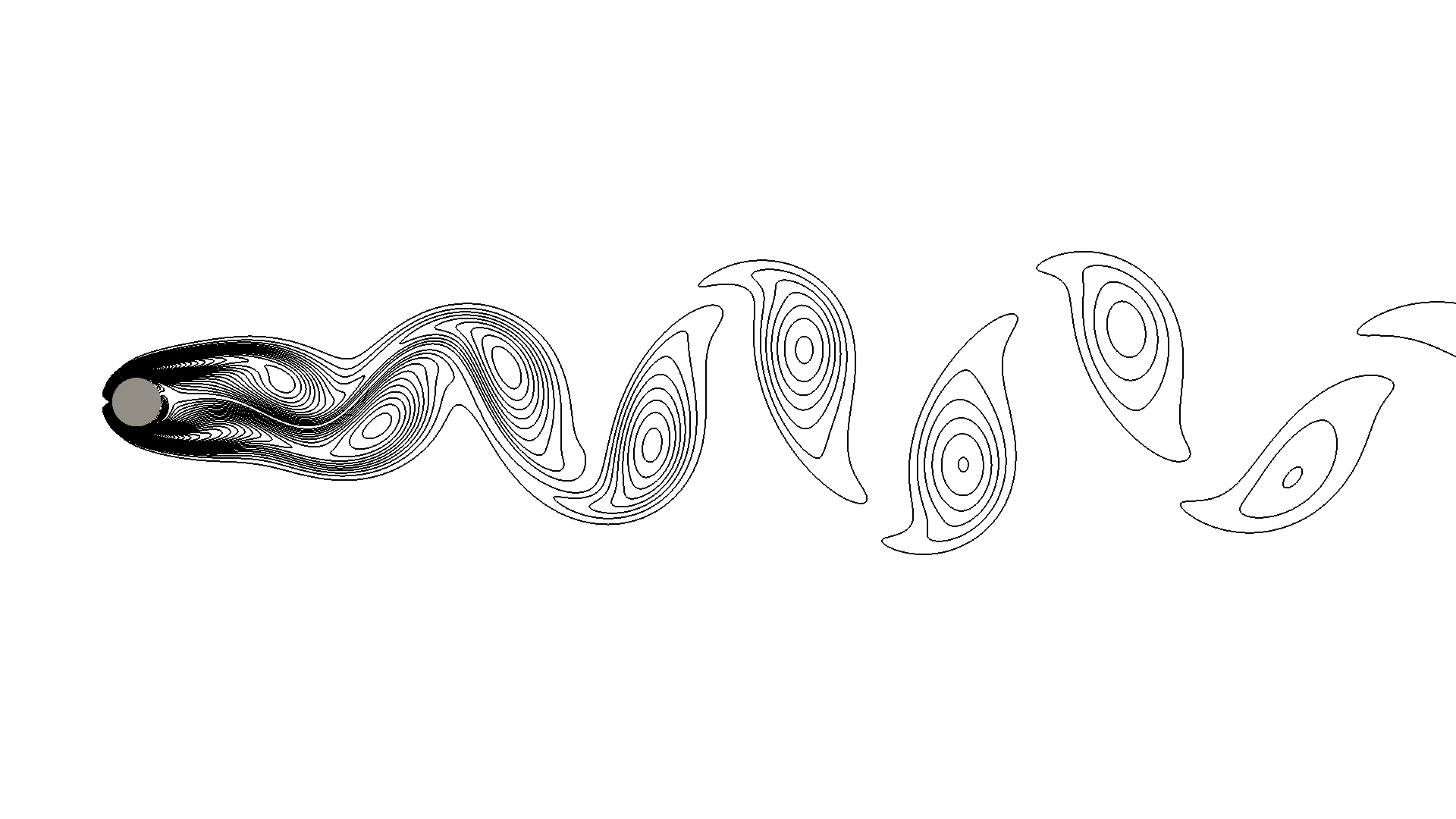}
        \caption{$1.5 \Unit{s}$}
        \label{fig:1_cyn_vortex_re100_t1500ms}
    \end{subfigure}%
    ~
    \begin{subfigure}[b]{0.48\textwidth}
        \includegraphics[trim = 0mm 80mm 0mm 80mm, clip, width=\textwidth]{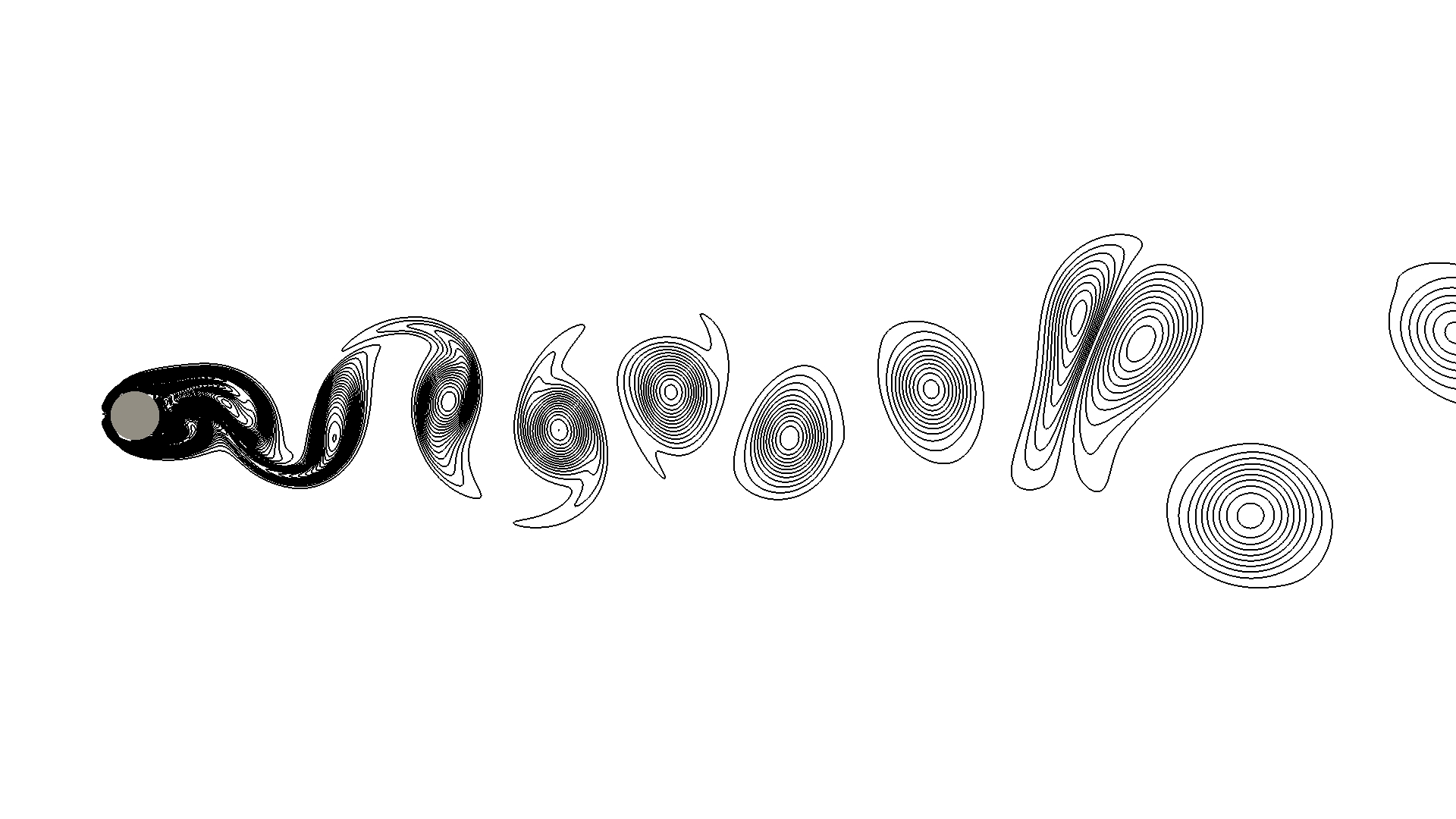}
        \caption{$1.5 \Unit{s}$}
        \label{fig:1_cyn_vortex_re200_t1500ms}
    \end{subfigure}%
    \\
    \begin{subfigure}[b]{0.48\textwidth}
        \includegraphics[trim = 0mm 80mm 0mm 80mm, clip, width=\textwidth]{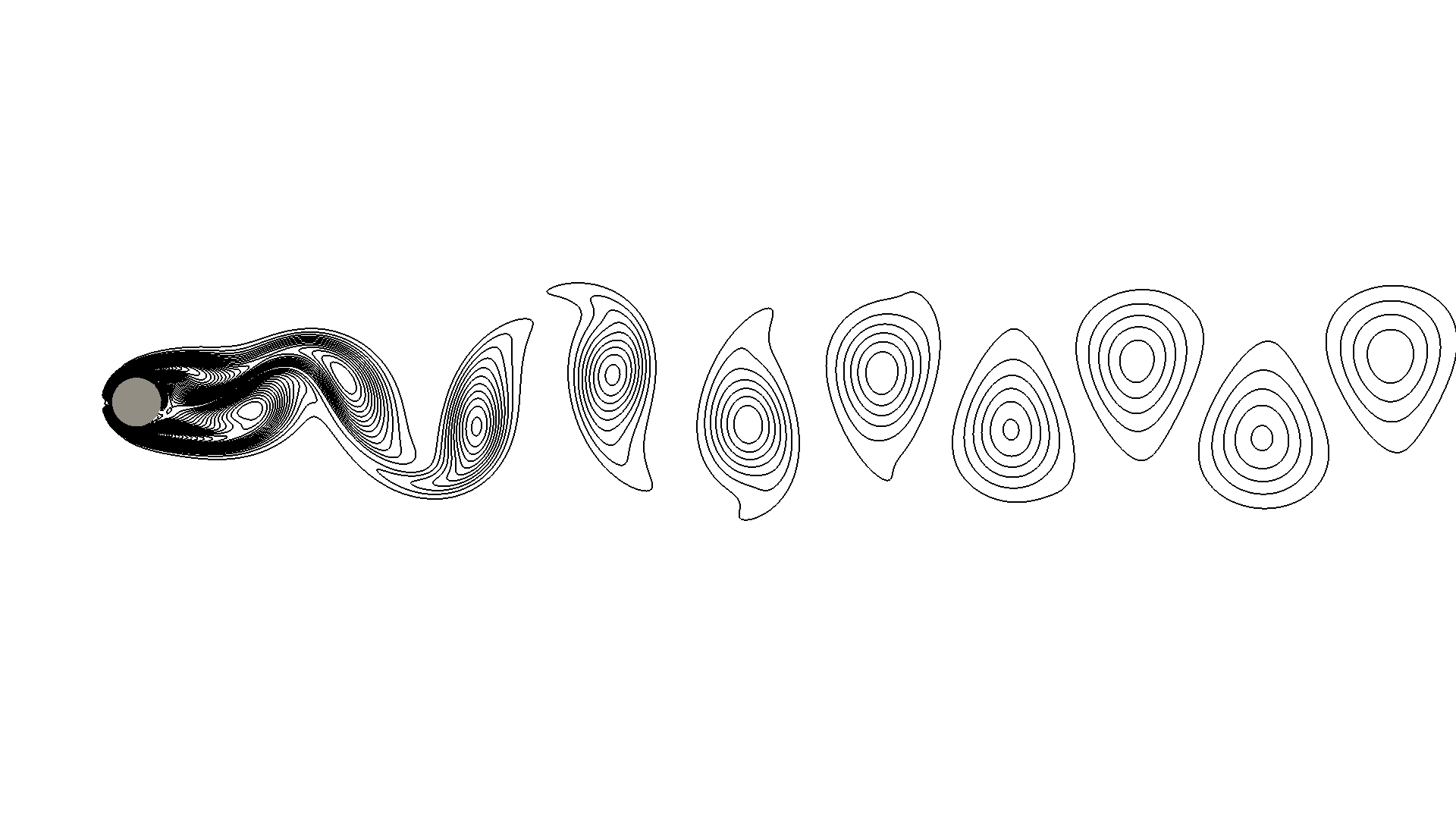}
        \caption{$3.0 \Unit{s}$}
        \label{fig:1_cyn_vortex_re100_t3000ms}
    \end{subfigure}%
    ~
    \begin{subfigure}[b]{0.48\textwidth}
        \includegraphics[trim = 0mm 80mm 0mm 80mm, clip, width=\textwidth]{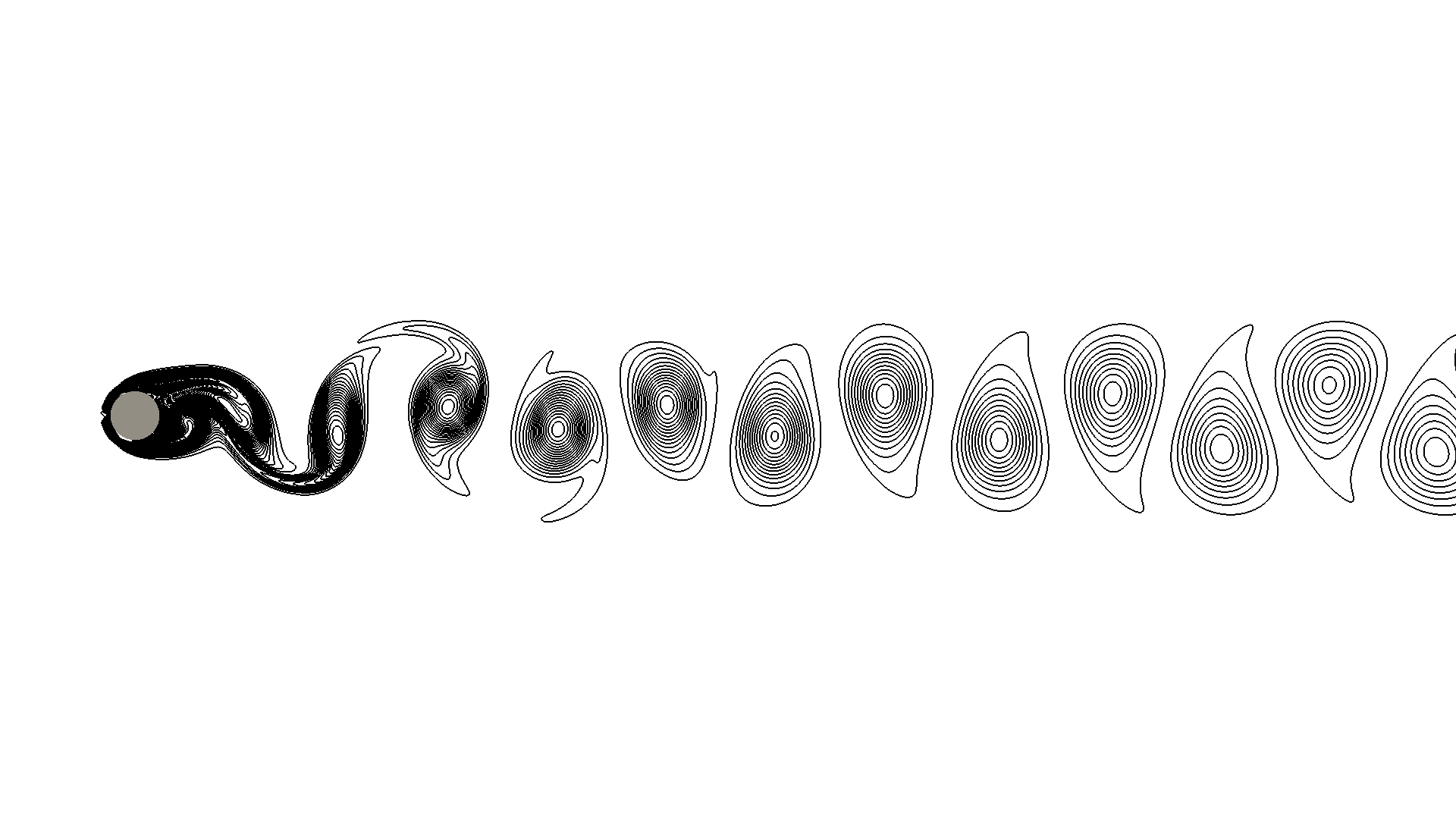}
        \caption{$3.0 \Unit{s}$}
        \label{fig:1_cyn_vortex_re200_t3000ms}
    \end{subfigure}%
    \\
    \begin{subfigure}[b]{0.48\textwidth}
        \includegraphics[trim = 0mm 80mm 0mm 80mm, clip, width=\textwidth]{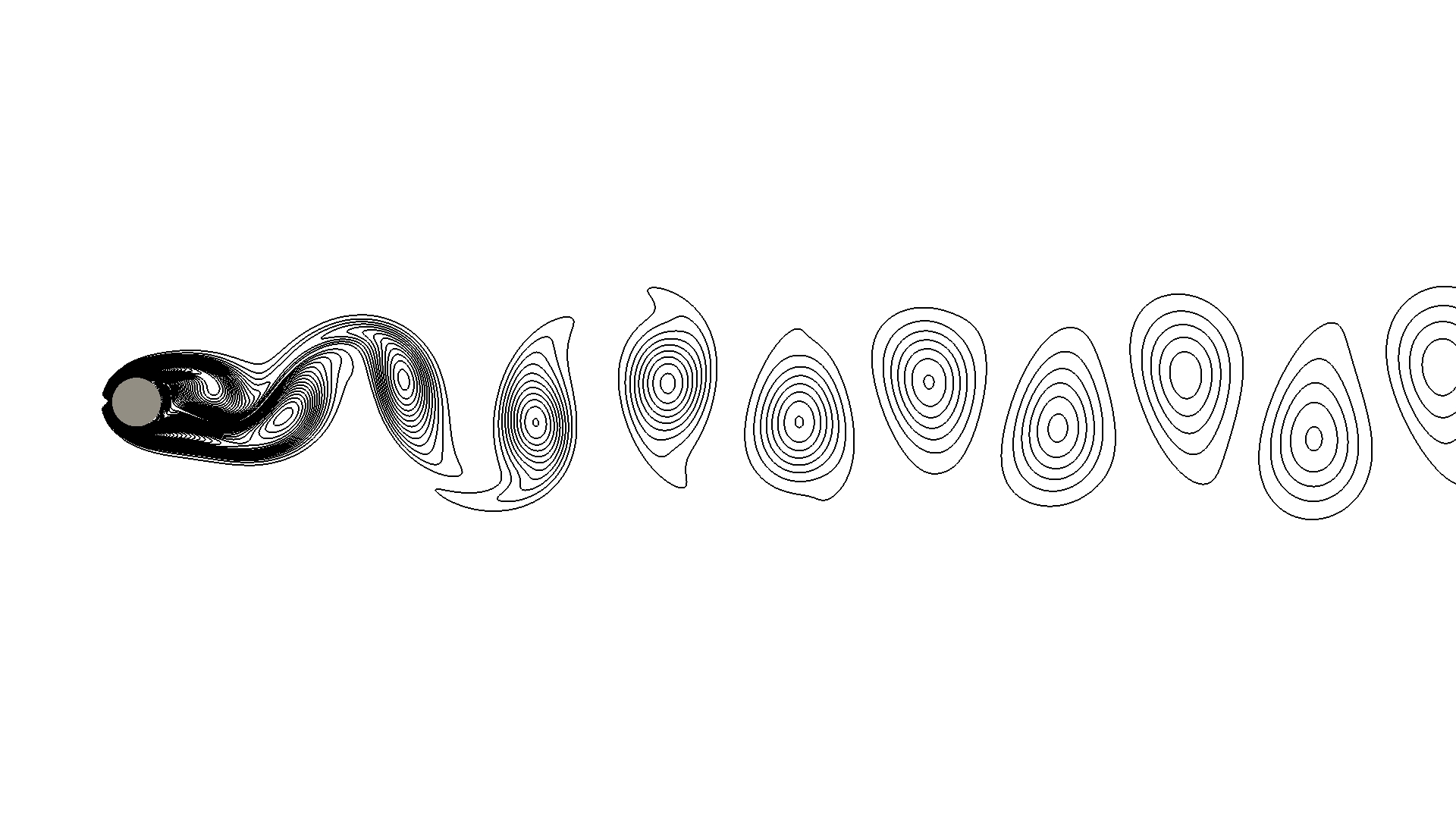}
        \caption{$5.0 \Unit{s}$}
        \label{fig:1_cyn_vortex_re100_t5000ms}
    \end{subfigure}%
    ~
    \begin{subfigure}[b]{0.48\textwidth}
        \includegraphics[trim = 0mm 80mm 0mm 80mm, clip, width=\textwidth]{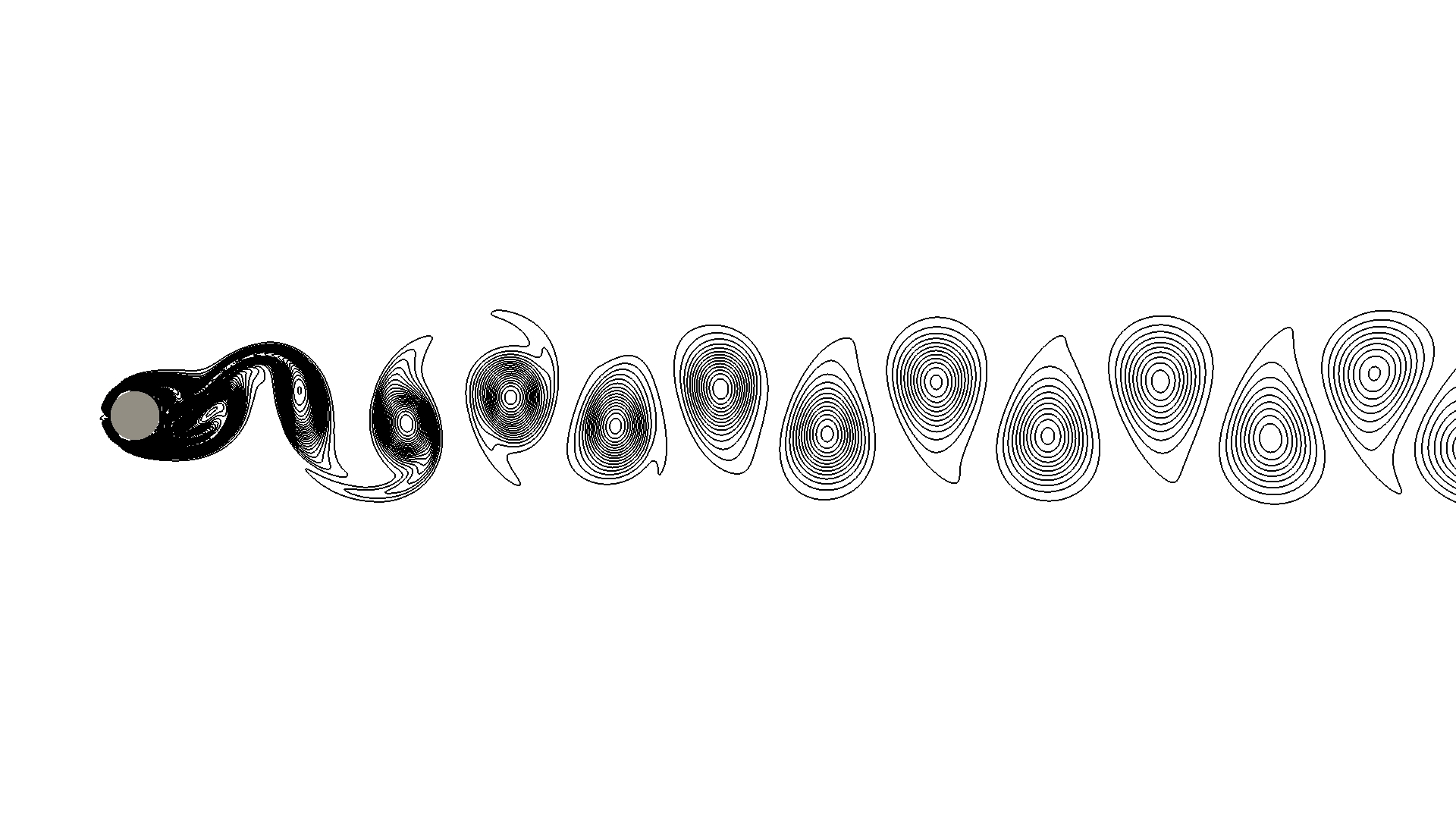}
        \caption{$5.0 \Unit{s}$}
        \label{fig:1_cyn_vortex_re200_t5000ms}
    \end{subfigure}%
    \caption{Instantaneous vorticity contours of the unsteady flows around a cylinder at different Reynolds numbers. (a), (c), (e), (g), (i) $Re=100$. (b), (d), (f), (h), (j) $Re=200$.}
    \label{fig:1_cyn_vortex_stream_unsteady}
\end{figure}

The temporal evolution of the unsteady flows at $Re=100$ and $200$ are shown in Fig.~\ref{fig:1_cyn_vortex_stream_unsteady}, in which the development of von K\'{a}rm\'{a}n vortex shedding in the wake region is captured in detail. The unsteady flows can be quantified via several flow parameters, such as the Strouhal number $St = f D / u_{\infty}$, drag coefficient $C_{\Des{D}}$, and lift coefficient $C_{\Des{L}} = F_y / (0.5\rho_{\infty} u_{\infty}^2 D)$, where $f$ is the vortex shedding frequency (in this study, two point probes are positioned at $(15D, 15D)$ and $(20D, 15D)$ to record the time signals of the $y$-component of the velocity, whose FFT are then used to determine the vortex shedding frequency) and $F_y$ is the $y$-component of the total force acting on the cylinder, respectively. A comparison of flow results is shown in Table~\ref{tab:1_cyn_vortex_para_unsteady}, in which good agreement is obtained.
\begin{table}[!htbp]
    \centering
    \begin{tabular}{lccc}
        \hline\hline
         & $St$ & $C_{\Des{D}}$ & $C_{\Des{L}}$\\
        \hline
        $Re=100$ & & & \\
        \citet{berger1972periodic}$^*$ & $0.160-0.170$ & $-$ & $-$\\
        \citet{linnick2005high} & $0.165$ & $1.320$ & $\pm 0.320$\\
        \citet{brehm2015locally} & $0.170$ & $1.380$ & $\pm 0.342$\\
        \emph{Present study} & $0.163$ & $1.304$ & $\pm0.310$\\
        $Re=200$ & & & \\
        \citet{berger1972periodic}$^*$ & $0.180-0.190$ & $-$ & $-$\\
        \citet{linnick2005high} & $0.192$ & $1.300$ & $\pm 0.660$\\
        \citet{brehm2015locally} & $0.198$ & $1.380$ & $\pm 0.700$\\
        \emph{Present study} & $0.187$ & $1.286$ & $\pm 0.613$\\
        \hline\hline
    \end{tabular}
    \caption{Comparison of results for unsteady flows around a cylinder. [Nomenclature: $St$, Strouhal number; $C_{\Des{D}}$, drag coefficient; $C_{\Des{L}}$, lift coefficient; $*$, experimental results.]}
    \label{tab:1_cyn_vortex_para_unsteady}
\end{table}

\subsection{Supersonic shock-sphere interaction}

\begin{figure}[!htbp]
    \centering
    \begin{subfigure}[b]{0.40\textwidth}
        \includegraphics[width=\textwidth]{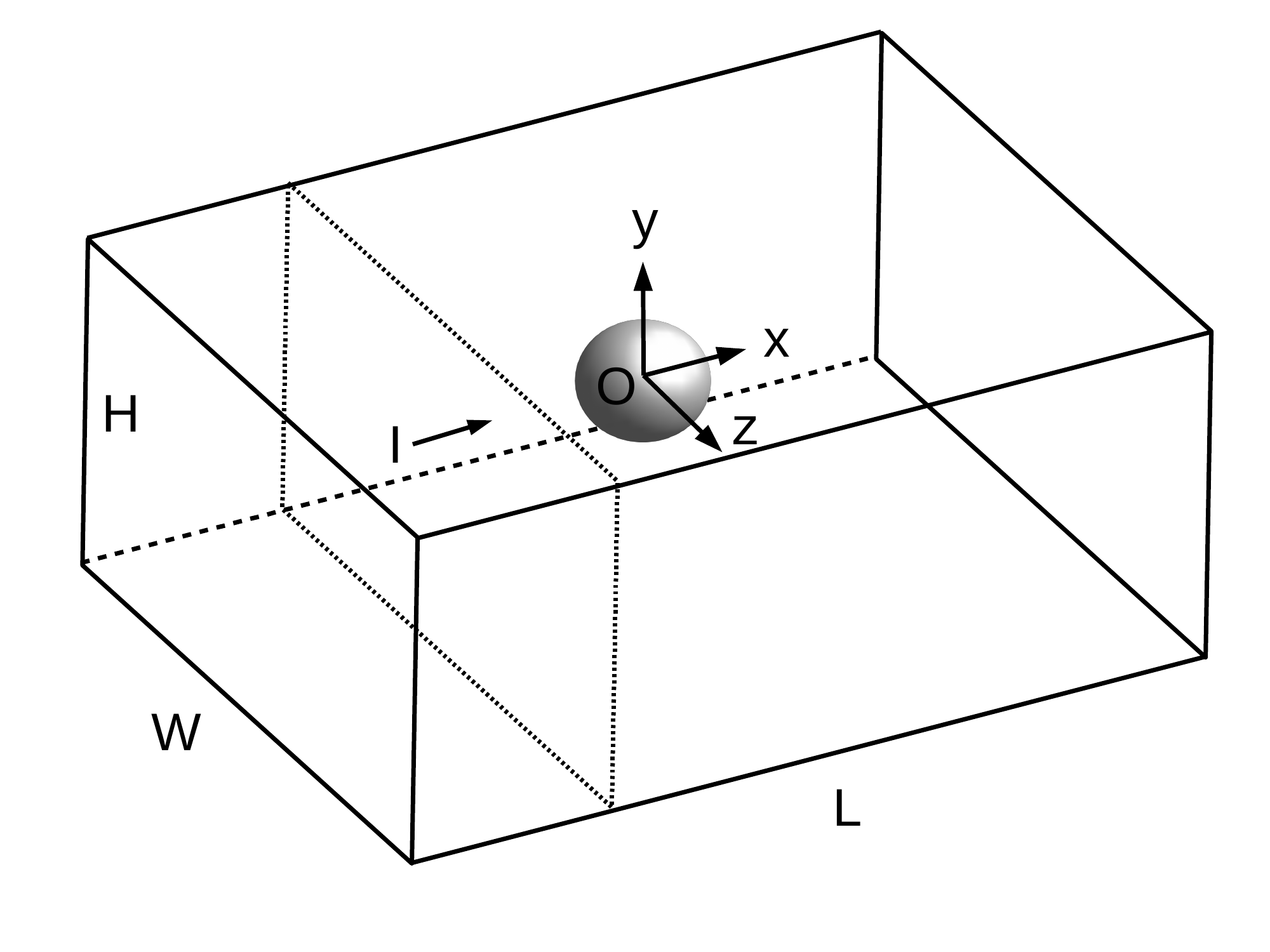}
        \caption{}
        \label{fig:shock_sphere_demo_a}
    \end{subfigure}%
    ~
    \begin{subfigure}[b]{0.40\textwidth}
        \includegraphics[width=\textwidth]{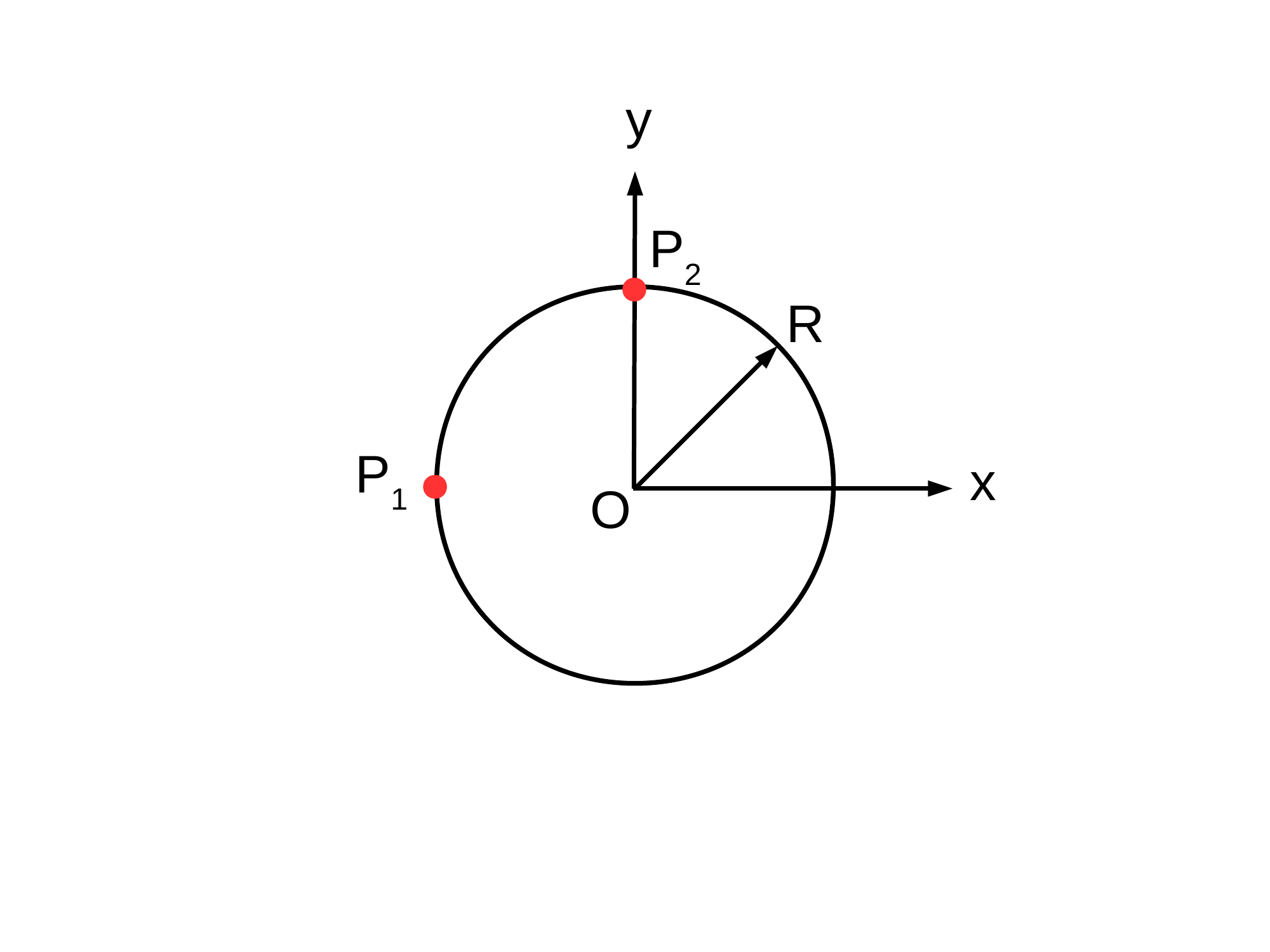}
        \caption{}
        \label{fig:shock_sphere_demo_b}
    \end{subfigure}%
    \caption{Schematic diagrams illustrating the shock-sphere interaction problem. (a) The $3D$ problem domain. (b) Pressure probe locations. [Nomenclature: $L$, domain length; $H$, domain height; $W$, domain width; $I$, incident shock; $R$, sphere radius; $P_1$ and $P_2$, pressure probes at sphere surface.]}
    \label{fig:shock_sphere_demo}
\end{figure}

As illustrated in Fig.~\ref{fig:shock_sphere_demo}, the unsteady drag force and pressure history acting on a radius $R=0.04 \Unit{m}$ sphere suspended in a $L \times H \times W=0.5 \Unit{m} \times 0.3 \Unit{m} \times 0.3 \Unit{m}$ test region and impacted by a Mach $1.22$ planar incident shock are studied to validate the proposed field function applied to supersonic fluid-solid interactions. The center of the sphere overlaps with the center of the test region and is at the origin position $O(0,0,0)$, and the incident shock is initially positioned at $x=-1.5R$. The pre-shock and post-shock states are $(\rho_1, u_1, v_1, w_1, p_1)=(1.205 \Unit{kg/m^3}, 0, 0, 0, 101325 \Unit{Pa})$ and $(\rho_2, u_2, v_2, w_2, p_2)=(1.658 \Unit{kg/m^3}, 114.477 \Unit{m/s}, 0, 0, 159060 \Unit{Pa})$, respectively. The drag coefficient is computed as $C_{\Des{D}} = F_x / (0.5\rho_2 u_2^2 \pi R^2)$, where $F_x$ is the $x$-component of the total force acting on the sphere.

\citet{tanno2003interaction} experimentally measured the drag coefficient and pressure history using a shock tube facility. In addition, employing a $2D$ curvilinear grid and the axisymmetric Navier--Stokes equations, they also numerically simulated the flow with the no-slip wall boundary condition and reported the obtained drag coefficient. In this study, $3D$ Cartesian grids are used. To achieve an affordable computational cost, the Euler equations with the slip-wall boundary condition are applied instead. Since the viscous effect is very limited in this supersonic flow, and the pressure force dominants the shock-sphere interaction \citep{tanno2003interaction}, the inviscid flow assumption is suggested to be adequate. Three levels of grids, $300\times180\times180$, $400\times240\times240$, and $500\times300\times300$, were used to test the asymptotic range of convergence. The $400\times240\times240$ grid, which has a grid resolution of about $0.03D$ ($32$ nodes per diameter) was found to be sufficient, and its numerical results are reported below.
\begin{figure}[!htbp]
    \centering
    \begin{subfigure}[b]{0.24\textwidth}
        \includegraphics[trim = 75mm 0mm 75mm 0mm, clip, width=\textwidth]{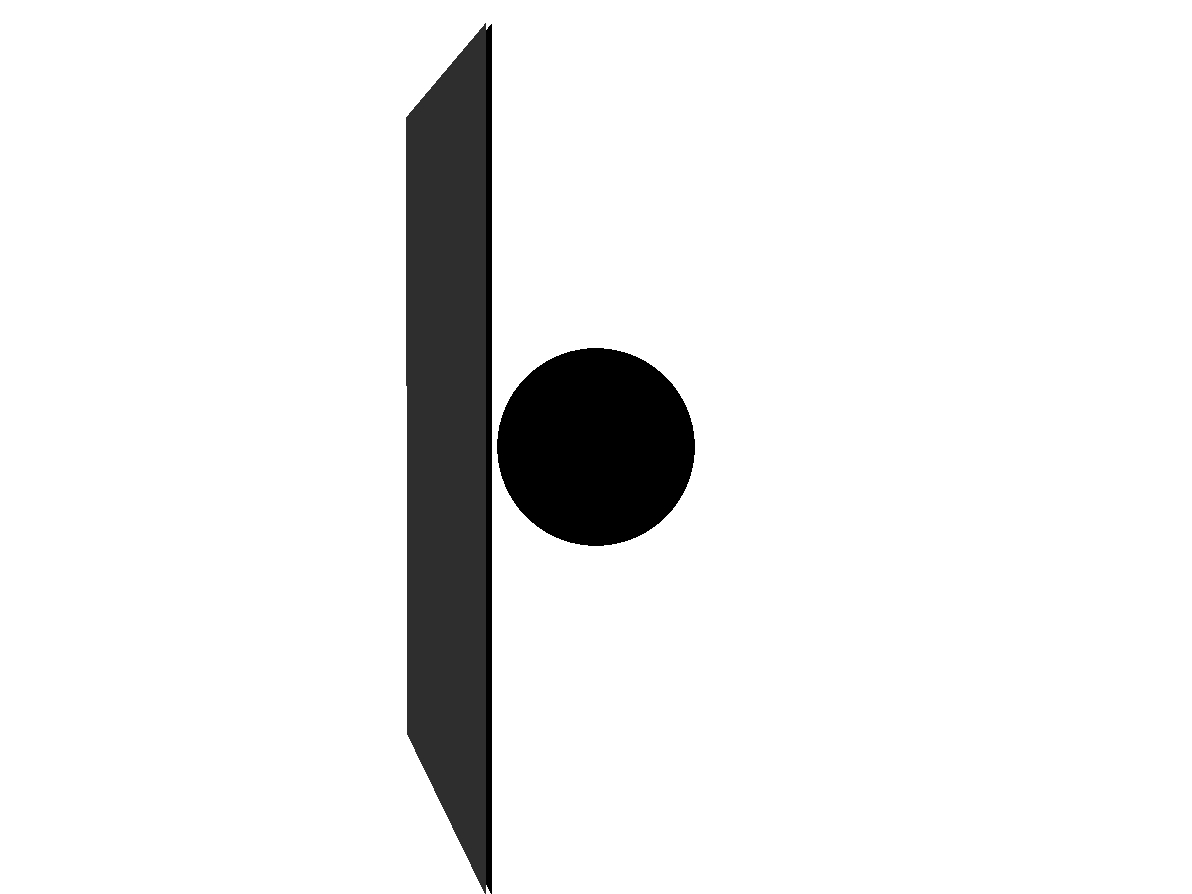}
        \caption{$0 \Unit{\mu s}$}
        \label{fig:1_cyn_force_schilieren_iso_t00}
    \end{subfigure}%
    ~
    \begin{subfigure}[b]{0.24\textwidth}
        \includegraphics[trim = 75mm 0mm 75mm 0mm, clip, width=\textwidth]{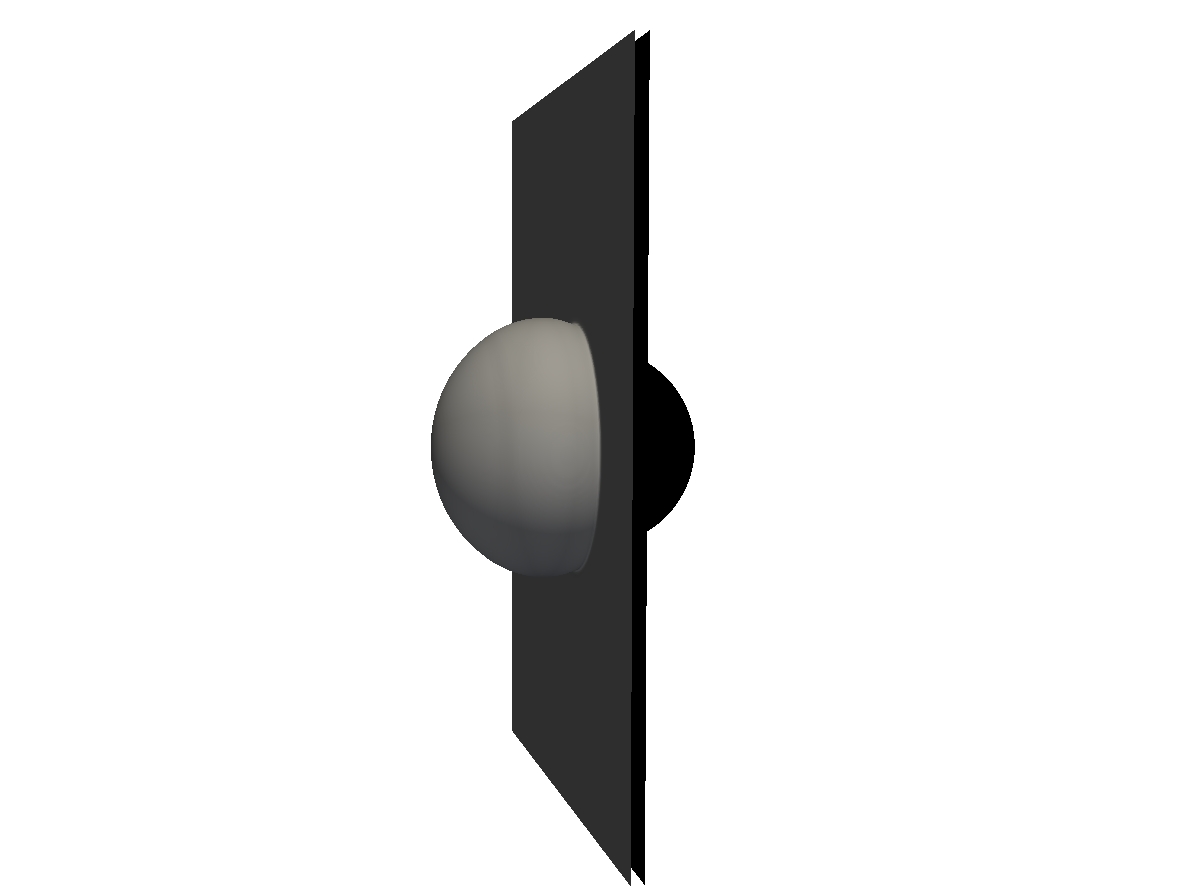}
        \caption{$126 \Unit{\mu s}$}
        \label{fig:1_cyn_force_schilieren_iso_t10}
    \end{subfigure}%
    ~
    \begin{subfigure}[b]{0.24\textwidth}
        \includegraphics[trim = 75mm 0mm 75mm 0mm, clip, width=\textwidth]{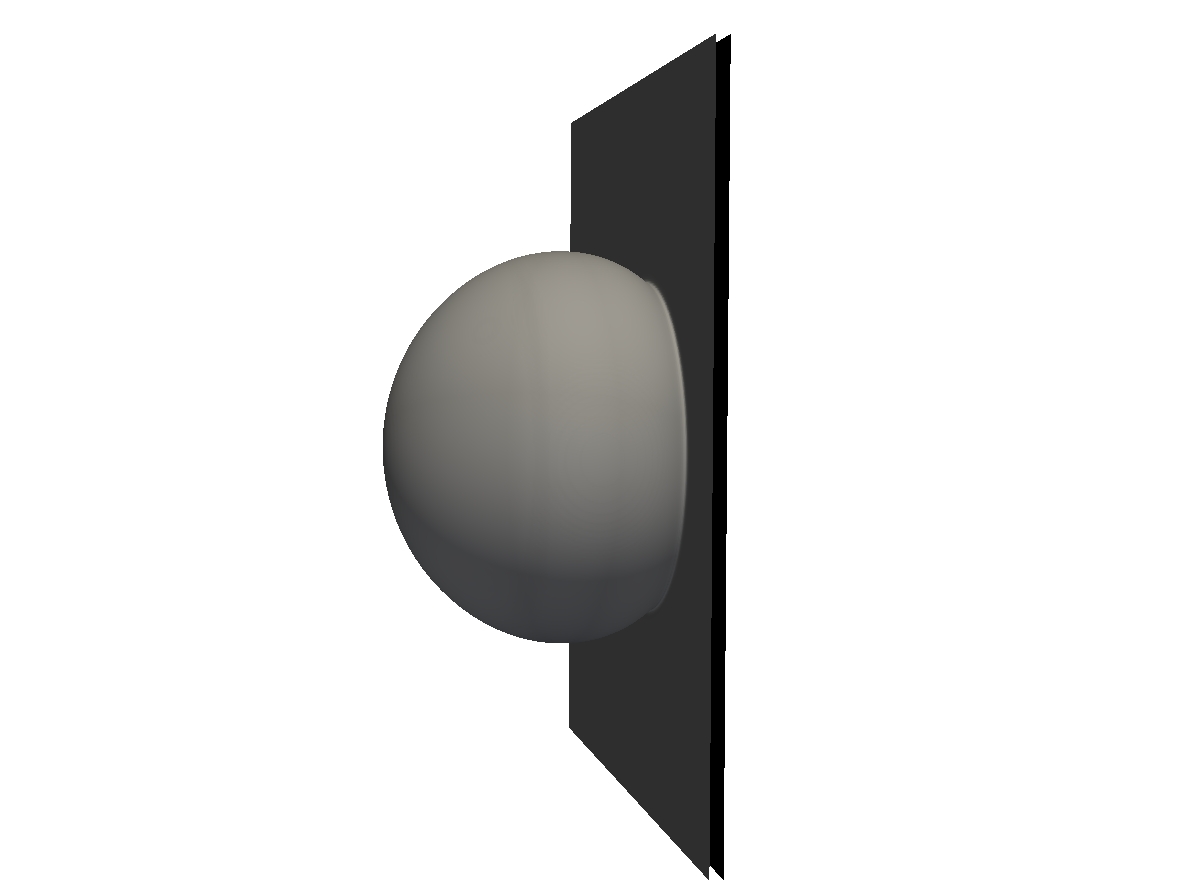}
        \caption{$196 \Unit{\mu s}$}
        \label{fig:1_cyn_force_schilieren_iso_t15}
    \end{subfigure}%
    ~
    \begin{subfigure}[b]{0.24\textwidth}
        \includegraphics[trim = 75mm 0mm 75mm 0mm, clip, width=\textwidth]{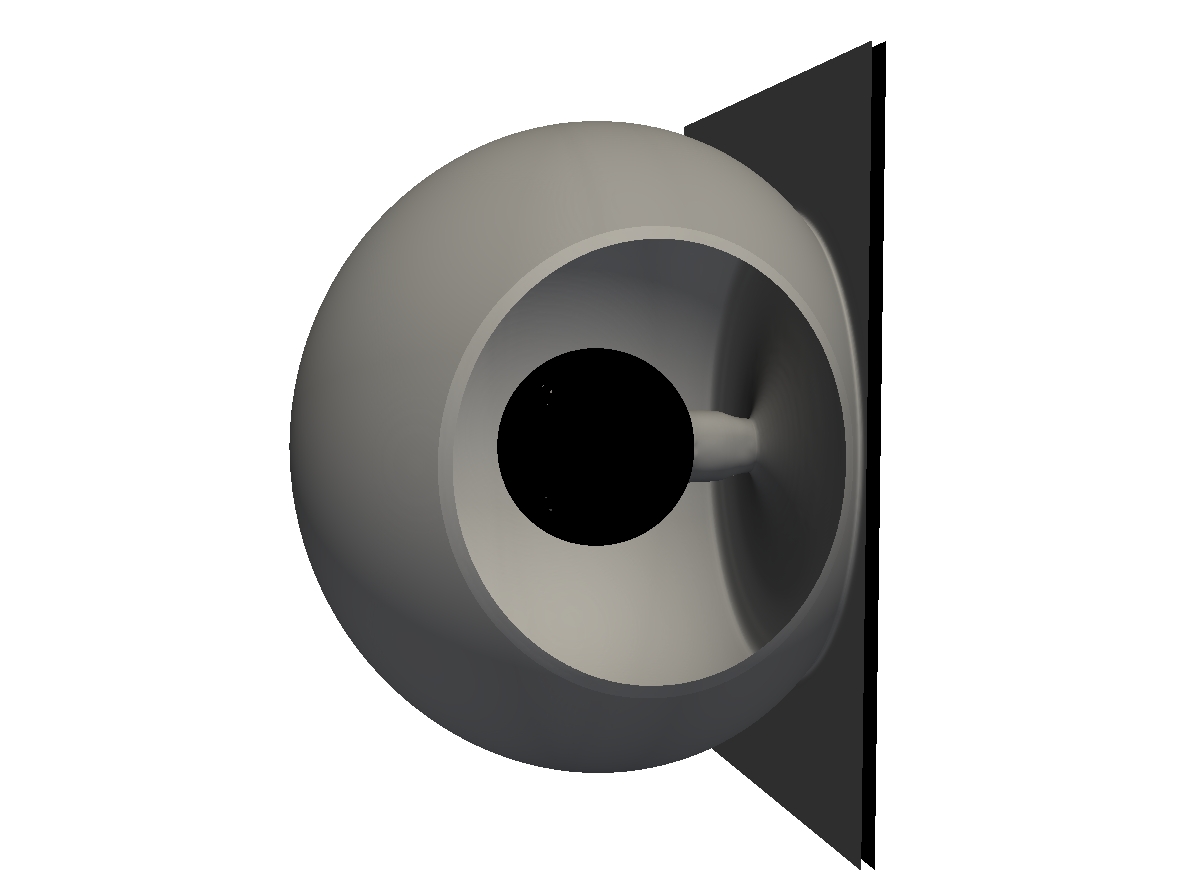}
        \caption{$332 \Unit{\mu s}$}
        \label{fig:1_cyn_force_schilieren_iso_t25}
    \end{subfigure}%
    \caption{Numerical Schlieren iso-surface of shock-sphere interaction.}
    \label{fig:1_cyn_force_schilieren_iso}
\end{figure}
\begin{figure}[!htbp]
    \centering
    \begin{subfigure}[b]{0.40\textwidth}
        \includegraphics[width=\textwidth]{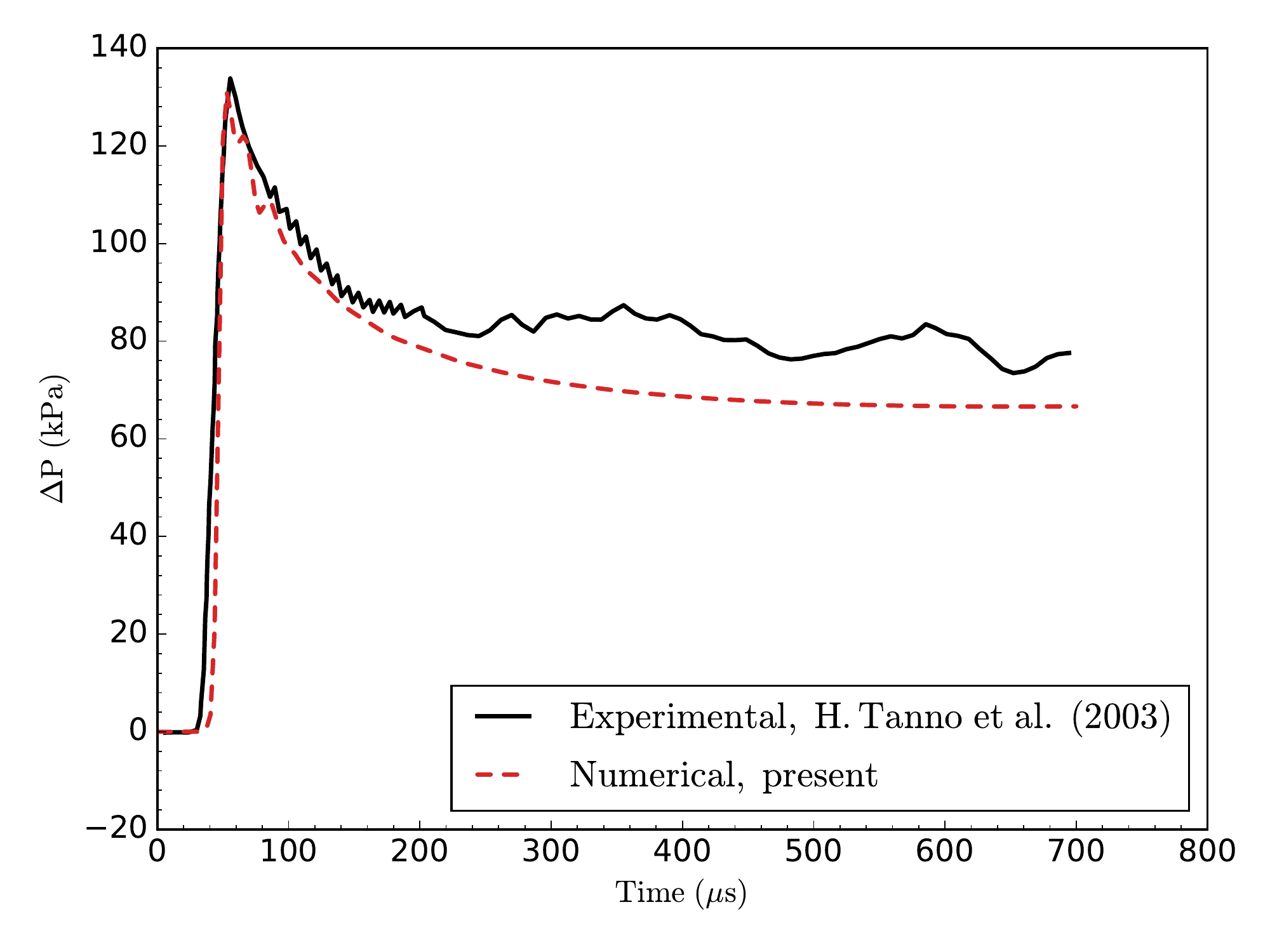}
        \caption{}
        \label{fig:1_cyn_force_p1}
    \end{subfigure}%
    ~
    \begin{subfigure}[b]{0.40\textwidth}
        \includegraphics[width=\textwidth]{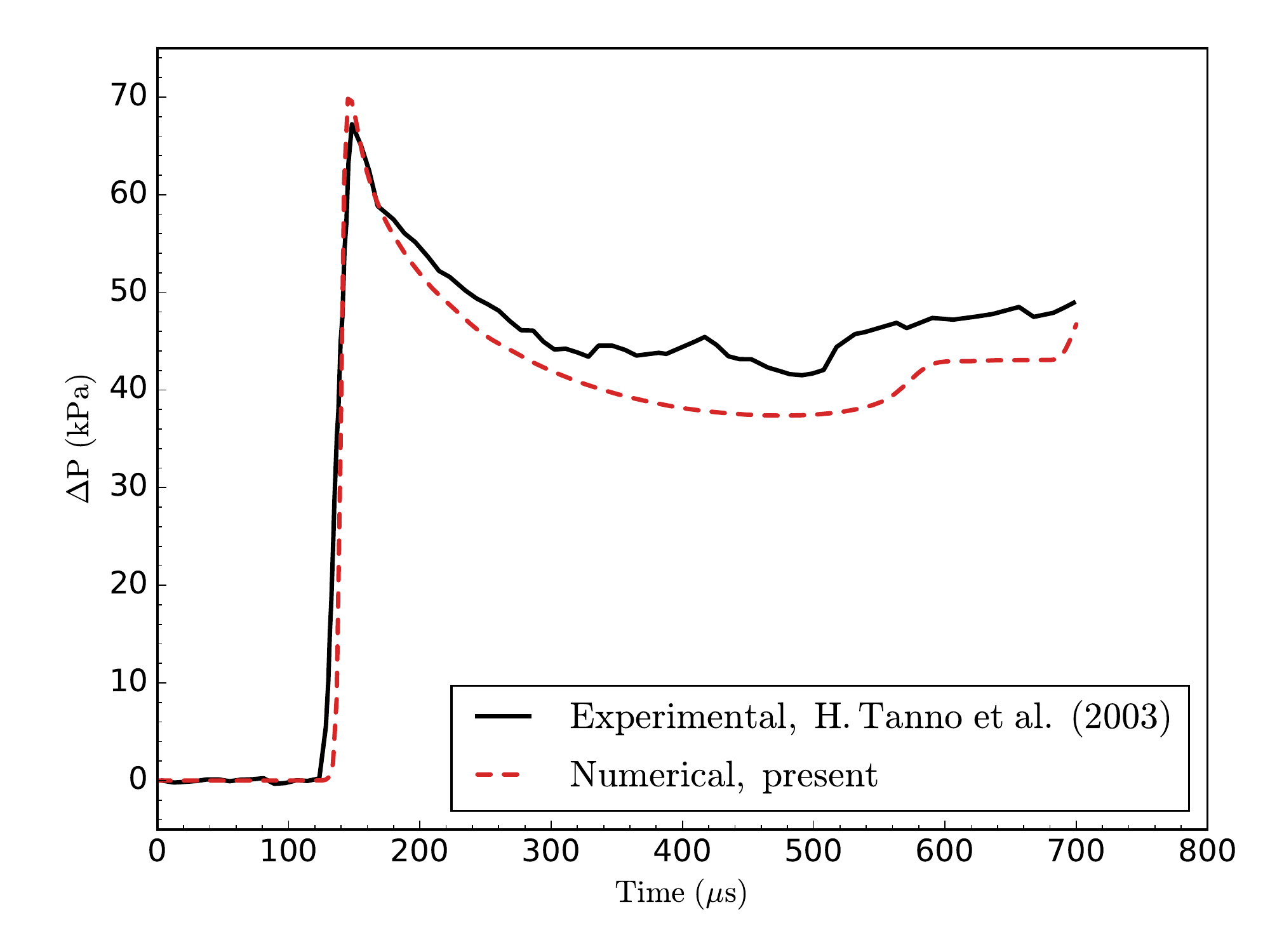}
        \caption{}
        \label{fig:1_cyn_force_p7}
    \end{subfigure}%
    \caption{Pressure variation over time at the probe locations. (a) Probe $P_1$. (b) Probe $P_2$.}
    \label{fig:1_cyn_force_p}
\end{figure}
\begin{figure}[!htbp]
    \centering
    \includegraphics[width=0.5\textwidth]{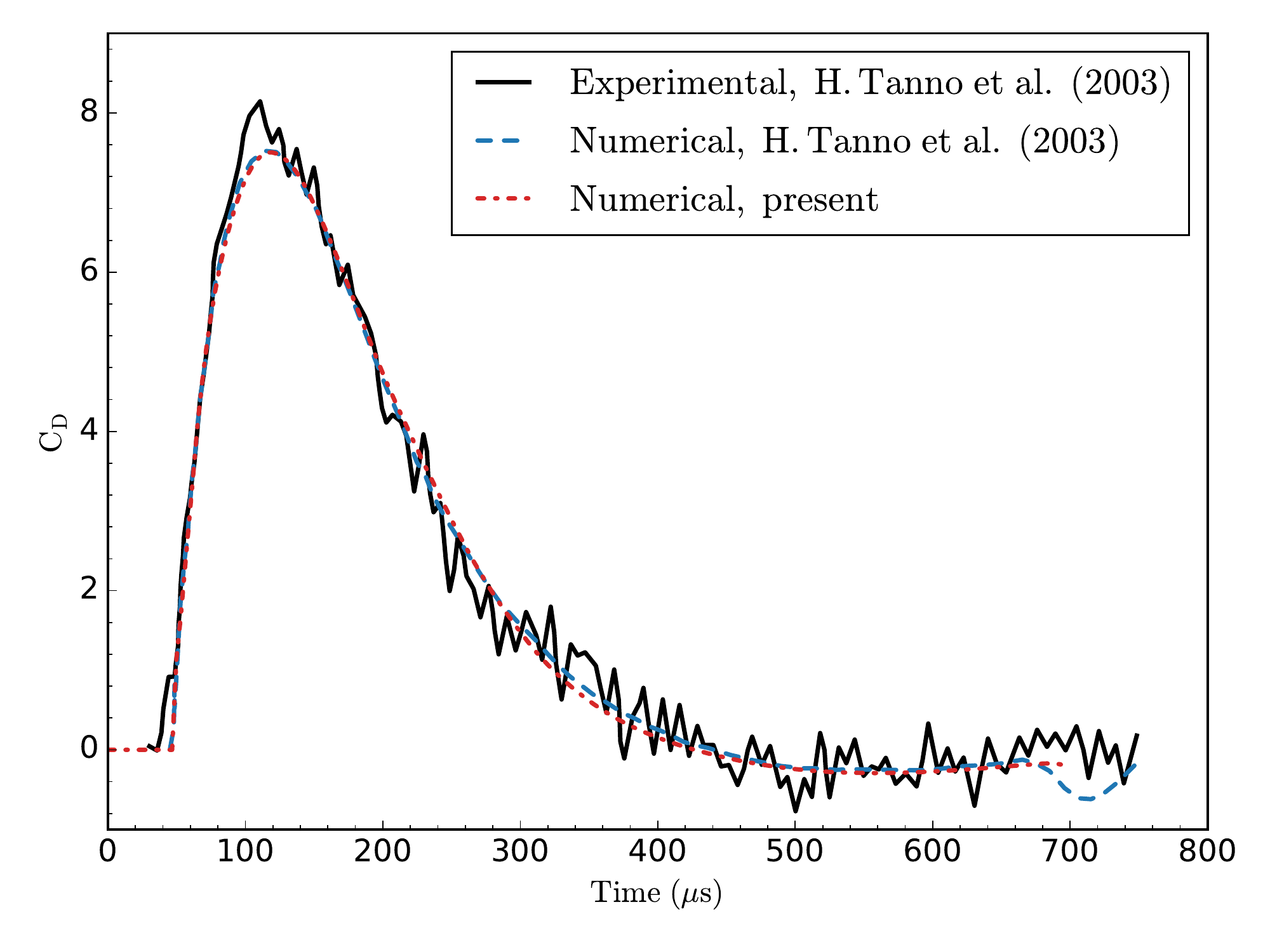}
    \caption{Comparison of drag coefficient for shock-sphere interaction.}
    \label{fig:1_cyn_force_cd}
\end{figure}

The evolution of the shock front through an iso-surface of the numerical Schlieren field is captured in Fig.~\ref{fig:1_cyn_force_schilieren_iso}, in which the reflection and diffraction of the shock wave along sphere surface and the formation of wake by shock collision are clearly illustrated. A comparison of the predicted pressure variation $\Delta p = p - p_1$ at the two probe locations with the experimental measurements in reference \citep{tanno2003interaction} is shown in Fig.~\ref{fig:1_cyn_force_p}. For $t \in [0, 160 \Unit{\mu s}]$, the numerical pressure data closely resemble the experimental ones, including both the arrival time and the value of the peak pressures. At the later stage, the numerical and experimental results present moderate discrepancies. A further comparison of the predicted drag coefficient with the experimental and numerical data in reference \citep{tanno2003interaction} is captured in Fig.~\ref{fig:1_cyn_force_cd}. The obtained drag coefficient herein is consistent with the numerical result in \citep{tanno2003interaction}, and the two numerical results both agree very well with the experimental measurement. It is worth noting that, in the reference \citep{tanno2003interaction}, compared to the pressure measurement, a model configuration less affecting the flow was used in the drag measurement, which might be one of the reasons leading to the different levels of agreement between the numerical and experimental data on pressure and drag coefficient at the late stage of evolution.

\subsection{A multibody contact and collision system}

\begin{figure}[!htbp]
    \centering
    \includegraphics[width=0.45\textwidth]{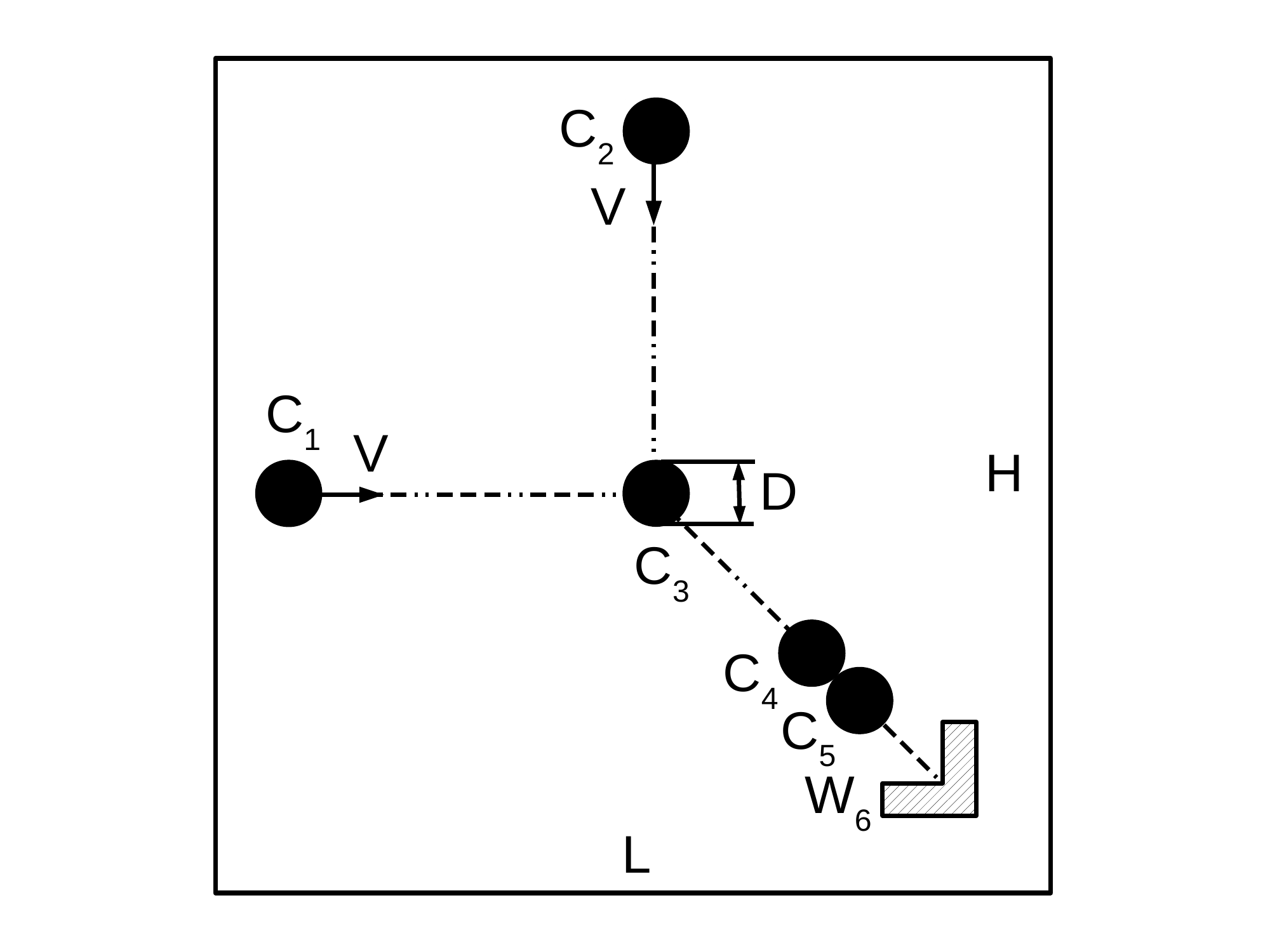}
    \caption{A schematic diagram of a fluid-solid system with analytically solvable multibody contact and collisions.}
    \label{fig:multibody_collision_demo}
\end{figure}

As illustrated in Fig.~\ref{fig:multibody_collision_demo}, a fluid-solid system is designed to examine the applicability of the field function for multicontact and collision scenarios. In a $L \times H = [-5D, 5D] \times [-5D, 5D]$ domain, the gas filling the domain has the initial state $(\rho_0, u_0, v_0, p_0)=(1.4 \Unit{kg/m^3}, 0, 0, 400 \Unit{Pa})$, in which the speed of sound is $a_0=20 \Unit{m/s}$, and the flow is assumed to be inviscid. In addition, five identical and cylindrical solids with diameter $D=1 \Unit{m}$ and a $90^{\circ}$-angled wall are placed in the domain. The centers of the solids are $C_1(-4D,0)$, $C_2(0,4D)$, $C_3(0,0)$, $C_4(2D, -2D)$, and $C_5(2D+1/\sqrt{2}D, -2D-1/\sqrt{2}D)$, respectively. The inner corner of the wall locates at $W_6(2.5D+\sqrt{2}D, -2.5D-\sqrt{2}D)$.

The solids $C_1$ and $C_2$ have an initial velocity magnitude $V=\Unit{50m/s}$, which corresponds to a Mach number of $M=V/a_0=2.5$. The solids $C_3$, $C_4$, and $C_5$ are initially stationary, and the wall $W_6$ is fixed in space. If all the collisions are assumed to be elastic and fluid forces acting on solids are neglected, then the motions of the solids are analytically solvable. More specifically, $C_1$ and $C_2$ will collide with $C_3$ simultaneously at $t=60 \Unit{ms}$ and completely transfer their momentum to $C_3$. At $t=100 - 10\sqrt{2} \Unit{ms}$, $C_3$ with the velocity state $(u,v)=(50 \Unit{m/s}, 50 \Unit{m/s})$ will collide with $C_4$ to form a solid string consisting of $C_3$, $C_4$, and $C_5$ and instantly transfer its momentum to $C_5$. At $t=100 \Unit{ms}$, $C_5$ will collide with the wall $W_6$ and bounce back, causing the collision sequence to be inverted. At $t=100 + 10\sqrt{2} \Unit{ms}$, $C_5$ with the velocity state $(u,v)=(50 \Unit{m/s}, 50 \Unit{m/s})$ will collide with $C_4$ to form a solid string consisting of $C_5$, $C_4$, and $C_3$ and instantly transfer its momentum to $C_3$. At $t=140 \Unit{ms}$, $C_3$ will collide with $C_1$ and $C_2$ simultaneously and completely transfer its $x$-momentum to $C_1$ and $y$-momentum to $C_2$. At $t= 200 \Unit{ms}$, $C_1$ and $C_2$ travel back to their initial positions.

The evolution process of this designed system involves multibody collisions with momentum transfer at both aligned and angled directions. In addition, the collision between $C_5$ and $W_6$ has two contact regions occurring simultaneously, which represents a multicontact collision and, in general, is a difficult problem to solve in rigid-body dynamics. Therefore, this fluid-solid system is solved to $t=200 \Unit{ms}$ on a $1200\times1200$ grid to test the proposed field function.

\begin{figure}[!htbp]
    \centering
    \begin{subfigure}[b]{0.22\textwidth}
        \includegraphics[trim = 40mm 0mm 40mm 0mm, clip, width=\textwidth]{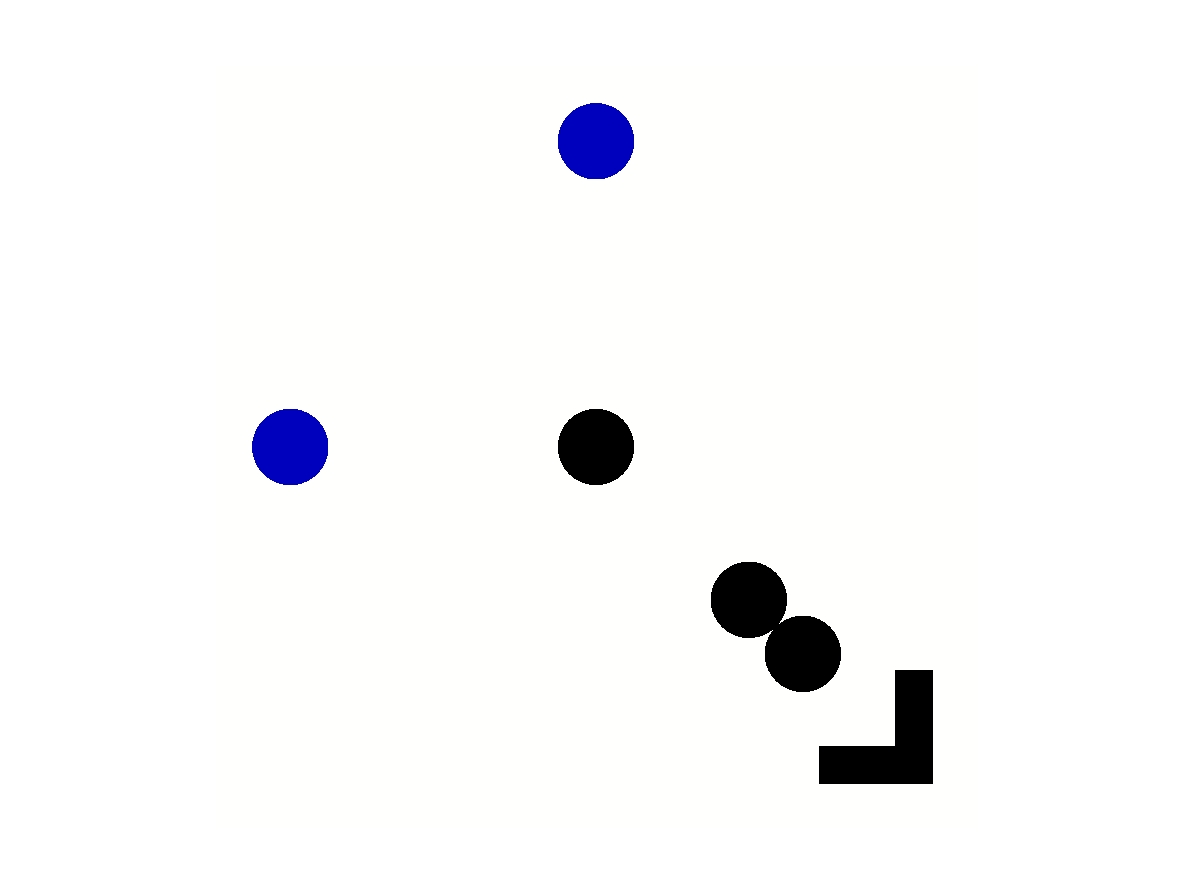}
        \caption{$0 \Unit{ms}$}
        \label{fig:multi_col_wall_m1200_t000ms}
    \end{subfigure}%
    ~
    \begin{subfigure}[b]{0.22\textwidth}
        \includegraphics[trim = 40mm 0mm 40mm 0mm, clip, width=\textwidth]{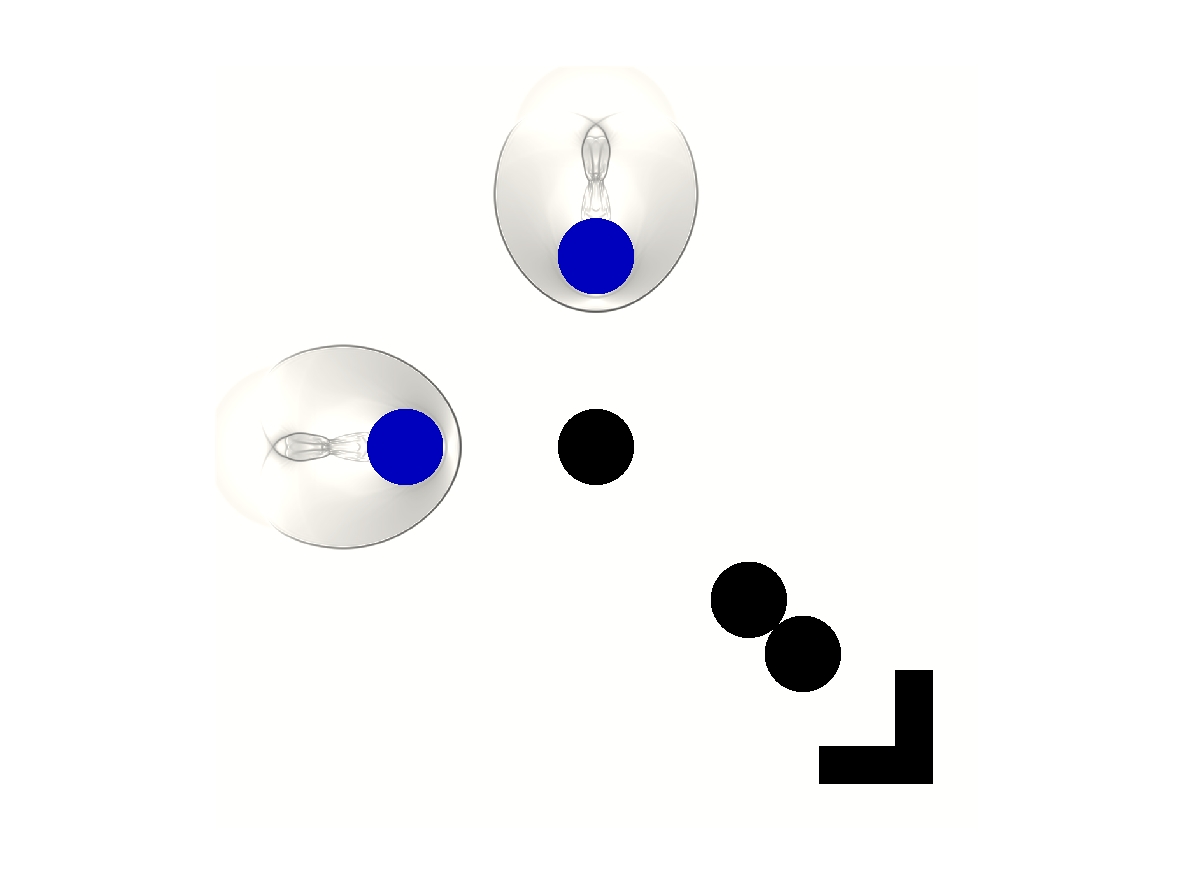}
        \caption{$30 \Unit{ms}$}
        \label{fig:multi_col_wall_m1200_t030ms}
    \end{subfigure}%
    ~
    \begin{subfigure}[b]{0.22\textwidth}
        \includegraphics[trim = 40mm 0mm 40mm 0mm, clip, width=\textwidth]{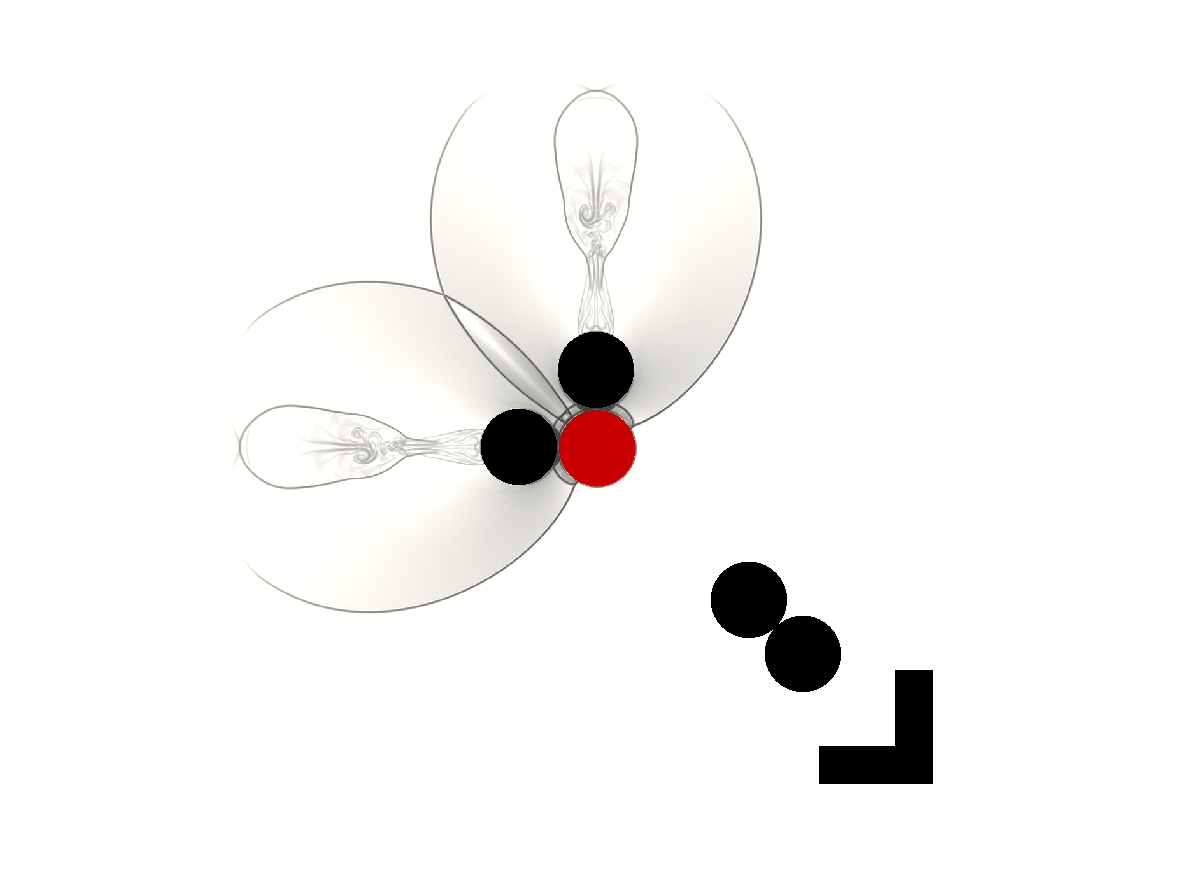}
        \caption{$60 \Unit{ms}$}
        \label{fig:multi_col_wall_m1200_t060ms}
    \end{subfigure}%
    \\
    \begin{subfigure}[b]{0.22\textwidth}
        \includegraphics[trim = 40mm 0mm 40mm 0mm, clip, width=\textwidth]{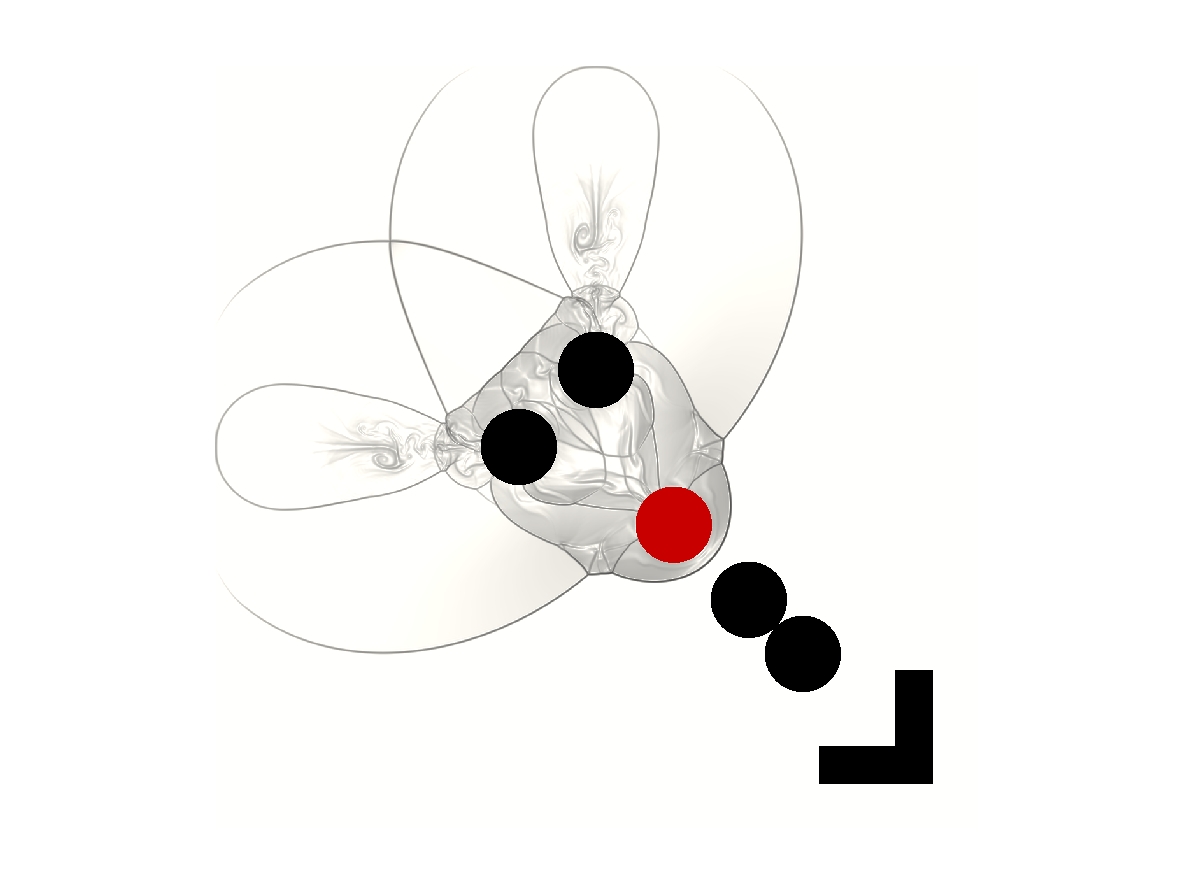}
        \caption{$80 \Unit{ms}$}
        \label{fig:multi_col_wall_m1200_t080ms}
    \end{subfigure}%
    ~
    \begin{subfigure}[b]{0.22\textwidth}
        \includegraphics[trim = 40mm 0mm 40mm 0mm, clip, width=\textwidth]{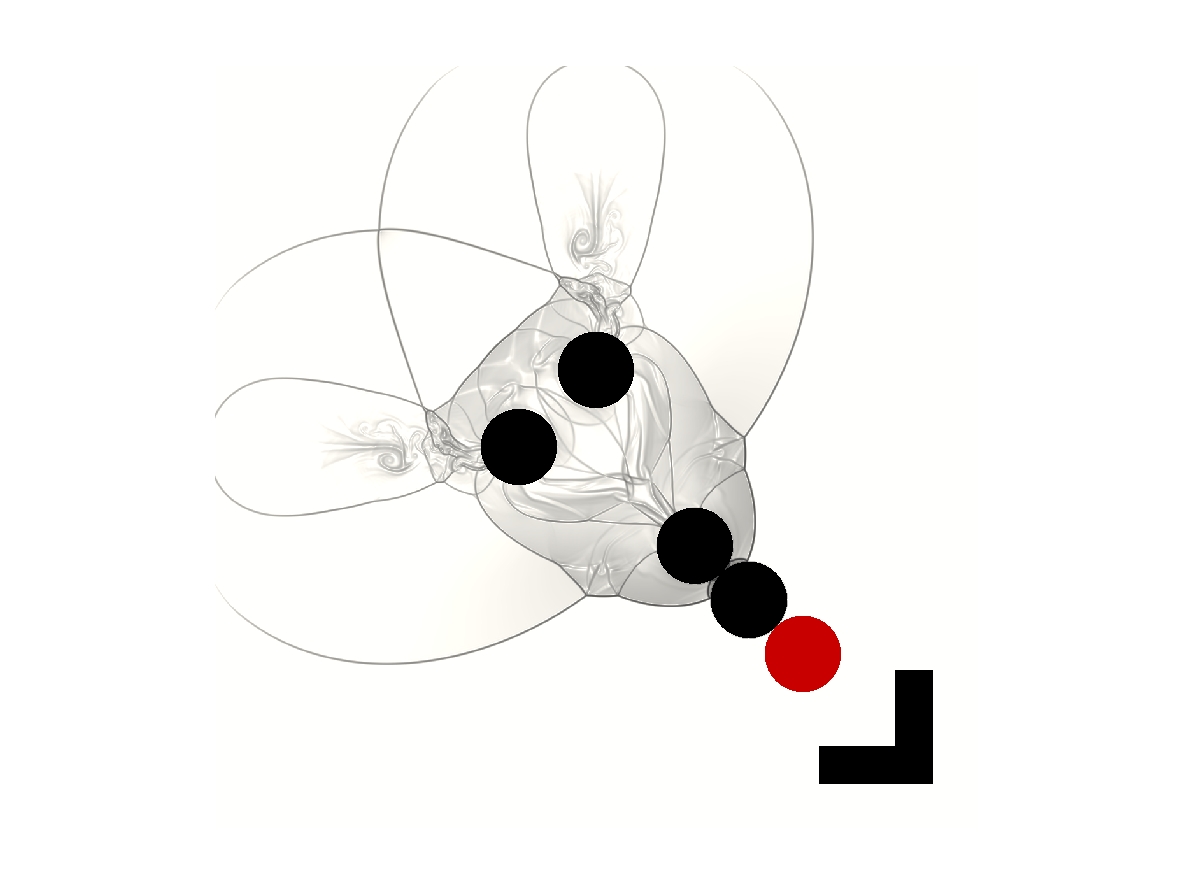}
        \caption{$86 \Unit{ms}$}
        \label{fig:multi_col_wall_m1200_t086ms}
    \end{subfigure}%
    ~
    \begin{subfigure}[b]{0.22\textwidth}
        \includegraphics[trim = 40mm 0mm 40mm 0mm, clip, width=\textwidth]{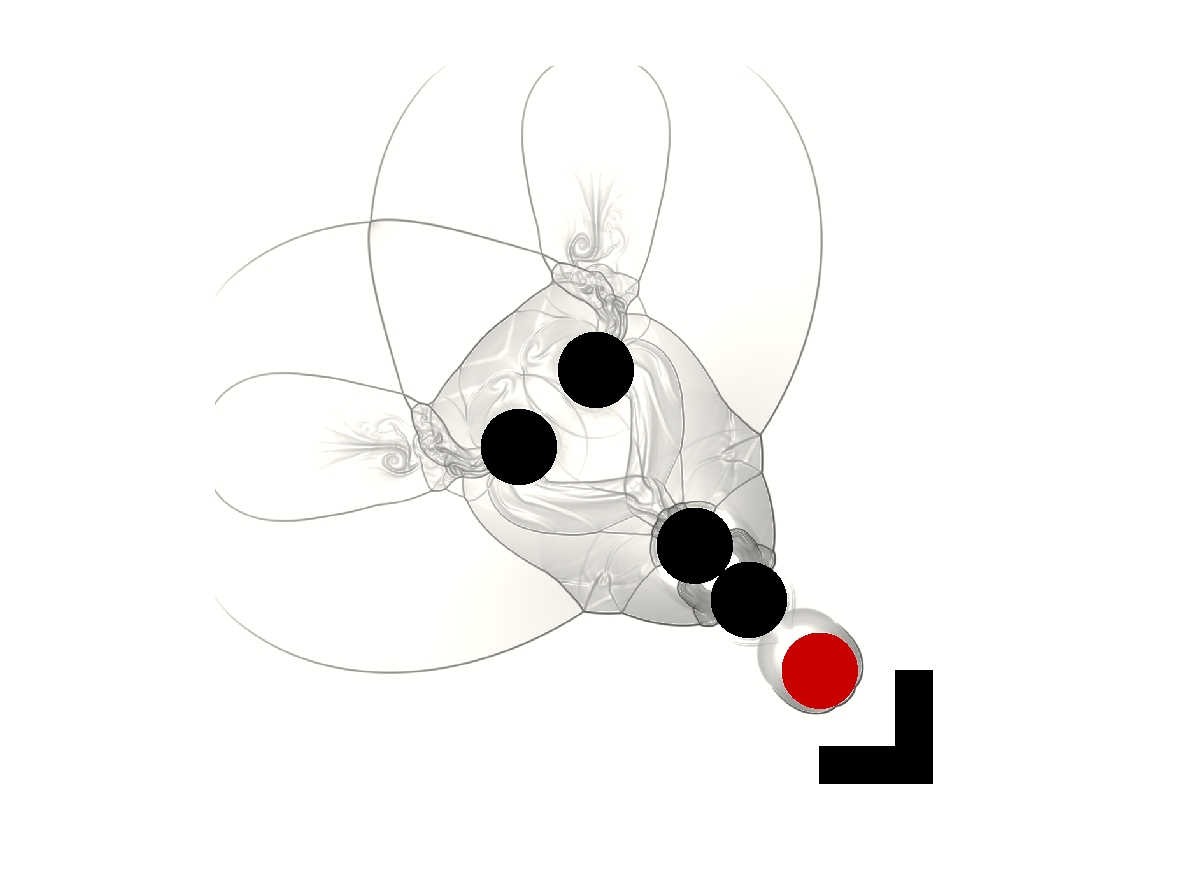}
        \caption{$90 \Unit{ms}$}
        \label{fig:multi_col_wall_m1200_t090ms}
    \end{subfigure}%
    \\
    \begin{subfigure}[b]{0.22\textwidth}
        \includegraphics[trim = 40mm 0mm 40mm 0mm, clip, width=\textwidth]{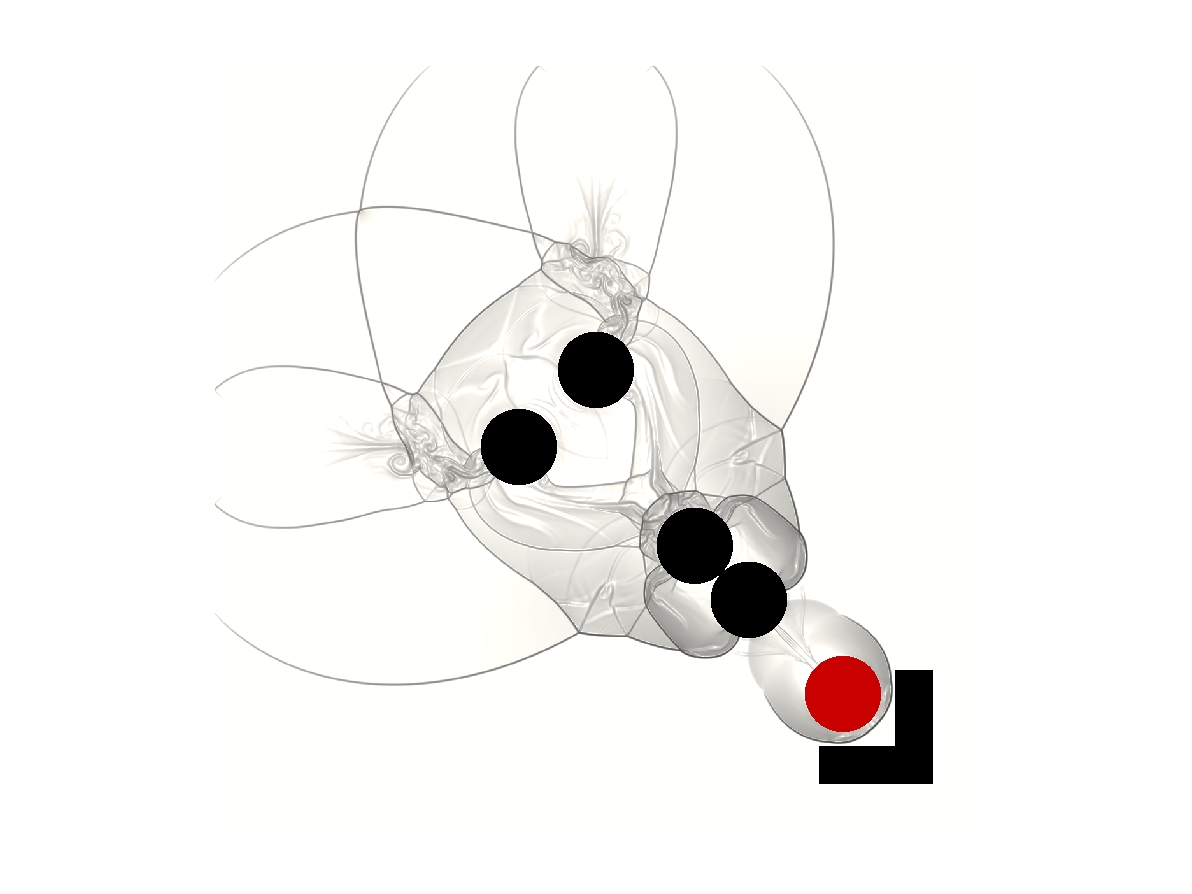}
        \caption{$96 \Unit{ms}$}
        \label{fig:multi_col_wall_m1200_t096ms}
    \end{subfigure}%
    ~
    \begin{subfigure}[b]{0.22\textwidth}
        \includegraphics[trim = 40mm 0mm 40mm 0mm, clip, width=\textwidth]{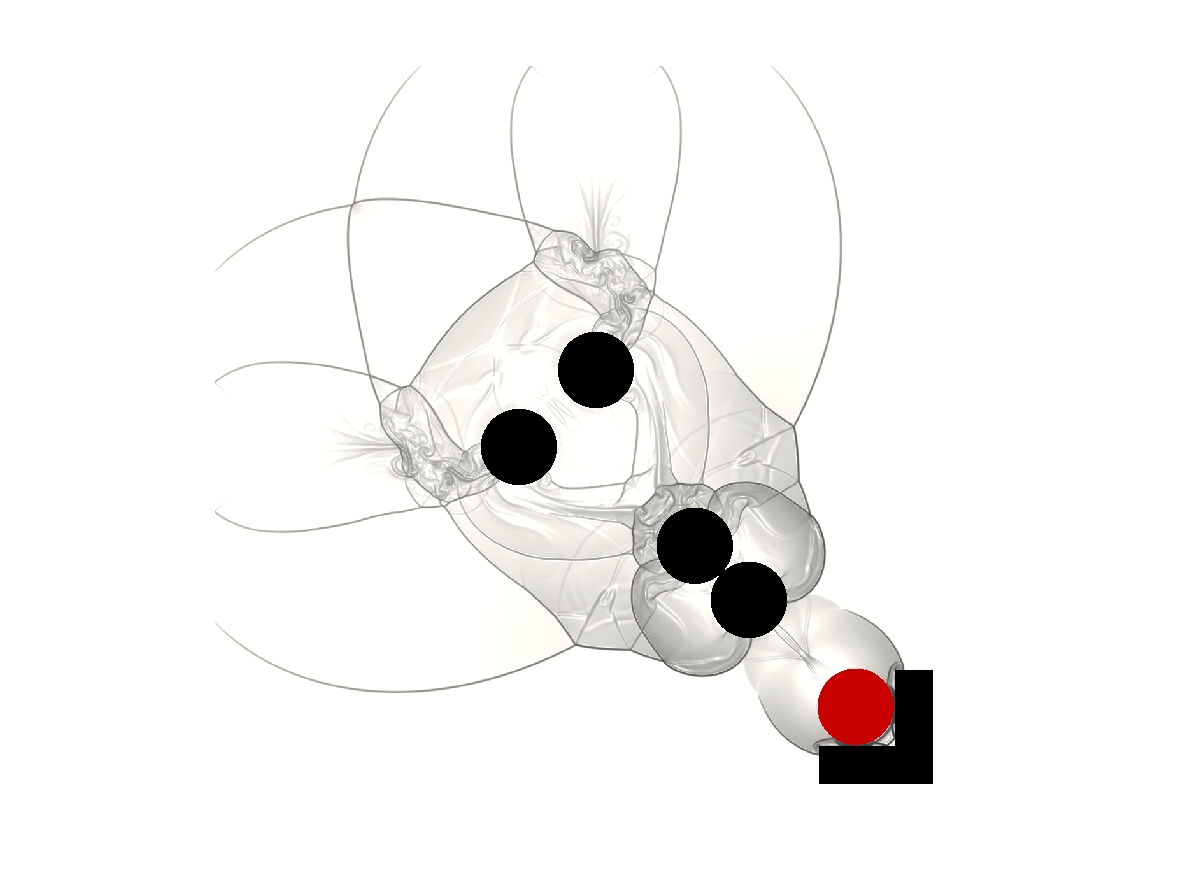}
        \caption{$100 \Unit{ms}$}
        \label{fig:multi_col_wall_m1200_t100ms}
    \end{subfigure}%
    ~
    \begin{subfigure}[b]{0.22\textwidth}
        \includegraphics[trim = 40mm 0mm 40mm 0mm, clip, width=\textwidth]{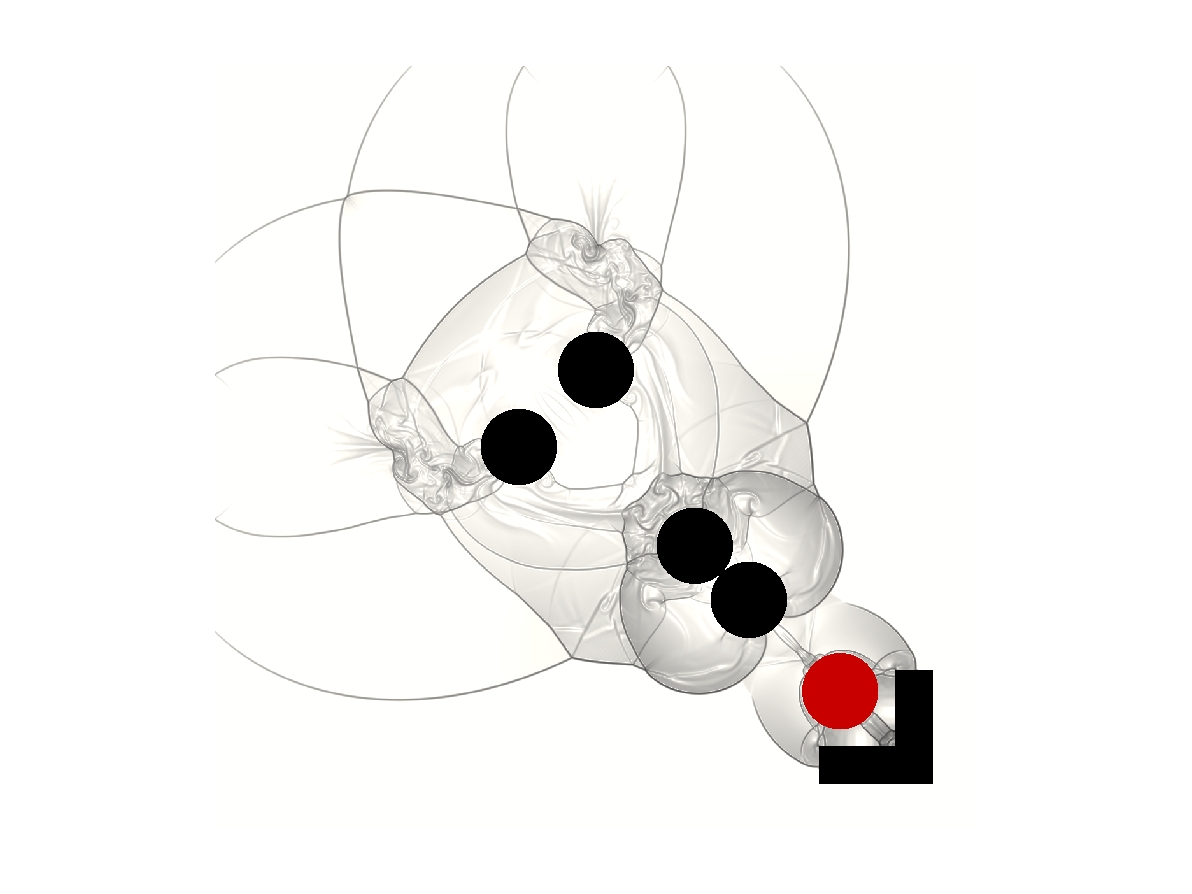}
        \caption{$104 \Unit{ms}$}
        \label{fig:multi_col_wall_m1200_t104ms}
    \end{subfigure}%
    \\
    \begin{subfigure}[b]{0.22\textwidth}
        \includegraphics[trim = 40mm 0mm 40mm 0mm, clip, width=\textwidth]{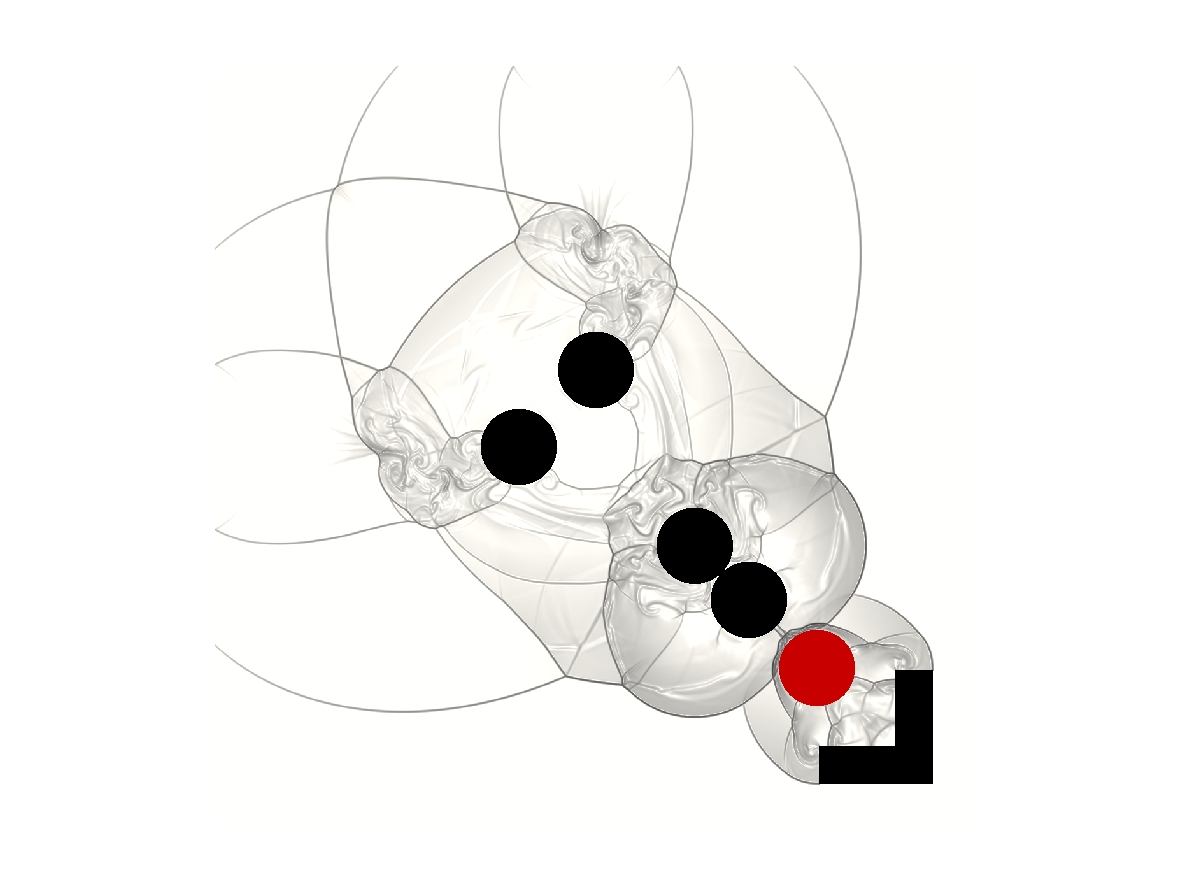}
        \caption{$110 \Unit{ms}$}
        \label{fig:multi_col_wall_m1200_t110ms}
    \end{subfigure}%
    ~
    \begin{subfigure}[b]{0.22\textwidth}
        \includegraphics[trim = 40mm 0mm 40mm 0mm, clip, width=\textwidth]{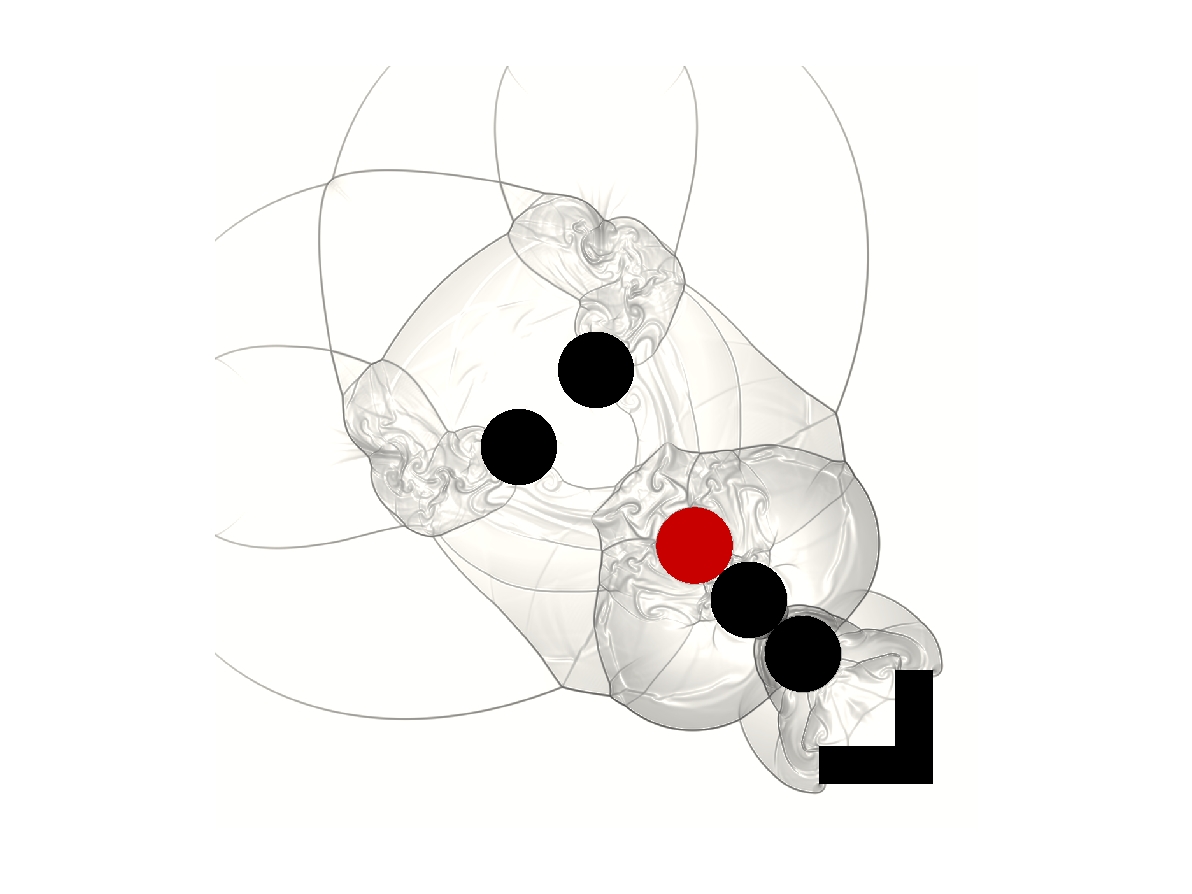}
        \caption{$114 \Unit{ms}$}
        \label{fig:multi_col_wall_m1200_t114ms}
    \end{subfigure}%
    ~
    \begin{subfigure}[b]{0.22\textwidth}
        \includegraphics[trim = 40mm 0mm 40mm 0mm, clip, width=\textwidth]{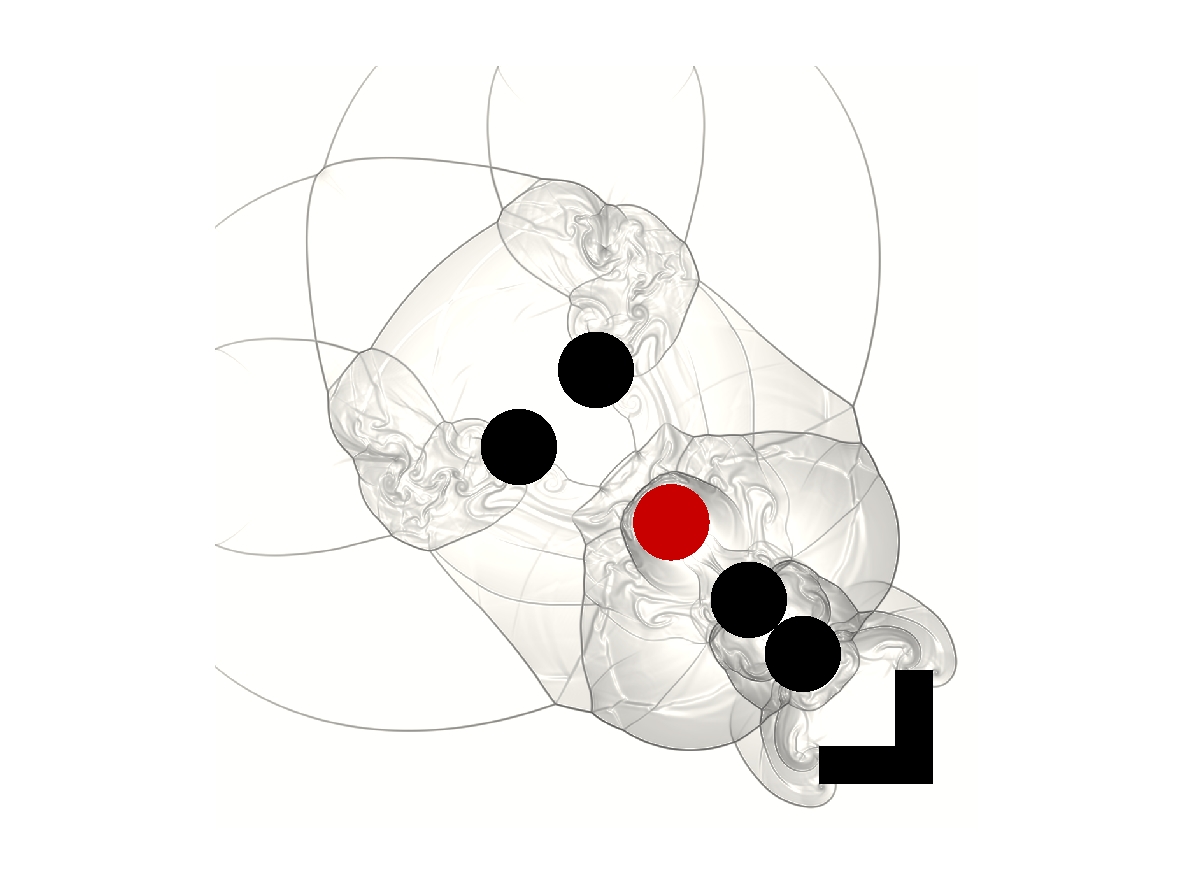}
        \caption{$120 \Unit{ms}$}
        \label{fig:multi_col_wall_m1200_t120ms}
    \end{subfigure}%
    \\
    \begin{subfigure}[b]{0.22\textwidth}
        \includegraphics[trim = 40mm 0mm 40mm 0mm, clip, width=\textwidth]{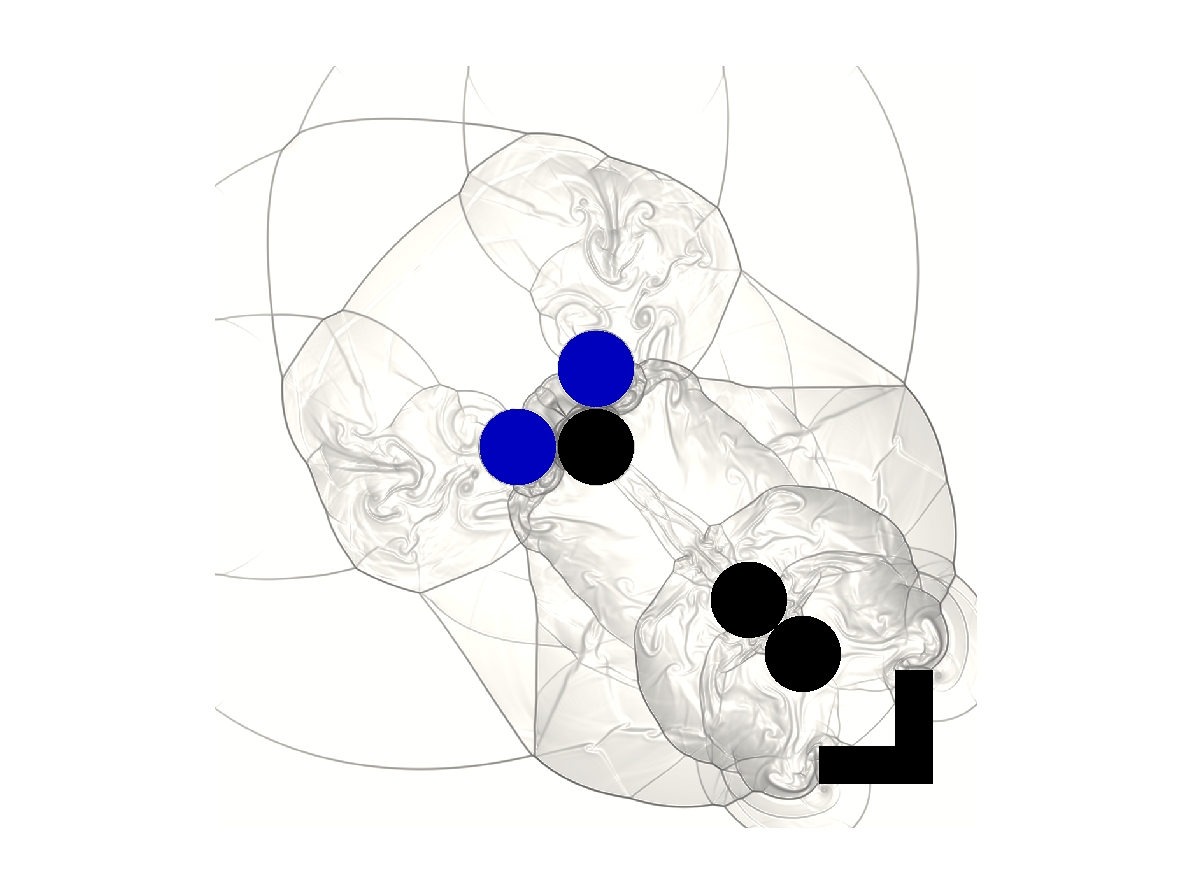}
        \caption{$140 \Unit{ms}$}
        \label{fig:multi_col_wall_m1200_t140ms}
    \end{subfigure}%
    ~
    \begin{subfigure}[b]{0.22\textwidth}
        \includegraphics[trim = 40mm 0mm 40mm 0mm, clip, width=\textwidth]{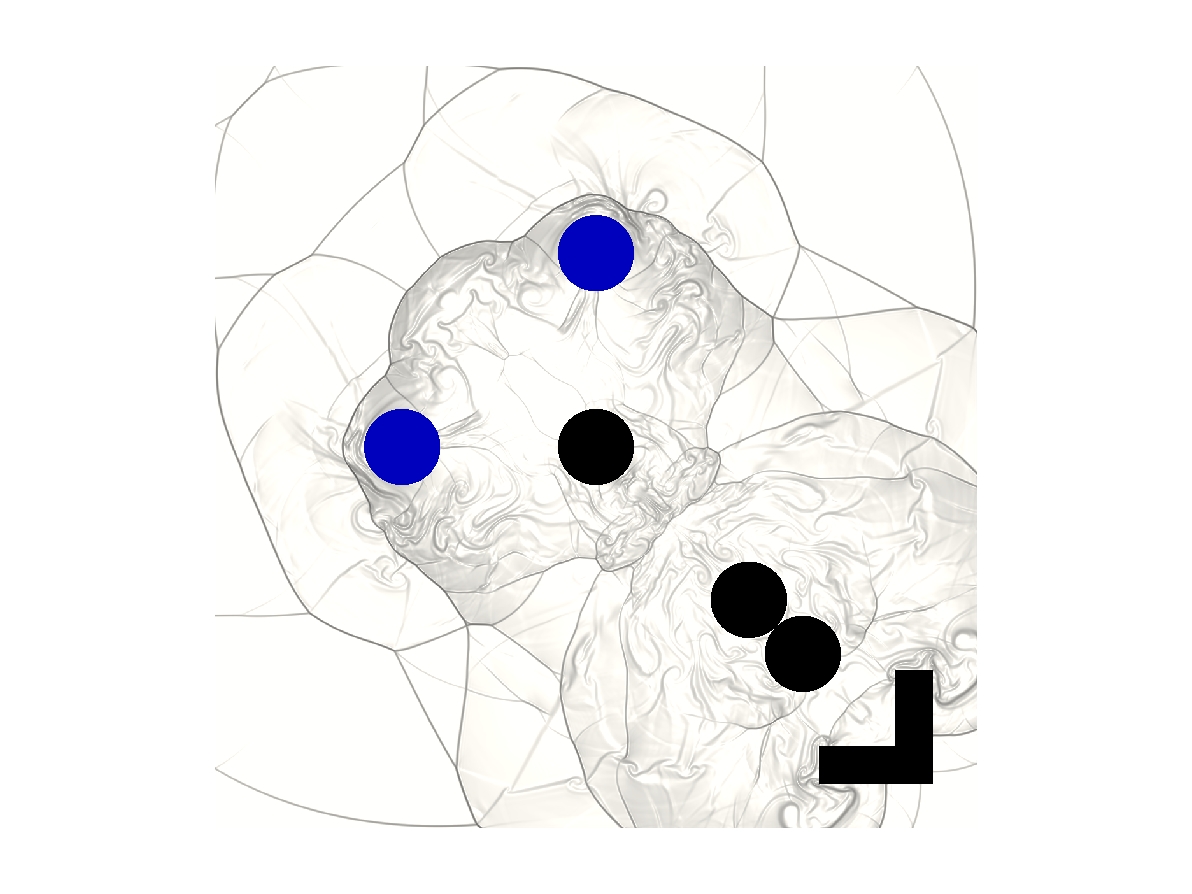}
        \caption{$170 \Unit{ms}$}
        \label{fig:multi_col_wall_m1200_t170ms}
    \end{subfigure}%
    ~
    \begin{subfigure}[b]{0.22\textwidth}
        \includegraphics[trim = 40mm 0mm 40mm 0mm, clip, width=\textwidth]{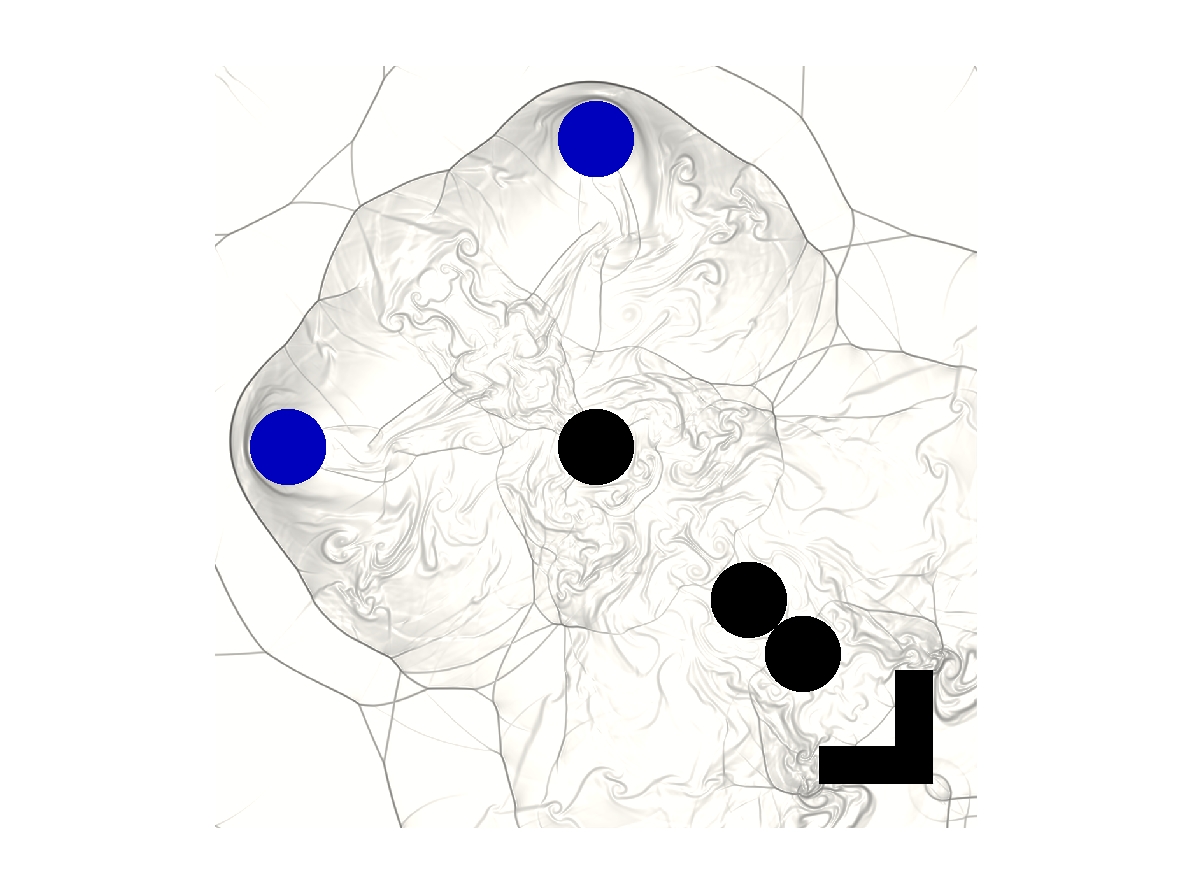}
        \caption{$200 \Unit{ms}$}
        \label{fig:multi_col_wall_m1200_t200ms}
    \end{subfigure}%
    \\
    \begin{subfigure}[b]{0.50\textwidth}
        \includegraphics[trim = 40mm 150mm 40mm 120mm, clip, width=\textwidth]{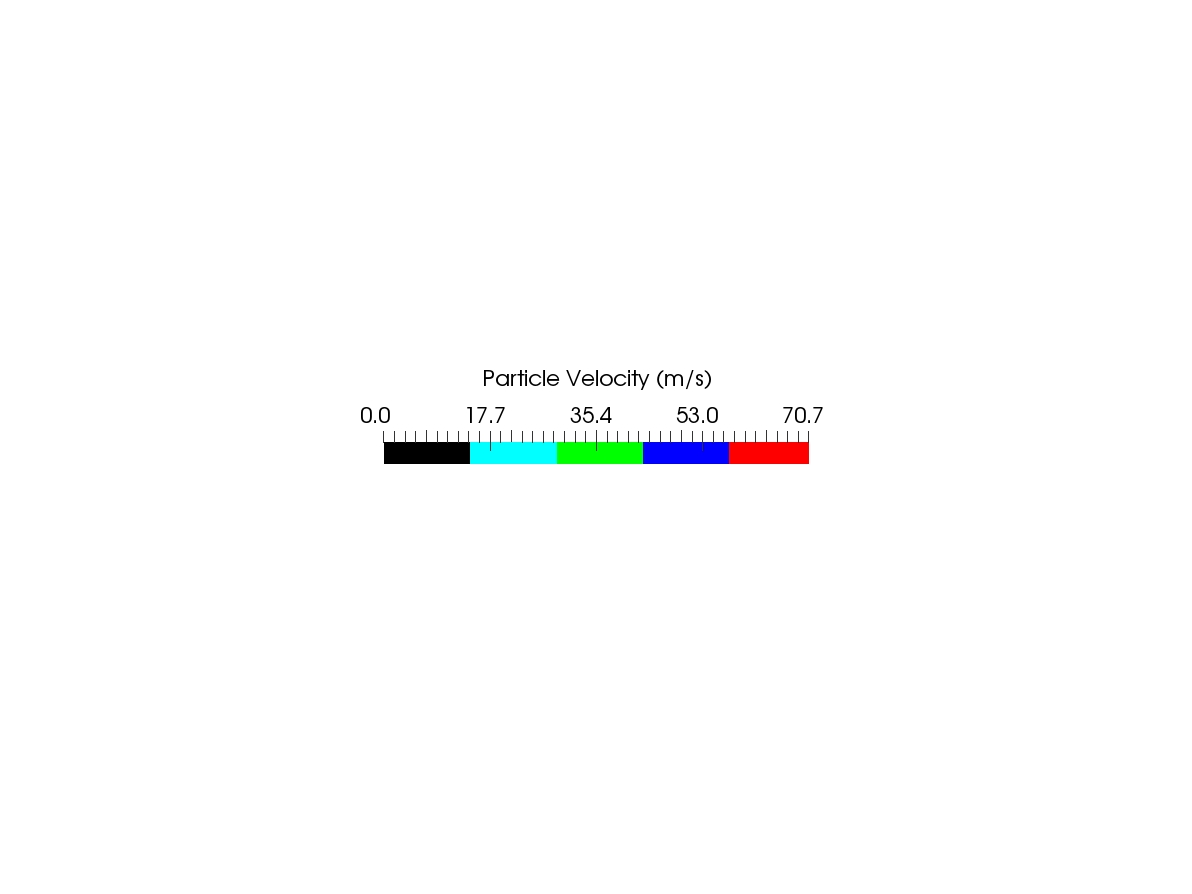}
    \end{subfigure}%
    \caption{Numerical solution of a fluid-solid system with analytically solvable multibody contact and collisions.}
    \label{fig:multibody_collision}
\end{figure}

As captured in Fig.~\ref{fig:multibody_collision}, the proposed field function can successfully facilitate the solution of the fluid-solid system, and the collision detection and response algorithms based on the field function can correctly resolve the multibody contact and collisions. In the numerical results recorded in Table~\ref{tab:multibody_collision}, the velocity states of all the solids are predicted exactly. The maximum position error of the solid centers happens at $C_1$ and $C_2$ and is $2.9\%$ relative to the diameter $D$. During the solid collisions, although solid states instantly switch between $M=0$ and $M=5\sqrt{2}$, the presented framework manages these challenging conditions successfully. As an ill-posed problem, multibody collision in general is an unsolvable problem unless additional assumptions are imposed \citep{ivanov1995multiple}. Nonetheless, the current collision model provides a deterministic approach for approximating multibody collision response with parameterized elasticity and friction and, more importantly, effectively supports the validation of the collision detection capability of the proposed field function.
\begin{table}[!htbp]
    \centering
    \begin{tabular}{lcccc}
        \hline\hline
         & $\Vector{x}_{\Des{c}}^{t=0 \Unit{ms}} (\Unit{D})$ & $\Vector{x}_{\Des{c}}^{t=200 \Unit{ms}} (\Unit{D})$ & $\Vector{V}^{t=0 \Unit{ms}} (\Unit{m/s})$ & $\Vector{V}^{t=200 \Unit{ms}} (\Unit{m/s})$\\
        \hline
         $C_1$ & $(-4, 0)$ & $(0, -4 - 2.9\mathrm{e}{-2})$ & $(50, 0)$ & $(-50, 0)$\\
         $C_2$ & $(0, 4)$ & $(0, 4 + 2.9\mathrm{e}{-2})$ & $(0, -50)$ & $(0, 50)$\\
         $C_3$ & $(0, 0)$ & $(0 - 3.0\mathrm{e}{-5}, 0 + 3.0\mathrm{e}{-5})$ & $(0, 0)$ & $(0, 0)$\\
         $C_4$ & $(2, -2)$ & $(2 + 2.0\mathrm{e}{-5}, -2 - 2.0\mathrm{e}{-5})$ & $(0, 0)$ & $(0, 0)$\\
         $C_5$ & $(2.7, -2.7)$ & $(2.7 + 4.9\mathrm{e}{-4}, -2.7 - 4.9\mathrm{e}{-4})$ & $(0, 0)$ & $(0, 0)$\\
        \hline\hline
    \end{tabular}
    \caption{Comparison of initial and end data regarding solid position and velocity for the fluid-solid system.}
    \label{tab:multibody_collision}
\end{table}

\subsection{Supersonic wedge penetrating a particle bed}

\begin{figure}[!htbp]
    \centering
    \includegraphics[width=0.8\textwidth]{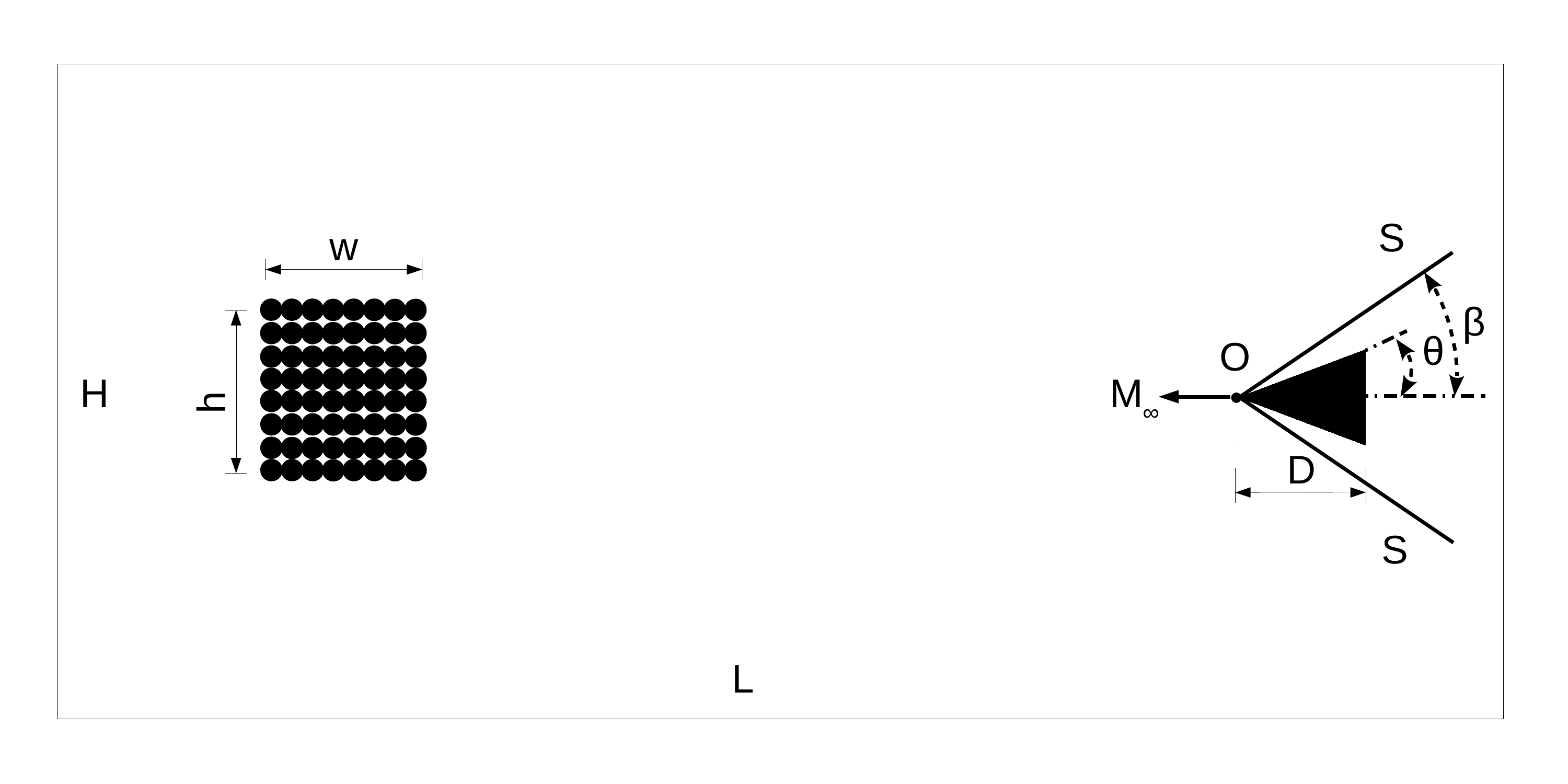}
    \caption{A schematic diagram for the supersonic wedge penetrating a particle bed problem. [Nomenclature: $M_{\infty}$, Mach number of the moving wedge; $S$, oblique shock; $\theta$, deflection angle; $\beta$, shock angle; $D$, length of wedge; $O$, the front vertex of wedge; $L$, domain length; $H$, domain height; $w$, particle bed width; $h$, particle bed height.]}
    \label{fig:wedge_penetrating_demo}
\end{figure}

A supersonic wedge penetrating a particle bed is simulated to further demonstrate the applicability of the field function for solving complex and dynamic fluid-solid systems. As illustrated in Fig.~\ref{fig:wedge_penetrating_demo}, in a $L \times H = [-0.5D, 13.5D] \times [-3.5D, 3.5D]$ domain, a wedge with length $D=1 \Unit{m}$ and deflection angle $\theta=15^{\circ}$ is horizontally positioned in the domain, and the front vertex of the wedge locates at $O(12D, 0)$. In addition, in the $w \times h = [2D, 4D] \times [-1D, 1D]$ region, $64$ identical cylindrical particles with diameter $d=0.25D$ are tightly packed.

Initially, the gas in the domain has the state $(\rho_0, u_0, v_0, p_0)=(1.4 \Unit{kg/m^3}, 0, 0, 400 \Unit{Pa})$, in which the speed of sound is $a_0=20 \Unit{m/s}$. The wedge has a density $\rho_{\Des{s}}=2700 \Unit{kg/m^3}$, a coefficient of restitution $C_R=0.5$, and an initial velocity $M_{\Des{\infty}}=3$. The particles have a density $\rho_{\Des{s}}$, a coefficient of restitution $C_R=0.0$, and zero initial velocity. The flow inside the domain is assumed to be inviscid. The slip-wall condition is imposed at the top and bottom domain boundaries as well as the wedge and particle surfaces, while the outflow condition is enforced at the left and right domain boundaries. The evolution of this fluid-solid system is solved to $t=0.25 \Unit{s}$ on a $2800\times1400$ Cartesian grid.

During the solution process, before the wedge collides with the particle bed (for $t \le 4/30 \Unit{s}$), the fluid forces acting on the wedge are deactivated such that the wedge can move with a constant supersonic speed. As a result, oblique shock waves generated at the nose of the moving wedge can reach a steady state with a constant shock angle $\beta$. The $M_{\infty}-\theta-\beta$ relation satisfies the following analytical formulation \citep{anderson2010fundamentals}:
\begin{equation}
    \tan \theta = \frac{2}{\tan \beta} \frac{M_{\infty}^2 \sin^2 \beta - 1}{M_{\infty}^2 (\gamma + \cos(2\beta)) + 2}
\end{equation}
where $\gamma$ is the heat capacity ratio.

\begin{figure}[!htbp]
    \centering
    \begin{subfigure}[b]{0.48\textwidth}
        \includegraphics[width=\textwidth]{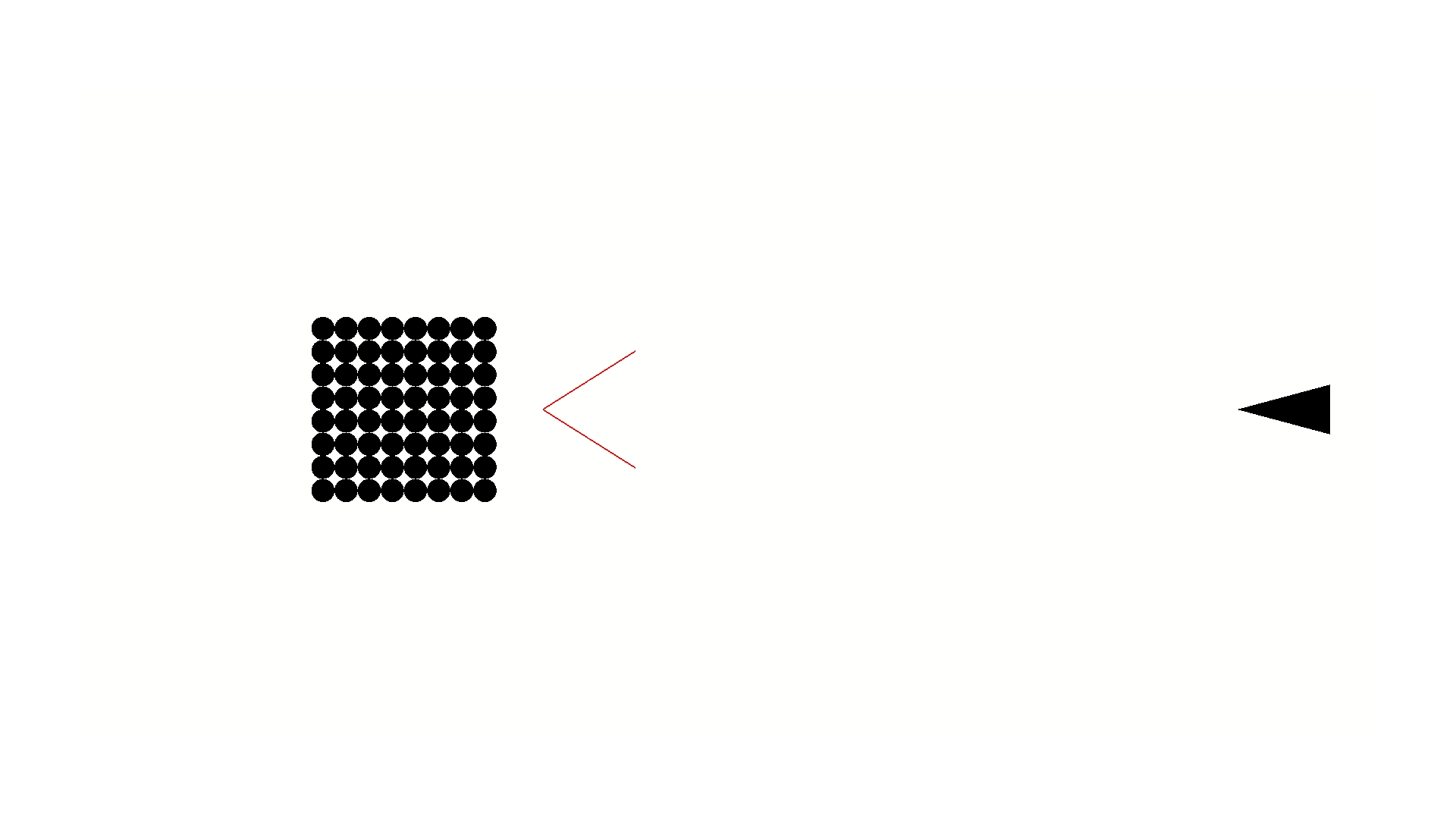}
        \caption{$0 \Unit{ms}$}
        \label{fig:1_wedge_impact_deg15_mach3_cr0d00_t0d000}
    \end{subfigure}%
    ~
    \begin{subfigure}[b]{0.48\textwidth}
        \includegraphics[width=\textwidth]{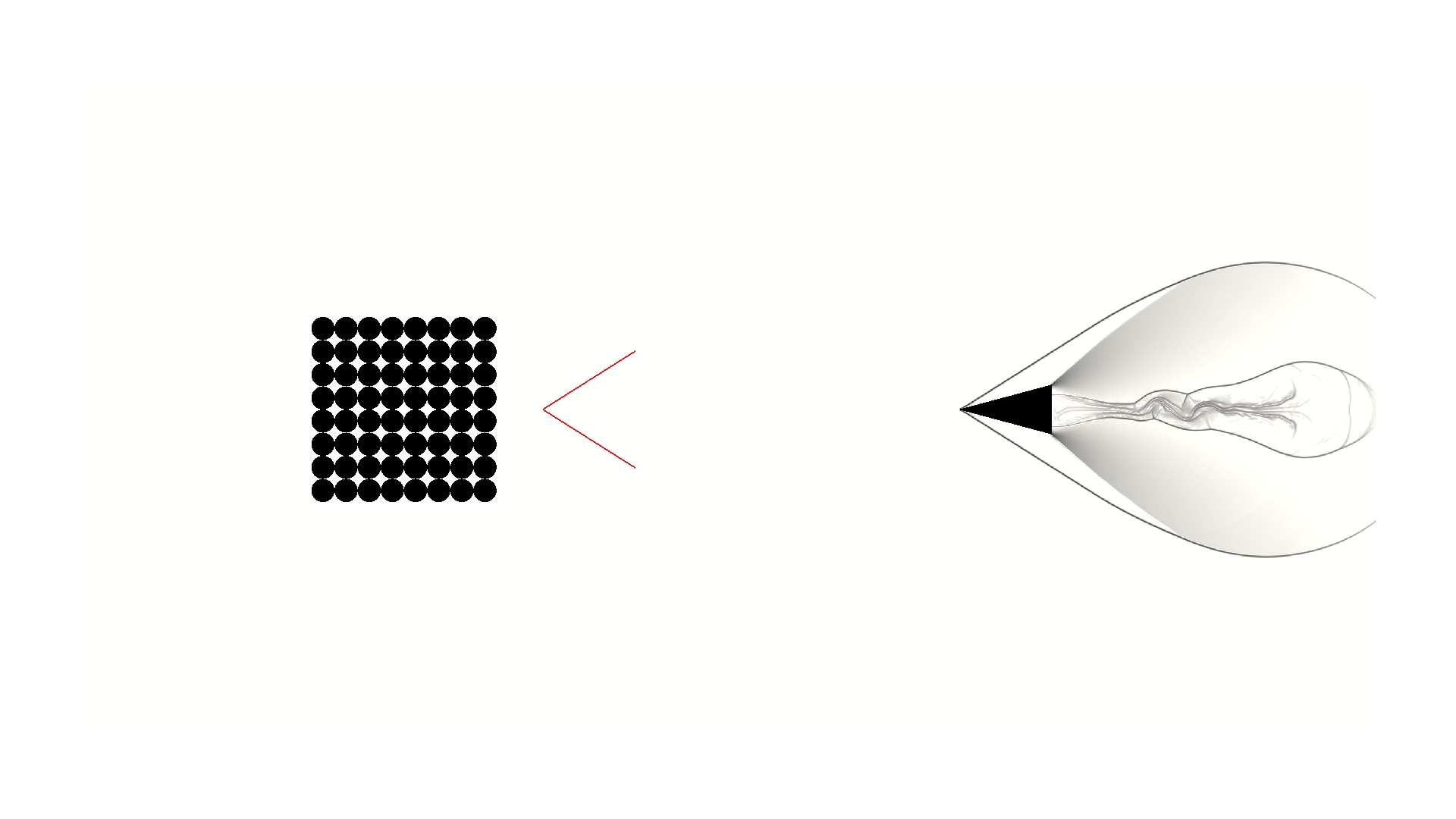}
        \caption{$50 \Unit{ms}$}
        \label{fig:1_wedge_impact_deg15_mach3_cr0d00_t0d050}
    \end{subfigure}%
    \\
    \begin{subfigure}[b]{0.48\textwidth}
        \includegraphics[width=\textwidth]{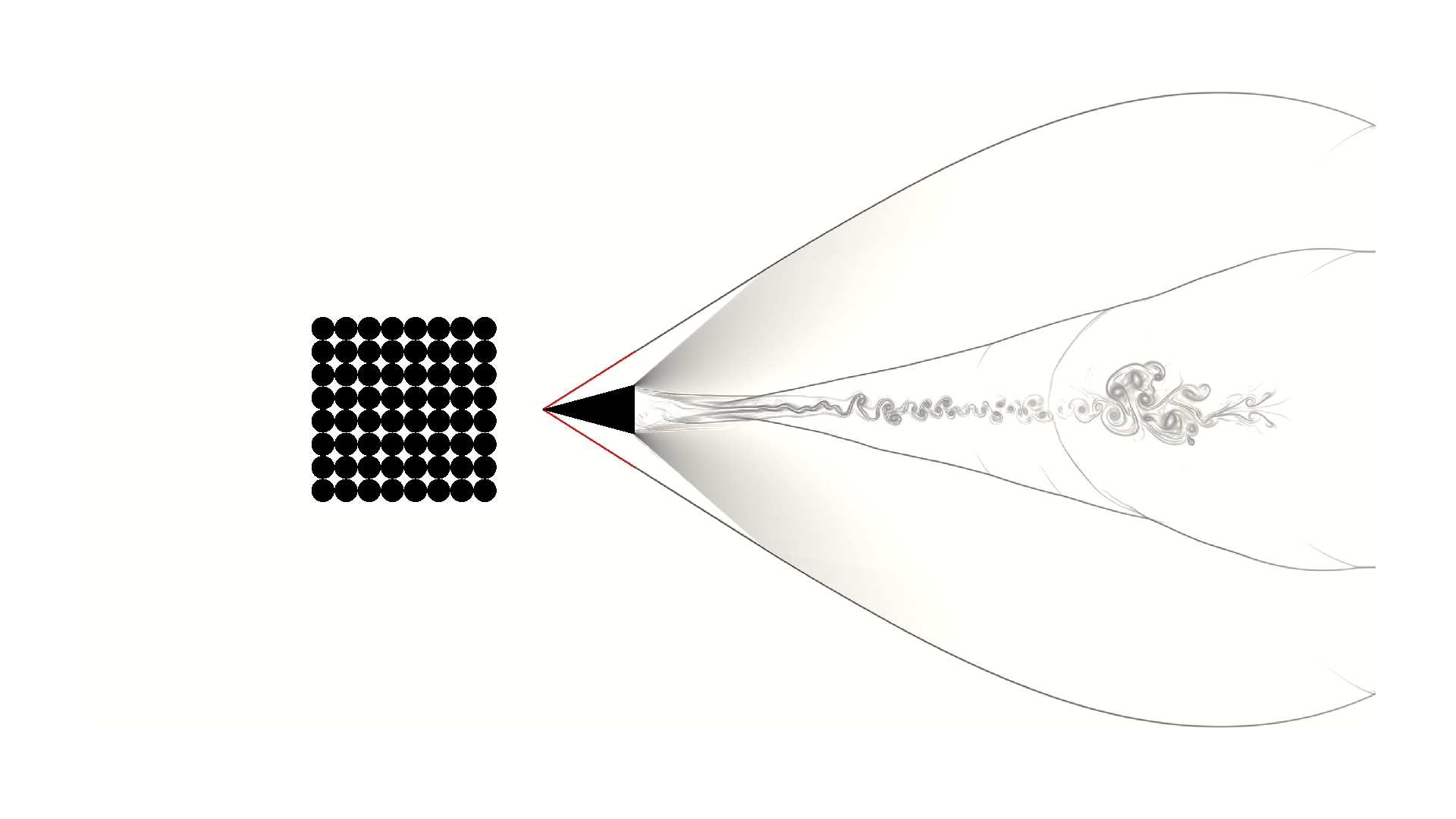}
        \caption{$125 \Unit{ms}$}
        \label{fig:1_wedge_impact_deg15_mach3_cr0d00_t0d125}
    \end{subfigure}%
    ~
    \begin{subfigure}[b]{0.48\textwidth}
        \includegraphics[width=\textwidth]{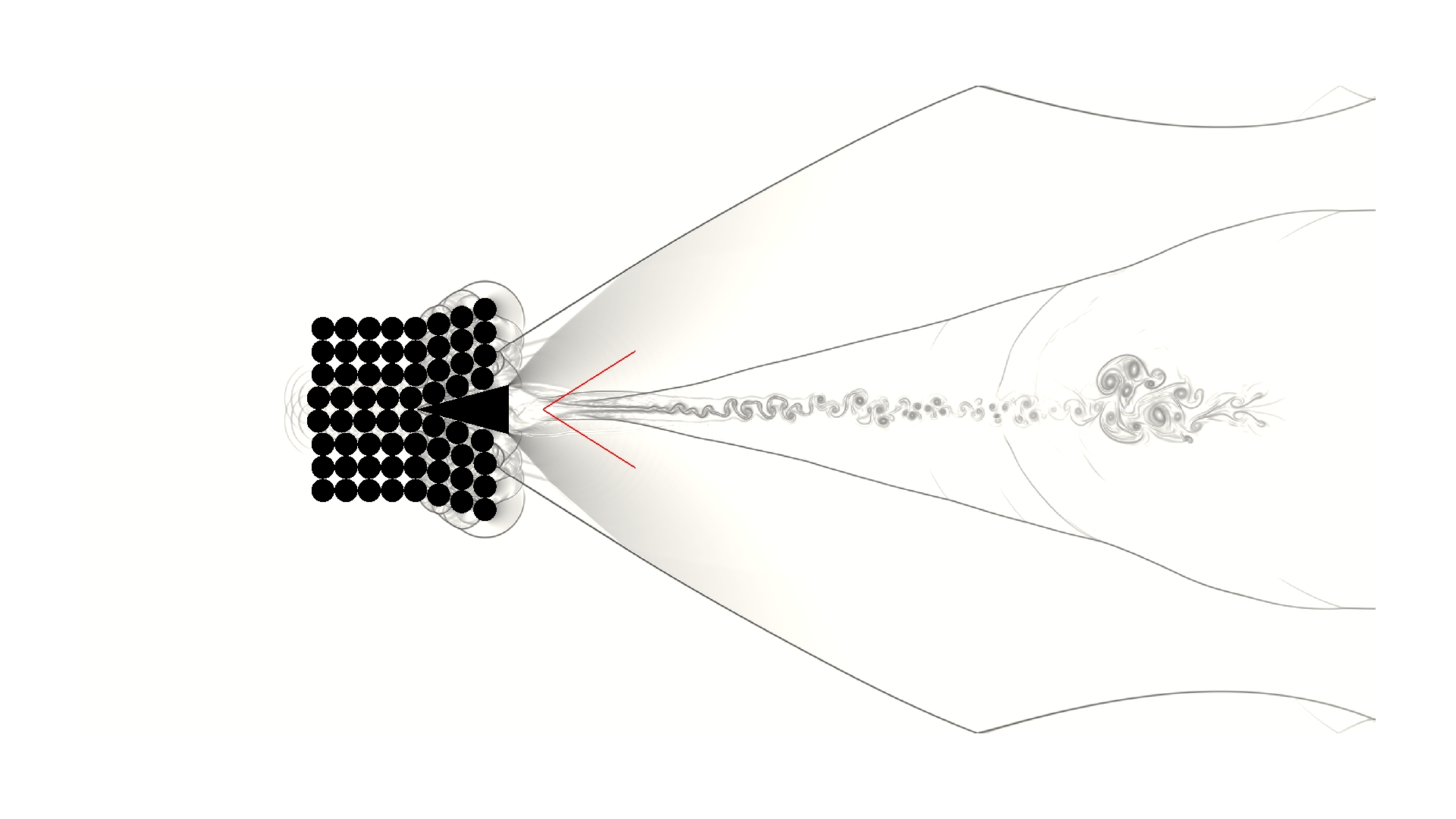}
        \caption{$150 \Unit{ms}$}
        \label{fig:1_wedge_impact_deg15_mach3_cr0d00_t0d150}
    \end{subfigure}%
    \\
    \begin{subfigure}[b]{0.48\textwidth}
        \includegraphics[width=\textwidth]{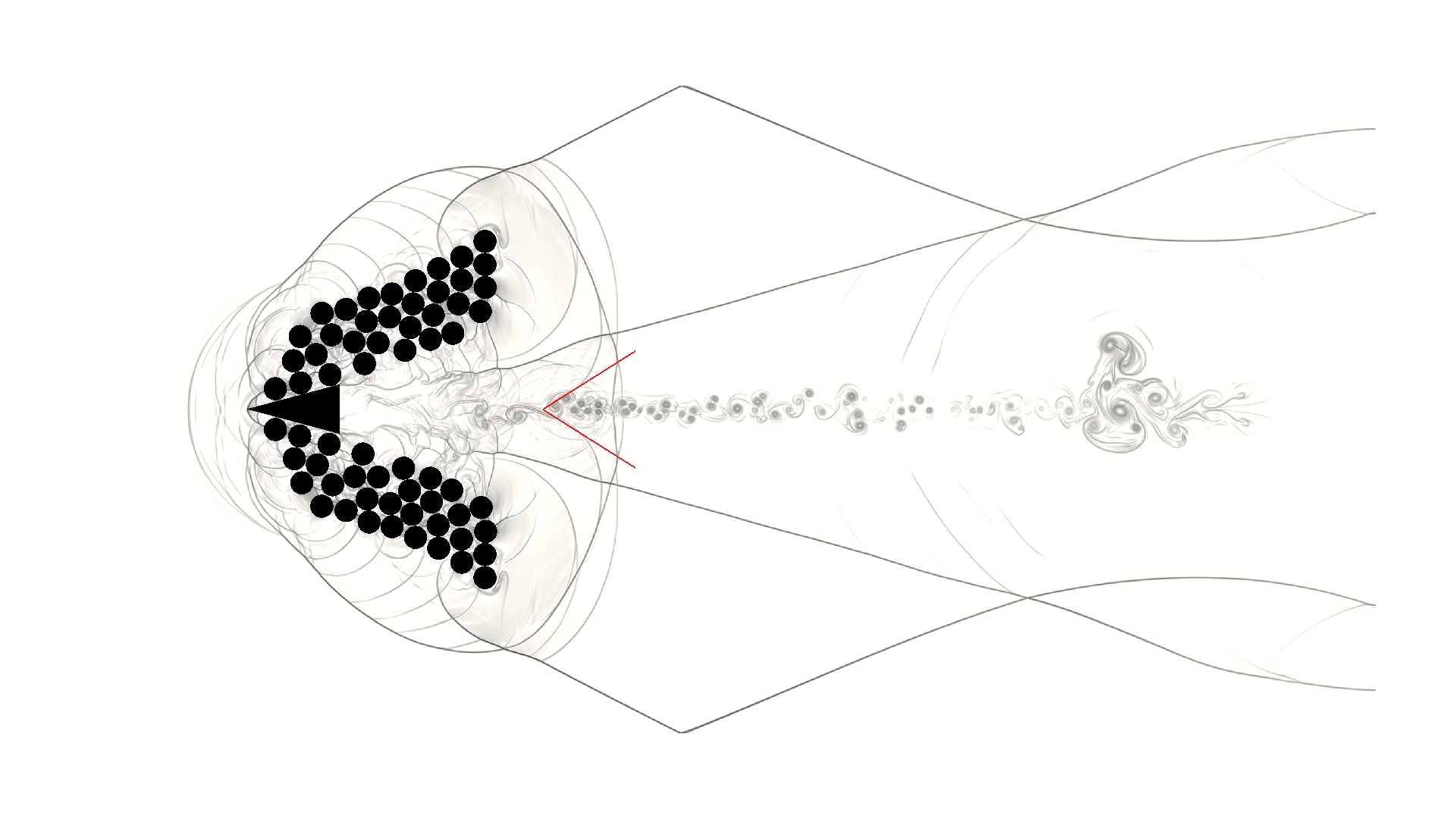}
        \caption{$200 \Unit{ms}$}
        \label{fig:1_wedge_impact_deg15_mach3_cr0d00_t0d200}
    \end{subfigure}%
    ~
    \begin{subfigure}[b]{0.48\textwidth}
        \includegraphics[width=\textwidth]{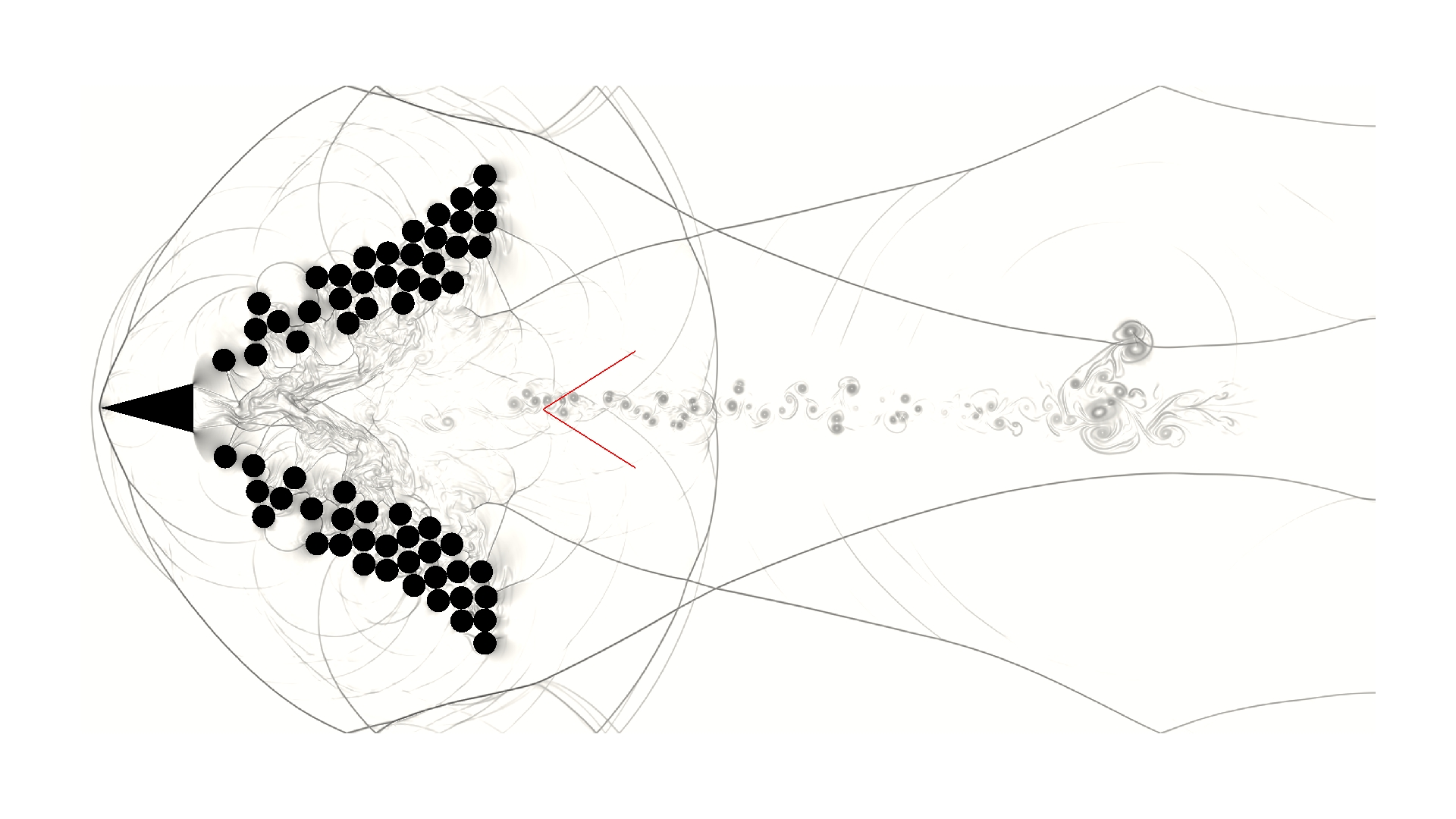}
        \caption{$250 \Unit{ms}$}
        \label{fig:1_wedge_impact_deg15_mach3_cr0d00_t0d250}
    \end{subfigure}%
    \caption{Time evolution of a supersonic wedge penetrating a particle bed. Red lines represent the analytical solutions of the oblique shocks.}
    \label{fig:1_wedge_impact_deg15_mach3_cr0d00_t}
\end{figure}

The simulated time evolution of the system is captured in Fig.~\ref{fig:1_wedge_impact_deg15_mach3_cr0d00_t}, in which the red lines represent the analytical solutions of the shock angles of the oblique shocks at position $(4.5D, 0)$. As shown in Fig.~\ref{fig:1_wedge_impact_deg15_mach3_cr0d00_t0d125}, the predicted oblique shock angle $\beta_n = 32.259^{\circ}$ agrees very well with the analytical solution $\beta_{e}= 32.240^{\circ}$. After the wedge collides with the particle bed, a force chain within the contacted particles is created due to the penetrating wedge. This force chain accelerates the particles and fractures the particle bed. The suddenly destabilized particle bed generates strong flow disturbances at the surrounding area, which interact with the wedge generated shocks and waves, forming complex wave diffraction and interference patterns in space. During the wedge penetrating the particle bed, intensive multibody contact and collisions are successfully simulated, and an intuitive dynamic process is presented in the solution. Moreover, the $y$-plane symmetry is well preserved for the entire penetrating process. These results illustrate the ability of the presented field function for facilitating the solution of complex and dynamic fluid-solid systems involving coupled fluid-fluid, fluid-solid, and solid-solid interactions.

\section{Conclusion}

An integer-type field function has been developed to facilitate the solution of complex and dynamic fluid-solid systems on Cartesian grids with interface-resolved fluid-fluid, fluid-solid, and solid-solid interactions. The main conclusions are summarized as the following:
\begin{itemize}
    \item For a Cartesian-grid-discretized computational domain segmented by a set of solid bodies, the developed field function explicitly tracks each subdomain with multiple resolved interfacial node layers. Therefore, it is straightforward to enforce designated governing equations, constitutive models, numerical schemes, and interface conditions for each subdomain.

    \item The presented field function enables low-memory-cost multidomain node mapping, efficient node remapping, fast collision detection, and expedient surface force integration for computing fluid-solid systems. Easy-to-implement algorithms for the field function and its described functionalities are also presented. The proposed node mapping algorithm effectively solves a generalized point-in-polyhedron problem represented by a set of points together with a set of polyhedrons. In addition, this algorithm successfully unifies the procedures of the initial multidomain node mapping and the subsequent node remapping, which can simplify the code structure and reduce the complexity of implementation.

    \item Equipped with a deterministic multibody collision model, the applicability of the developed field function for solving complex and dynamic fluid-solid systems is validated and illustrated through numerical experiments ranging from subsonic to supersonic flows, such as the subsonic flow around a cylinder, supersonic shock-sphere interaction, a multibody contact and collision system, and a supersonic wedge penetrating a particle bed. The obtained numerical results are all in close agreement with the corresponding published numerical data, experimental observations, or analytical solutions.
\end{itemize}

\section*{Acknowledgements}

Financial support of this work was provided by Natural Sciences and Engineering Research Council of Canada (NSERC) and Defence Research and Development Canada (DRDC). This work was made possible by the facilities of the Shared Hierarchical Academic Research Computing Network (SHARCNET: www.sharcnet.ca) and Compute/Calcul Canada.


\bibliography{ref}

\end{document}